\documentclass[twocolumn]{aastex61}
\usepackage{graphicx, graphics, amsmath,amssymb, amsfonts}
\usepackage{dcolumn}
\usepackage{bm}
\usepackage{natbib}
\usepackage{enumerate}

\begin{document}

\title{Active modes and dynamical balances in MRI-turbulence of Keplerian disks with a net vertical magnetic field}

\correspondingauthor{G. Mamatsashvili}
\email{g.mamatsashvili@hzdr.de}

\author{D. Gogichaishvili}
\affiliation{Department of Physics, The University of Texas at
Austin, Austin, Texas 78712, USA}

\author{G. Mamatsashvili}
\affiliation{Niels Bohr International Academy, Niels Bohr Institute,
Blegdamsvej 17, 2100 Copenhagen, Denmark}
\affiliation{Helmholtz-Zentrum Dresden-Rossendorf, Dresden D-01328,
Germany} \affiliation{Abastumani Astrophysical Observatory, Ilia
State University, Tbilisi 0162, Georgia} \affiliation{Institute of
Geophysics, Tbilisi State University, Tbilisi 0193, Georgia}

\author{W. Horton}
\affiliation{Institute for Fusion Studies, The University of Texas
at Austin, Austin, Texas 78712, USA} \affiliation{Space and
Geophysics Laboratory, The University of Texas at Austin, 10000
Burnet Rd, Austin, Texas 78758, USA}

\author{G. Chagelishvili}
\affiliation{Abastumani Astrophysical Observatory, Ilia State
University, Tbilisi 0162, Georgia}\affiliation{Institute of
Geophysics, Tbilisi State University, Tbilisi 0193, Georgia}

\begin{abstract}
We studied dynamical balances in magnetorotational instability (MRI)
turbulence with net vertical field in the shearing box model of
disks. Analyzing the turbulence dynamics in Fourier (${\bf
k}$-)space, we identified three types of active modes that define
turbulence characteristics. These modes have lengths similar to the
box size, i.e., lie in the small wavenumber region in Fourier space
labeled \emph{the vital area} and are: \emph{(i)} the channel mode
-- uniform in the disk plane with the smallest vertical wavenumber,
\emph{(ii)} the zonal flow mode -- azimuthally and vertically
uniform with the smallest radial wavenumber and \emph{(iii)}
\emph{the rest modes}. The rest modes comprise those harmonics in
the vital area whose energies reach more than $50 \%$ of the maximum
spectral energy. The rest modes individually are not so significant
compared to the channel and zonal flow modes, however, the combined
action of their multitude is dominant over these two modes. These
three mode types are governed by interplay of the linear and
nonlinear processes, leading to their interdependent dynamics. The
linear processes consist in disk flow nonmodality-modified classical
MRI with a net vertical field. The main nonlinear process is
transfer of modes over wavevector angles in Fourier space --
\textit{the transverse cascade}. The channel mode exhibits episodic
bursts supplied by linear MRI growth, while the nonlinear processes
mostly oppose this, draining the channel energy and redistributing
it to the rest modes. As for the zonal flow, it does not have a
linear source and is fed by nonlinear interactions of the rest
modes.
\end{abstract}

\section{Introduction}
\label{sec:Introduction}

Anisotropic turbulence offers a means of enhanced transport of
angular momentum in astrophysical disks
\citep{Shakura_Sunyaev73,Lynden-Bell_Pringle74}, whereas isotropic
turbulence is unable to ensure such a transport. It is not
surprising that from the 1980s, research in astrophysical disks
focused on identifying sources of turbulence and understanding its
statistical characteristics. The turning point was the beginning of
the 1990s, when a linear instability mediated by a weak vertical
magnetic field in differentially rotating conducting fluids
\citep{Velikhov59,Chandrasekhar60} -- subsequently named as the
magnetorotational instability (MRI) -- was rediscovered for
sufficiently ionized astrophysical disks by \citet{Balbus_Hawley91}.
MRI is a robust dynamical instability that leads to the exponential
growth of axisymmetric perturbations, gives rise to and steadily
supplies with energy magnetohydrodynamic (MHD) turbulence, as was
demonstrated in earlier numerical simulations shortly after the
significance of linear MRI in disks had been realized
\citep[e.g.,][]{Hawley_Balbus91,Hawley_Balbus92,Hawley_etal95,Brandenburg_etal95,Balbus_Hawley98}.

The generic nonnormality (non-self-adjointness) of a strongly
sheared Keplerian flow of astrophysical disks provides an additional
important linear mechanism of energy supply to the turbulence -
transient, or nonmodal growth of perturbations
\citep{Lominadze_etal88,
Chagelishvili_etal03,Yecko04,Afshordi_etal05,Tevzadze_etal08,
Shtemler_etal11,Salhi_etal12,Pessah_Chan12,Mamatsashvili_etal13,
Zhuravlev_Razdoburdin14,Squire_Bhattacharjee14,Razdoburdin_Zhuravlev17}.
Although the nonmodal growth is generally less powerful than the
classical (exponentially growing) MRI, it is nevertheless capable of
driving quite robust MHD turbulence when the latter is absent, for
instance, in disks threaded by a purely azimuthal/toroidal magnetic
field \citep[e.g.,][]{Hawley_etal95,Fromang_Nelson06,
Simon_Hawley09,Guan_etal09,Guan_Gammie11,Flock_etal12a,Nauman_Blackman14,
Meheut_etal15,Gogichaishvili_etal17}. In such spectrally, or modally
stable magnetized disk flows lacking exponentially growing MRI
modes, nonlinear processes are vital for turbulence sustenance. In
this situation, the canonical -- direct and inverse -- cascade
processes in classical Kolmogorov or Iroshnikov-Kraichnan
phenomenologies are not capable of sustaining the turbulence -- this
role is taken over by a new kind of the cascade process, so-called
\emph{the nonlinear transverse cascade}
\citep{Horton_etal10,Mamatsashvili_etal14}. What is the physical
reason for the emergence of the transverse cascade in shear flows
and its specific nature? The thing is that the anisotropy of the
nonnormality/shear-induced linear nonmodal dynamics entails the
anisotropy of the nonlinear processes -- transverse, or angular
redistribution of perturbation harmonics in Fourier (wavenumber)
space. In spectrally stable shear flows, in which the only mechanism
for the energy supply to perturbations is the linear transient
growth process, the nonlinear transverse cascade, by transferring
the harmonics over wavevector angles (i.e., changing the orientation
of their wavevectors), continually replenishes those areas in
Fourier space where they can experience transient amplification
\citep{Mamatsashvili_etal14,Mamatsashvili_etal16}. In this way, the
transverse cascade guarantees a long-time sustenance of the
turbulence. In the absence of such a feedback, non-axisymmetric
modes that undergo in this case the most effective nonmodal
transient growth, get eventually sheared away and decay. In the
context of disks, this process was studied in detail in our recent
paper \citet{Gogichaishvili_etal17} (Paper I) for the
above-mentioned case of the Keplerian shear flow with a net
azimuthal magnetic field by combining direct simulations of the
turbulence and, based on them, the subsequent analysis of the
dynamical processes in Fourier space.

In this paper, we consider a Keplerian disk threaded by a net
vertical/poloidal magnetic field, where there exists classical MRI
with an exponential growth of axisymmetric modes
\citep{Balbus_Hawley91,Balbus_Hawley98,Wardle99,
Pessah_etal06,Lesur_Longaretti07,Pessah_Chan08,Longaretti_Lesur10,
Latter_etal15,Shakura_Postnov15}. Because of this, there is no
deficit in energy supply and the role of nonlinearity in the
sustenance of perturbations is not as vital as in the case of
azimuthal field. However, at the same time, MRI also owes its
existence to the shear of disk flow and therefore is inevitably
subject to nonmodal effects
\citep{Mamatsashvili_etal13,Squire_Bhattacharjee14}. As demonstrated
in this paper, this has two important consequences. First, the
nonmodally-modified, or for short nonmodal MRI growth of both
axisymmetric and non-axisymmetric modes during finite times are in
fact more important in the energy supply process of turbulence,
because the main time scales involved are of the order of
dynamical/orbital time \citep{Walker_etal16}, than the modal
(exponential) growth of axisymmetric modes prevalent at large times
\citep[see also][]{Squire_Bhattacharjee14}. Second, underlying
nonlinear processes are necessarily anisotropic in Fourier space as
a result of the inherent anisotropy of linear nonmodal dynamics due
to the shear. This anisotropy first of all gives rise to the
nonlinear transverse cascade and one of the main goals of the
present paper is to vividly demonstrate its importance in forming
the overall dynamical picture of net vertical field MRI-turbulence.

Recently, \citet{Murphy_Pessah15} investigated the saturation of MRI
in disks with a net vertical field and the properties of the ensuing
MHD turbulence both in physical and Fourier space. Their study was
mainly devoted to characterizing the anisotropic nature of this
turbulence. They also pointed out a general lack of analysis of the
anisotropy in the existing literature on MRI-turbulence: ``Although
there have been many studies of the linear phase of the MRI and its
nonlinear evolution, only a fraction have explored the mechanism
responsible for its saturation in detail, and none have focused
explicitly on the evolution of the degree of anisotropy exhibited by
the magnetized flow as it evolves from the linear regime of the
instability to the ensuing turbulent state''. The main reason for
overlooking the anisotropic nature of MRI-driven turbulence in most
of previous works that focused on its spectral dynamics
\citep[e.g.,][]{Fromang_Papaloizou07,Simon_etal09,Davis_etal10,Lesur_Longaretti11}
was a somewhat misleading mathematical treatment, specifically,
spherical shell-averaging procedure in Fourier space \citep[borrowed
from forced MHD turbulence studies without shear flow, see
e.g.,][]{Verma04,Alexakis_etal07}, which had been employed to
extract statistical information about the properties of
MRI-turbulence. Obviously, the use of the shell-averaging technique,
which in fact smears out the transverse cascade, is, strictly
speaking, justified for isotropic turbulence, but by no means for
shear flow turbulence and therefore for, its special case,
MRI-turbulence ``nourished'' in a sheared environment of disk flow.
Thus, the shell-averaging is not an optimal tool for analyzing
spectra as well as dynamical processes in Fourier space that
underlie MRI-driven turbulence and are far from isotropic due to the
shear \citep[see also][Paper
I]{Hawley_etal95,Nauman_Blackman14,Lesur_Longaretti11,Murphy_Pessah15}.
This fact also calls into question those investigations based on the
shell-averaging that aim to identify a power-law character and
associated slopes of turbulent energy spectrum, because the
anisotropy of the energy spectrum itself, a direct consequence of
the transverse cascade, is wiped out in these cases. This is perhaps
the reason why the kinetic and magnetic energy spectra do not
generally exhibit a well-defined power-law behavior in
MRI-turbulence in disk flows with a net vertical field
\citep{Simon_etal09,Lesur_Longaretti11,Meheut_etal15,Walker_etal16}.

In view of the above, in this work we focus on the dynamics and
balances in MRI-turbulence with a net vertical field. We adopt the
local shearing box model of the disk with constant vertical thermal
stratification. The analysis is performed in three-dimensional (3D)
Fourier space in full, i.e., without doing the averaging of spectral
quantities over spherical shells of given wavenumber magnitude $k =
|{\bf k}|$. This allows us to capture the spectral anisotropy of the
MRI-turbulence due to the shear and the resulting nonlinear angular
redistribution of perturbation modes in Fourier space, i.e.,
transverse cascade, thereby getting a deeper understanding of
spectral and statistical properties of the turbulence. In previous
relevant studies on a net vertical field MRI
\citep[e.g.,][]{Goodman_Xu94,Hawley_etal95,Sano_Inutsuka01,Lesur_Longaretti07,
Bodo_etal08,Latter_etal09,Latter_etal10,Simon_etal09,Pessah_Goodman09,
Longaretti_Lesur10,Pessah10,Bai_Stone14,Murphy_Pessah15}, this
redistribution (scatter) of modes over wavevector orientations
(angles) in Fourier space was attributed to secondary, or
\emph{parasitic} instabilities. Specifically, the most unstable,
exponentially growing axisymmetric MRI modes (channels) are subject
to secondary instabilities of non-axisymmetric modes with growth
rates proportional to the amplitude of these channel solutions. In
this way, the parasitic instabilities redistribute the energy from
the primary axisymmetric channel modes to non-axisymmetric parasitic
ones, halting the exponential growth of the former and leading to
the saturation of MRI. Several approximations are made in this
description: 1. the large amplitude channel mode is a
time-independent background on which the small-amplitude parasitic
modes feed and 2. the effects of the imposed vertical field, the
Coriolis force, and the basic Keplerian shear are all usually
neglected. These assumptions clearly simplify the analysis of the
excitation and dynamics of the non-axisymmetric parasitic modes,
but, more importantly, because of neglecting the basic flow shear,
omit independent from the primary MRI modes source of their support
- the nonmodal growth of non-axisymmetric modes, which, as discussed
above, is an inevitable linear process in shear (disk) flows.

To keep an analysis general and self-consistent, together with the
nonmodal growth process, we employ the concept of the nonlinear
transverse cascade in the present problem of MRI with a net vertical
magnetic field that naturally encompasses the secondary
instabilities too. This unifying framework enables us to correctly
describe the interaction between the channel and non-axisymmetric
``parasitic'' modes when the above assumptions break down -- the
amplitudes of these two mode types become comparable, so that it is
no longer possible to clearly distinguish between the channel as a
primary background and parasites as small perturbations on top of
that. This is the case in the developed turbulent state of vertical
field MRI, where channels undergo recurrent amplifications (bursts)
and decays
\citep[e.g.,][]{Sano_Inutsuka01,Lesur_Longaretti07,Bodo_etal08,Simon_etal09,Murphy_Pessah15}.
This decay phase is usually attributed to the linear
non-axisymmetric parasitic instabilities, however, in the fully
developed turbulent state it is more likely governed by
nonlinearity. As is shown in this paper, the transverse cascade
accounts for the transfer of energy from the axisymmetric channel
modes to a broad spectrum of non-axisymmetric ones (referred to as
the rest modes here) as well as to the axisymmetric zonal flow mode,
which appears to commonly accompany MRI-turbulence
\citep{Johansen_etal09,Simon_etal12,Bai_Stone14}. The nonlinear
transverse cascade may not be the vital source of the energy supply
to turbulence in the presence of purely exponentially growing MRI,
but still it shapes the dynamics, sets the saturation level and
determines the overall ``design'' of the turbulence the ``building
blocks'' of which are these three types of perturbation modes
analyzed in this paper.

In the spirit of our recent works \citep[][Paper I]{Horton_etal10,
Mamatsashvili_etal14,Mamatsashvili_etal16}, here we investigate in
detail the roles of underlying different anisotropic linear and
nonlinear dynamical processes shaping the net vertical field
MRI-turbulence. Namely, we first perform numerical simulations of
the turbulence and then, using the simulation data, explicitly
calculate individual linear and nonlinear terms and explore their
action in Fourier space. The underlying physics, active modes and
dynamical balances of the net vertical field MRI-turbulence in disks
are, however, entirely different from those of the net azimuthal
field one studied in Paper I, primarily because energy sources for
the turbulence in these two field configurations differ in essence:
in the first case, the turbulence is mostly supplied by the
(nonmodally-modified) exponentially growing MRI, while in the second
case just by transient growth of non-axisymmetric modes.

The present study can be regarded as a generalization of the related
works by \citet{Simon_etal09,Lesur_Longaretti11}, where the dynamics
of vertical field MRI-turbulence -- the spectra of energy, injection
and nonlinear transfers -- were analyzed in Fourier space, however,
using a restrictive approach of shell-averaging, which misses out
the shear-induced anisotropy of the turbulence and hence the
interaction of axisymmetric channel and non-axisymmetric modes. It
also extends the study of \citet{Murphy_Pessah15}, who focused on
the anisotropy of vertical field MRI-turbulence in both physical and
Fourier space and examined, in particular, anisotropic spectra of
the magnetic energy and Maxwell stress during the linear growth
stage of the channel solutions and after saturation, but not the
action of the various linear and nonlinear terms governing their
evolution.

The paper is organized as follows. The physical model and main
equations in Fourier space is given in Section
\ref{sec:Basicequations}. The linear nonmodal growth of MRI is
analyzed in Section \ref{sec:Optimal}. Simulations of the ensuing
MRI-turbulence and its general characteristics, such as
time-development, energy spectra and the classification of
dynamically active modes are given in Section \ref{sec:DNS}. In this
section we also give the main analysis of the individual dynamics of
the active modes and their interdependence in Fourier space that
underlie the dynamics of the turbulence. Summary and discussions are
given in Section \ref{sec:Conclusion}.

\section{Physical model and basic equations}
\label{sec:Basicequations}

We investigate the essence of MHD turbulence driven by the classical
MRI in Keplerian disks using a shearing box model, which is located
at a fiducial radius and corotates with the disk at orbital
frequency $\Omega$ \citep{Hawley_etal95}. In this model, the basic
equations of incompressible non-ideal MHD are written in a Cartesian
coordinate frame $(x,y,z)$ with the unit vectors ${\bf e}_x, {\bf
e}_y, {\bf e}_z$, respectively, in the radial, azimuthal and
vertical directions. The fluid is assumed to be thermally stratified
in the vertical direction. Using the Boussinesq approximation for
the stratification, these equations have the form,
\begin{multline}\label{eq:mom}
\frac{\partial {\bf U}}{\partial t}+({\bf U}\cdot \nabla) {\bf
U}=-\frac{1}{\rho}\nabla P +\frac{\left({\bf B}\cdot\nabla
\right){\bf B}}{4\pi \rho} - 2{\bf \Omega}\times{\bf
U}\\+2q{\Omega}^2 x {\bf e}_x - g \theta {\bf e}_z+\nu\nabla^2 {\bf
U},
\end{multline}
\begin{equation}\label{eq:dens}
\frac{\partial \theta}{\partial t}+{\bf U}\cdot \nabla \theta
=u_z\frac{N^2}{g}+\chi \nabla^2 \theta,
\end{equation}
\begin{equation}\label{eq:ind}
\frac{\partial {\bf B}}{\partial t}=\nabla\times \left( {\bf
U}\times {\bf B}\right)+\eta\nabla^2{\bf B},
\end{equation}
\begin{equation}\label{eq:divv}
\nabla\cdot {\bf U}=0,
\end{equation}
\begin{equation}\label{eq:divb}
\nabla\cdot {\bf B}=0,
\end{equation}
where $\rho$ is the density, ${\bf U}$ is the velocity, ${\bf B}$ is
the magnetic field, $P$ is the sum of the thermal and magnetic
pressures, $\theta \equiv {\delta \rho}/\rho$ is the perturbation of
density logarithm, or entropy. The fluid has constant kinematic
viscosity $\nu$, thermal diffusivity $\chi$ and Ohmic resistivity
$\eta$. The shear parameter $q=-d\ln\Omega/d\ln r=3/2$ for Keplerian
rotation considered here. $N^2$ is the standard Brunt-V$\ddot{\rm
a}$is$\ddot{\rm a}$l$\ddot{\rm a}$ frequency squared characterizing
vertical stratification in the Boussinesq approach. $g$ is the
vertical component of the central object's gravity, which drops out
from the main equations after using rescaling $g\theta \rightarrow
\theta$. In the adopted local disk model, the thermal stratification
is incorporated in a simple manner, that is, $N^2$ is taken to be
positive (i.e., convectively stable) and constant, with value
$0.25\Omega^2$ \citep[see also][]{Lesur_Ogilvie10}.

An equilibrium represents a stationary azimuthal flow with linear
shear in the radial $x$-direction, ${\bf U}_0=-q\Omega x{\bf e}_y$,
pressure $P_0$, density $\rho_0$ and threaded by a constant vertical
magnetic field, ${\bf B}_0=B_{0z}{\bf e}_z, B_{0z}=const > 0$.
Perturbations of the velocity, ${\bf u}={\bf U}-{\bf U}_0$, total
pressure, $p=P-P_0$, and magnetic field, ${\bf b}={\bf B}-{\bf
B}_0$, of arbitrary amplitude are imposed on top of this
equilibrium. Inserting them into Equations
(\ref{eq:mom})-(\ref{eq:divb}), we obtain a nonlinear system
(\ref{eq:App-ux})-(\ref{eq:App-divperb}) governing the dynamics of
perturbations, which is explicitly given in Appendix. These
equations form the basis for our simulations below to get a complete
data set of the perturbation evolution in the turbulent state.

For further use, we also introduce the perturbation kinetic,
magnetic and thermal energy densities, respectively, as
\[
E_K=\frac{1}{2}\rho_0{\bf u}^2,~~~E_M=\frac{{\bf
b}^2}{8\pi},~~~E_{th}=\frac{\rho_0\theta^2}{2N^2}.
\]

We normalize time by $\Omega^{-1}$, length by the disk thickness
$H$, velocities by $\Omega H$, magnetic field by $\Omega H\sqrt{4\pi
\rho_0}$ and the pressure and energies by $\rho_0\Omega^2H^2$.
Viscosity, thermal diffusivity and resistivity are characterized,
respectively, by the Reynolds number, ${\rm Re}$, P\'eclet number,
${\rm Pe}$, and magnetic Reynolds number, ${\rm Rm}$, given by
\[
{\rm Re}= \frac{\Omega H^2}{\nu},~~~{\rm Pe}=\frac{\Omega
H^2}{\chi},~~~{\rm Rm}=\frac{\Omega H^2}{\eta},
\]
which are taken to be equal ${\rm Re}={\rm Pe}={\rm Rm}=3000$ (i.e.,
the magnetic Prandtl number ${\rm Pm}={\rm Rm}/{\rm Re}=1$). The
background field is defined by the parameter
$\beta=2\Omega^2H^2/v_A^2$, which is fixed to $\beta=10^3$, where
$v_A=B_{0z}/(4\pi \rho_0)^{1/2}$ is the associated Alfv\'en speed.
In this non-dimensional units, $B_{0z}=\sqrt{2/\beta}=0.0447.$

We carry out numerical integration of the main Equations
(\ref{eq:App-ux})-(\ref{eq:App-divperb}) using the publicly
available pseudo-spectral code SNOOPY
\citep{Lesur_Longaretti07}\footnote{http://ipag.obs.ujf-grenoble.fr/~lesurg/snoopy.html}.
The code is based on the spectral implementation of the shearing box
model for both HD and MHD with the corresponding boundary
conditions: periodic in the $y$- and $z$-directions, but
shear-periodic in the $x$-direction. The fiducial box has sizes
$(L_x,L_y,L_z)=(4,4,1)$ (in units of $H$) and resolutions
$(N_x,N_y,N_z)=(256,256,128)$, respectively, in the
$x,y,z$-directions. The chosen box aspect ratio is most preferable,
as it is itself isotropic in the $(x,y)$-plane and avoids
``numerical deformation'' of the generic, shear-induced anisotropic
dynamics of MRI-turbulence. (Such a deformation at different aspect
ratios, $L_y/L_x$, is analyzed in detail in Paper I for disk flows
with a net azimuthal field). Boxes with a similar aspect ratio was
adopted also in a number of other related studies of net vertical
field MRI-turbulence
\citep{Bodo_etal08,Longaretti_Lesur10,Lesur_Longaretti11,Bai_Stone14,Meheut_etal15}
in order to diminish the recurrent bursts of the channel mode,
however, as we show below, they still play an important role in the
turbulence dynamics even in this extended box. This box also
satisfies the requirement of \citet{Pessah_Goodman09}, that is, its
vertical size is large enough to include the wavelength of the
exponentially/modally fastest growing MRI channel mode (see Figure
\ref{fig:nonmodal_modal_vskz} below) as well as has the aspect
ratios $L_x,L_y>2L_z$, thereby accommodating at the same time the
most unstable parasitic modes. Preference for this aspect ratio in
this study is further explained in Section 3 based on the nonmodal
stability analysis, where we show that this box actually comprises
transiently/nonmodally most amplified modes over finite dynamical
times and hence is able to fully represent the turbulence dynamics.
Small amplitude random perturbations of velocity are imposed
initially on the Keplerian flow and subsequent evolution was run up
to $t_f = 700$ (about 111 orbits). The data accumulated from the
simulations, in fact, represents complete information about the
MRI-turbulence in the considered flow system. Following our previous
works
\citet{Horton_etal10,Mamatsashvili_etal14,Mamatsashvili_etal16};
Paper I, we analyze this data in Fourier (${\bf k}$-) space in order
to grasp the interplay of various linear and nonlinear processes
underlying the turbulence dynamics. Obtaining the simulation data is
the first, preparatory stage for the main part of our study that
focuses on the dynamics in Fourier space. This second stage involves
derivation of evolution equations for physical quantities (for the
amplitudes of velocity and magnetic field) in Fourier space and
subsequent analysis of the right hand side terms of these spectral
dynamical equations.

\subsection{Equations in Fourier space}
\label{sec:Fourierspace}

To analyze spectral dynamics, we decompose the perturbations
$f\equiv ({\bf u},p,\theta,{\bf b})$ into Fourier harmonics/modes
\begin{equation}\label{eq:fourier}
f({\bf r},t)=\int \bar{f}({\bf k},t)\exp\left({\rm i}{\bf
k}\cdot{\bf r} \right)d^3{\bf k}
\end{equation}
where $\bar{f}\equiv (\bar{\bf u}, \bar{p}, \bar{\theta},\bar{\bf
b})$ are the corresponding Fourier transforms. The derivation of
spectral equations of perturbations is a technical task and
presented in Appendix: substituting decomposition (\ref{eq:fourier})
into perturbation Equations
(\ref{eq:App-ux})-(\ref{eq:App-divperb}), we obtain the equations
for the spectral velocity (Equations
\ref{eq:App-uxk1}-\ref{eq:App-uzk1}), logarithmic density (Equation
\ref{eq:App-thetak}) and magnetic field (Equations
\ref{eq:App-bxk}-\ref{eq:App-bzk}). Below we give the final set of
the equations for the quadratic forms of these quantities in Fourier
space.

Multiplying Equations (\ref{eq:App-uxk1})-(\ref{eq:App-uzk1}),
respectively, by $\bar{u}_x^{\ast}$, $\bar{u}_y^{\ast}$,
$\bar{u}_z^{\ast}$, and adding up with their complex conjugates, we
obtain
\begin{equation}\label{eq:uxk2}
\frac{\partial}{\partial t}\frac{|\bar{u}_x|^2}{2} = -
qk_y\frac{\partial}{\partial k_x}\frac{|\bar{u}_x|^2}{2} + {\cal
H}_x + {\cal I}_x^{(u\theta)} + {\cal I}_x^{(ub)}+{\cal
D}_x^{(u)}+{\cal N}_x^{(u)},
\end{equation}
\begin{equation}\label{eq:uyk2}
\frac{\partial}{\partial t} \frac{|\bar{u}_y|^2}{2} = -
qk_y\frac{\partial }{\partial k_x}\frac{|\bar{u}_y|^2}{2} + {\cal
H}_y + {\cal I}_y^{(u\theta)} + {\cal I}_y^{(ub)}+{\cal
D}_y^{(u)}+{\cal N}_y^{(u)},
\end{equation}
\begin{equation}\label{eq:uzk2}
\frac{\partial}{\partial t} \frac{|\bar{u}_z|^2}{2} = -
qk_y\frac{\partial}{\partial k_x}\frac{|\bar{u}_z|^2}{2} + {\cal
H}_z + {\cal I}_z^{(u\theta)} + {\cal I}_z^{(ub)}+{\cal
D}_z^{(u)}+{\cal N}_z^{(u)},
\end{equation}
where the terms of linear origin are
\begin{equation}\label{eq:Hx}
{\cal
H}_x=\left(1-\frac{k_x^2}{k^2}\right)(\bar{u}_x\bar{u}_y^{\ast}+\bar{u}_x^{\ast}\bar{u}_y)+2(1-q)\frac{k_xk_y}{k^2}|\bar{u}_x|^2,
\end{equation}
\begin{equation}\label{eq:Hy}
{\cal
H}_y=\frac{1}{2}\left[q-2-2(q-1)\frac{k_y^2}{k^2}\right](\bar{u}_x\bar{u}_y^{\ast}+\bar{u}_x^{\ast}\bar{u}_y)
- 2\frac{k_xk_y}{k^2}|\bar{u}_y|^2
\end{equation}
\begin{equation}\label{eq:Hz}
{\cal H}_z =
(1-q)\frac{k_yk_z}{k^2}(\bar{u}_x\bar{u}_z^{\ast}+\bar{u}_x^{\ast}\bar{u}_z)
-\frac{k_xk_z}{k^2}(\bar{u}_y\bar{u}_z^{\ast}+\bar{u}_y^{\ast}\bar{u}_z),
\end{equation}
\begin{equation}\label{eq:Iuthi}
{\cal I}_i^{(u\theta)}
=\left(\frac{k_ik_z}{k^2}-\delta_{iz}\right)\frac{\bar{\theta}\bar{u}_i^{\ast}+\bar{\theta}^{\ast}\bar{u}_i}{2},
\end{equation}
\begin{equation}\label{eq:Iubi}
{\cal I}_i^{(ub)} = \frac{\rm
i}{2}k_zB_{0z}(\bar{u}_i^{\ast}\bar{b}_i -
\bar{u}_i\bar{b}_i^{\ast}),
\end{equation}
\begin{equation}\label{eq:Dui}
{\cal D}_i^{(u)}=-\frac{k^2}{\rm Re}|\bar{u}_i|^2,
\end{equation}
and the modified nonlinear transfer functions for the quadratic
forms of the velocity components are
\begin{equation}\label{eq:Nui}
{\cal
N}^{(u)}_i=\frac{1}{2}(\bar{u}_iQ^{\ast}_i+\bar{u}_i^{\ast}Q_i).
\end{equation}
Here the index $i=x,y,z$ henceforth, $\delta_{iz}$ is the Kronecker
delta and $Q_i$ (given by Equation \ref{eq:App-Qi}) describes the
nonlinear transfers via triad interactions for the spectral
velocities $\bar{u}_i$ in Equations
(\ref{eq:App-uxk1})-(\ref{eq:App-uzk1}). It is readily shown that
the sum of ${\cal H}_i$ is equal to the spectrum of the Reynolds
stress multiplied by the shear parameter $q$, ${\cal H}={\cal
H}_x+{\cal H}_y+{\cal
H}_z=q(\bar{u}_x\bar{u}_y^{\ast}+\bar{u}_x^{\ast}\bar{u}_y)/2$.

Next, multiplying Equation (\ref{eq:App-thetak}) by
$\bar{\theta}^{\ast}$ and adding up with its complex conjugate, we
get
\begin{equation}\label{eq:thk2}
\frac{\partial}{\partial
t}\frac{|\bar{\theta}|^2}{2}=-qk_y\frac{\partial}{\partial k_x}
\frac{|\bar{\theta}|^2}{2}+{\cal I}^{(\theta u)} + {\cal
D}^{(\theta)} + {\cal N}^{(\theta)},
\end{equation}
where the terms of linear origin are
\begin{equation}\label{eq:Ithu}
{\cal I}^{(\theta
u)}=\frac{N^2}{2}(\bar{u}_z\bar{\theta}^{\ast}+\bar{u}_z^{\ast}\bar{\theta}),
\end{equation}
\begin{equation}\label{eq:Dth} {\cal
D}^{(\theta)}=-\frac{k^2}{\rm Pe}|\bar{\theta}|^2
\end{equation}
and the modified nonlinear transfer function for the quadratic form
of the entropy is
\begin{equation}\label{eq:Nth1}
{\cal N}^{(\theta)}= \frac{\rm
i}{2}\bar{\theta}^{\ast}(k_xN^{(\theta)}_{x}+k_yN^{(\theta)}_{y}+k_zN^{(\theta)}_{z})+c.c.,
\end{equation}
where $N^{(\theta)}_i$ (given by Equation \ref{eq:App-Nth})
describes the nonlinear transfers for the entropy $\bar{\theta}$ in
Equation (\ref{eq:App-thetak}).

Finally, multiplying Equations
(\ref{eq:App-bxk})-(\ref{eq:App-bzk}), respectively, by
$\bar{b}_x^{\ast}$, $\bar{b}_y^{\ast}$, $\bar{b}_z^{\ast}$, and
adding up with their complex conjugates, we obtain
\begin{equation}\label{eq:bxk2}
\frac{\partial}{\partial
t}\frac{|\bar{b}_x|^2}{2}=-qk_y\frac{\partial}{\partial k_x}
\frac{|\bar{b}_x|^2}{2} + {\cal I}_x^{(bu)}+{\cal D}_x^{(b)}+{\cal
N}^{(b)}_x
\end{equation}
\begin{equation}\label{eq:byk2}
\frac{\partial}{\partial
t}\frac{|\bar{b}_y|^2}{2}=-qk_y\frac{\partial}{\partial k_x}
\frac{|\bar{b}_y|^2}{2}+{\cal M}+{\cal I}_y^{(bu)}+{\cal
D}_y^{(b)}+{\cal N}^{(b)}_y
\end{equation}
\begin{equation}\label{eq:bzk2}
\frac{\partial}{\partial
t}\frac{|\bar{b}_z|^2}{2}=-qk_y\frac{\partial}{\partial k_x}
\frac{|\bar{b}_z|^2}{2}+{\cal I}_z^{(bu)}+{\cal D}_z^{(b)}+{\cal
N}^{(b)}_z,
\end{equation}
where ${\cal M}$ is the spectrum of Maxwell stress multiplied by
$q$,
\begin{equation}\label{eq:M}
{\cal
M}=-\frac{q}{2}(\bar{b}_x\bar{b}_y^{\ast}+\bar{b}_x^{\ast}\bar{b}_y),
\end{equation}
\begin{equation}\label{eq:Ibui}
{\cal I}_i^{(bu)}= - {\cal I}_i^{(ub)}=\frac{\rm
i}{2}k_zB_{0z}(\bar{u}_i\bar{b}_i^{\ast} -
\bar{u}_i^{\ast}\bar{b}_i)
\end{equation}
\begin{equation}\label{eq:Dbi}
{\cal D}_i^{(b)}=-\frac{k^2}{\rm Rm}|\bar{b}_i|^2
\end{equation}
and the modified nonlinear terms for the quadratic forms of the
magnetic field components are
\begin{eqnarray}\label{eq:Nbx}
{\cal N}^{(b)}_x=\frac{\rm
i}{2}\bar{b}_x^{\ast}[k_y\bar{F}_z-k_z\bar{F}_y]+c.c., \\
{\cal N}^{(b)}_y=\frac{\rm
i}{2}\bar{b}_y^{\ast}[k_z\bar{F}_x-k_x\bar{F}_z]+c.c., \\
{\cal N}^{(b)}_z=\frac{\rm
i}{2}\bar{b}_z^{\ast}[k_x\bar{F}_y-k_y\bar{F}_x]+c.c.,
\end{eqnarray}
where $\bar{F}_x, \bar{F}_y, \bar{F}_z$ (given by Equations
\ref{eq:App-Fxk}-\ref{eq:App-Fzk}) are the fourier transforms of the
respective components of the perturbed electromotive force and
describe nonlinear transfers for the spectral magnetic field
components in Equations (\ref{eq:App-bxk})-(\ref{eq:App-bzk}).

Our analysis is based on Equations (\ref{eq:uxk2})-(\ref{eq:uzk2}),
(\ref{eq:thk2}) and (\ref{eq:bxk2})-(\ref{eq:bzk2}), which describe
the processes of linear [${\cal H}_i({\bf k},t)$, ${\cal
I}_i^{(u\theta)}({\bf k},t)$, ${\cal I}^{(\theta u)}({\bf k},t)$,
${\cal I}_i^{(ub)}({\bf k},t)$, ${\cal I}_i^{(bu)}({\bf k},t)$,
${\cal M}({\bf k},t)$] and nonlinear [${\cal N}_i^{(u)}({\bf k},t)$,
${\cal N}^{(\theta)}({\bf k},t)$, ${\cal N}_i^{(b)}({\bf k},t)$)
origin. The terms ${\cal D}_i^{(u)}({\bf k},t)$, ${\cal
D}^{(\theta)}({\bf k},t)$, ${\cal D}_i^{(b)}({\bf k},t)$ describe,
respectively, the effects of viscous, thermal and resistive
dissipation as a function of wavenumber and are negative definite.
Except for the orientation of the background field, these basic
dynamical equations in Fourier space and, hence the above terms, are
\emph{formally} analogous to the corresponding ones in the case of a
net azimuthal magnetic field derived in Paper I (Equations 11-17
there), which also gives the description/meaning of each dynamical
term in these spectral equations. The only, though principal,
difference lies in the kinetic-magnetic cross terms ${\cal
I}_i^{(ub)}({\bf k},t)=-{\cal I}_i^{(bu)}({\bf k},t)$, however, this
difference drastically changes the dynamical processes, resulting in
entirely different physics of the turbulence for azimuthal and
vertical orientations of the background field. These cross terms
describe, respectively, the influence of the $i$-component of the
magnetic field ($\bar{b}_i$) on the same component of the velocity
($\bar{u}_i$) and vice versa for each mode. They are of linear
origin and in the present case with vertical magnetic field, ${\cal
I}_i^{(ub)}({\bf k},t) \propto k_zB_{0z}$, while in the case with
azimuthal magnetic field, ${\cal I}_i^{(ub)}({\bf k},t) \propto
k_yB_{0y}$. Consequently, in the presence of net vertical field,
this mutual influence is nonzero for axisymmetric (channel) modes
with $k_y=0$, which are not sheared by the flow, leading to their
continual exponential growth, i.e., to the classical MRI in the
linear regime. As discussed in Introduction, in the given magnetized
disk shear flow, in addition to this exponential growth, which
dominates in fact at larger times, there exists shear-induced linear
nonmodal growth of non-axisymmetric (with $k_y\neq 0$) as well as
axisymmetric modes themselves, which dominates instead at finite
(dynamical) times
\citep{Mamatsashvili_etal13,Squire_Bhattacharjee14}. Assessment of
relative roles of the axisymmetric and non-axisymmetric modes in the
vertical field MRI problem is one of key questions of our study. We
note that despite the transient nature of the nonmodal growth, it is
not of secondary importance in forming spectral characteristics of
the turbulence.

The course of events in the presence of a net azimuthal magnetic
field, studied in Paper I, is essentially different: the cross terms
for axisymmetric modes vanish that leads to the absence of the
classical MRI and therefore of the exponential growth of the channel
mode -- a key ingredient in the dynamics of a net vertical field
MRI-turbulence. In this case, the only source of energy for the
turbulence remains linear transient growth of non-axisymmetric modes
\citep[see
also][]{Balbus_Hawley92,Hawley_etal95,Papaloizou_Terquem97,Brandenburg_Dintrans06,Simon_Hawley09}.
and its sustenance relies on the interplay between this linear
nonmodal transient growth and nonlinear transverse cascade. As a
result, as we will see below, the active modes and the dynamical
balances in the azimuthal and vertical field cases are also
fundamentally different -- it is the channel mode that mainly
provides energy for the rest modes via the transverse cascade in the
latter case.

Although our analysis is based on the above dynamical equations for
quadratic forms of physical quantities in Fourier space, for
completeness, we also present equation for the normalized spectral
kinetic energy of modes/harmonics, ${\cal E}_K =
(|\bar{u}_x|^2+|\bar{u}_y|^2+|\bar{u}_z|^2)/2$:
\begin{equation}\label{eq:ekspec}
\frac{\partial {\cal E}_K}{\partial t} = -qk_y \frac{\partial {\cal
E}_K}{\partial k_x} + {\cal H}_E + {\cal I}_E^{(u\theta)} + {\cal
I}_E^{(ub)} + {\cal D}_E^{(u)} + {\cal N}_E^{(u)},
\end{equation}
which is obtained by summing Equations
(\ref{eq:uxk2})-(\ref{eq:uzk2}). Right hand side terms are the sum
of the following terms:
\[
{\cal H}_E = \sum_i {\cal
H}_i=q(\bar{u}_x\bar{u}_y^{\ast}+\bar{u}_x^{\ast}\bar{u}_y)/2,
\]
\[
{\cal I}_E^{(u\theta)}=\sum_i {\cal I}_i^{(u\theta)},~~~~{\cal
I}_E^{(ub)}=\sum_i {\cal I}_i^{(ub)},
\]
\[
{\cal D}_E^{(u)}=\sum_i {\cal D}_i^{(u)}=-\frac{2k^2}{\rm Re}{\cal
E}_K,~~~{\cal N}_E^{(u)}=\sum_i {\cal N}^{(u)}_i.
\]
Similarly, we can get equation for the normalized spectral magnetic
energy of modes, ${\cal
E}_M=(|\bar{b}_x|^2+|\bar{b}_y|^2+|\bar{b}_z|^2)/2$, by summing
Equations (\ref{eq:bxk2})-(\ref{eq:bzk2}),
\begin{equation}\label{eq:emspec}
\frac{\partial {\cal E}_M}{\partial t} = -qk_y\frac{\partial {\cal
E}_M}{\partial k_x} + {\cal M}+{\cal I}_E^{(bu)}+{\cal
D}_E^{(b)}+{\cal N}_E^{(b)},
\end{equation}
where
\[
{\cal
M}=-q(\bar{b}_x\bar{b}_y^{\ast}+\bar{b}_x^{\ast}\bar{b}_y)/2,~~~~{\cal
I}_E^{(bu)}=\sum_i {\cal I}_i^{(bu)}=-{\cal I}_E^{(ub)},
\]
\[
{\cal D}_E^{(b)}=\sum_i {\cal D}_i^{(b)}=-\frac{2k^2}{\rm Rm}{\cal
E}_M,~~~{\cal N}_E^{(b)}=\sum_i {\cal N}^{(b)}_i.
\]
The equation of the normalized thermal energy of modes, ${\cal
E}_{th}=|\theta|^2/2N^2$, is readily derived by dividing Equation
(\ref{eq:thk2}) just by $N^2$. We do not give it here, because the
thermal energy is always small compared to the kinetic and magnetic
energies (see below).

One can easily write also the equation for the total normalized
spectral energy of modes, ${\cal E}={\cal E}_K+{\cal E}_{th}+{\cal
E}_M$,
\begin{multline}\label{eq:etotspec}
\frac{\partial {\cal E}}{\partial t} = -qk_y\frac{\partial {\cal
E}}{\partial k_x} + {\cal H}_E+{\cal M}+{\cal D}_E^{(u)}+N^2{\cal
D}^{(th)}\\+{\cal D}_E^{(b)}+{\cal N}_E^{(u)}+N^2{\cal
N}^{(\theta)}+{\cal N}_E^{(b)}.
\end{multline}

In the simulation box, wavenumbers are discrete and determined by
the box sizes $L_i$, $k_i=2\pi n_i/L_i$ where $n_i=0,\pm 1, \pm 2
...$ and the index $i=x,y,z$. For convenience, we normalize these
wavenumbers by the grid cell sizes of Fourier space, $\Delta
k_i=2\pi/L_i$, that is, $k_i/\Delta k_i\rightarrow k_i$ after which
all the wavenumbers become integers.

\section{Linear dynamics -- optimal nonmodal growth}
\label{sec:Optimal}

Before embarking on the nonlinear study, we first outline the
characteristic features of the linear dynamics of nonzero vertical
field MRI based on the linearized set of spectral Equations
(\ref{eq:App-uxk}-\ref{eq:App-divbk}). As distinct from the related
linear studies using classical modal treatment
\citep[e.g.,][]{Balbus_Hawley91,Lesur_Longaretti07,Pessah_Chan08,Longaretti_Lesur10},
the main goal of this preliminary linear analysis is to quantify the
nonmodal growth of axisymmetric and non-axisymmetric modes during
dynamical (characteristic nonlinear, of the order of orbital) time
and thereby to identify dynamically active modes which can have the
highest influence on the nonlinear dynamics -- on the dynamical
balances in the turbulence. These energy-carrying active modes
naturally form \textit{the vital area} in Fourier space, i.e., the
region of wavenumbers where most of the energy supply from the basic
flow to perturbation modes occurs. It is this nonmodal physics of
the MRI, taking place at finite times, that is in fact more relevant
for turbulence dynamics and its energy supply process, because these
have characteristic timescales of the order of orbital time
\citep[see also][]{Squire_Bhattacharjee14}.

Following Paper I, where we did an analogous nonmodal analysis for
the azimuthal field MRI, here for this purpose we also use the
nonmodal approach combined with the formalism of optimal
perturbations
\citep{Farrell_Ioannou96,Schmid_Henningson01,Zhuravlev_Razdoburdin14}.
This approach is more comprehensive/general than the modal one,
since it captures the evolution of perturbation modes at all times,
from intermediate (dynamical) times, when nonnormality/shear-induced
nonmodal effects (growth) dominate, to large times when modal
(exponential) MRI growth of axisymmetric modes prevails. The optimal
perturbations exhibit maximum nonmodal amplification during finite
times and therefore are able to draw most of the energy from the
disk flow among other modes.

\begin{figure}
\includegraphics[width=\columnwidth]{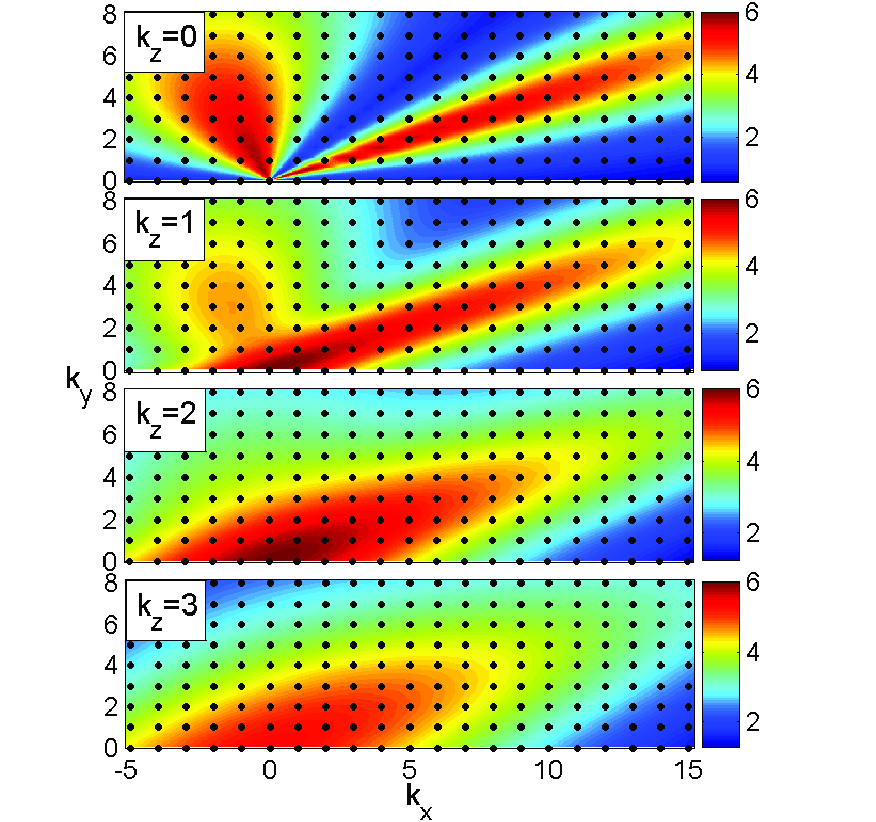}
\caption{Map of $G$ in $(k_x,k_y)$-plane at $t_d=1.33$ and various
$k_z=0,1,2,3$ calculated with the linearized version of the main
Equations (\ref{eq:App-ux})-(\ref{eq:App-divperb}). The black dots
represent discrete modes in the box. It is seen that the vital area
(red and yellow), where the largest nonmodal growth occurs, is
sufficiently densely populated with these modes.}
\label{fig:optimalgrowth}
\end{figure}
\begin{figure}
\includegraphics[width=\columnwidth]{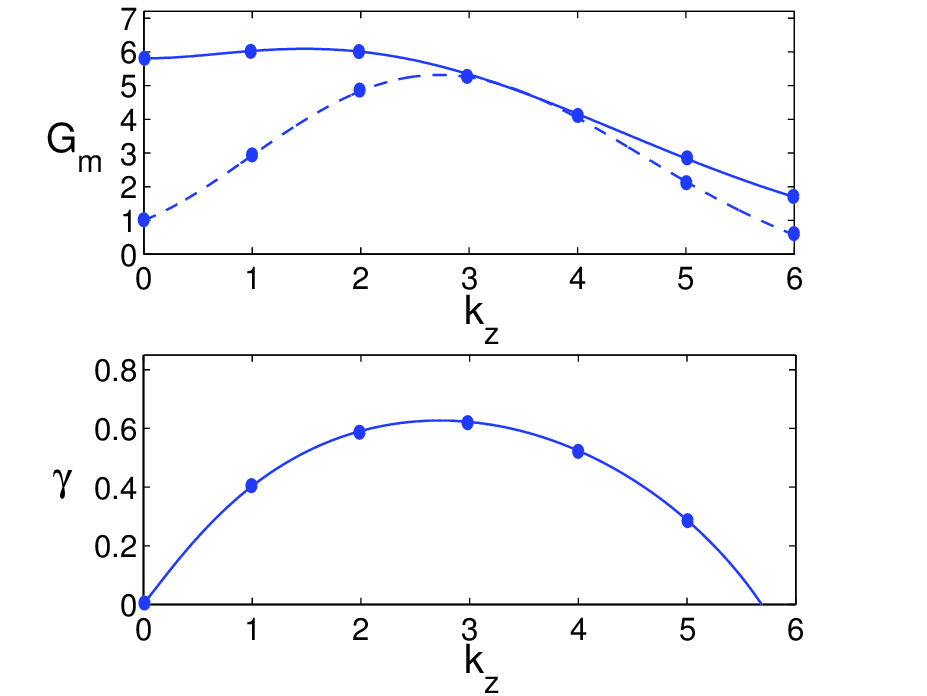}
\caption{The upper panel shows the nonmodal growth factor, $G_m$
(solid line), of the axisymmetric and radially uniform ($k_x=k_y=0$)
modes vs. $k_z$ at $t_d=1.33$ together with the modal growth factor
of the energy, $\exp(2\gamma t_d)$ (dashed line), of these modes for
the same time. Here $\gamma$ is the corresponding \emph{modal}
(exponential) growth rate at $k_x=k_y=0$ and the same values of
$\beta, {\rm Re}, {\rm Rm}$, which is plotted in the lower panel vs.
$k_z$. In both panels, $k_z$ is in units of $2\pi$ and dots indicate
the growth factors and growth rate at discrete $k_z=1,2,3,...$
present in the box.}\label{fig:nonmodal_modal_vskz}
\end{figure}

We quantify the nonmodal growth of individual modes in terms of the
maximum possible, or optimal growth factor, $G$, of the total
spectral energy, ${\cal E}$, by a specific time $t_d$, which was
used in \citet{Squire_Bhattacharjee14} and Paper I. The reference
time, $t_d$, is usually of the order of the orbital time. For
definiteness, here we take it as the $e$-folding time of the most
unstable MRI channel mode in the ideal case,
$t_d=1/\gamma_{max}=1.33$ (in units of $\Omega^{-1}$), where
$\gamma_{max}=0.75\Omega$ is its growth rate
\citep{Balbus_Hawley91}, because this time is also of the order of
the dynamical time.

\begin{figure}
\includegraphics[width=\columnwidth]{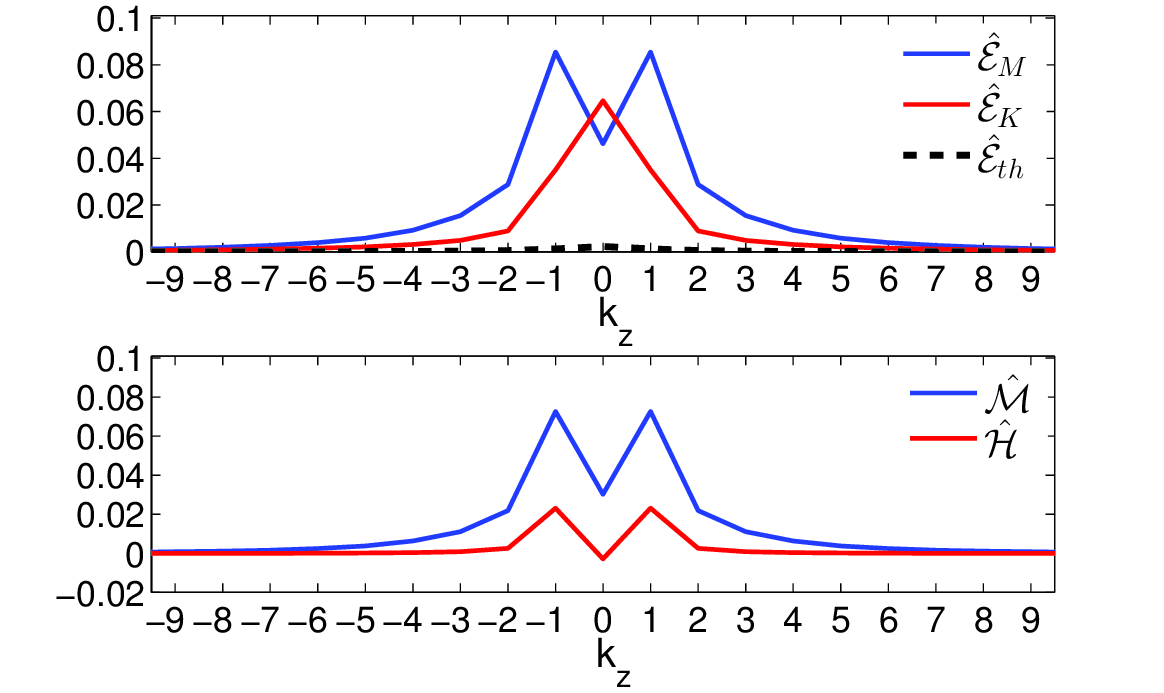}
\caption{Integrated in $(k_x,k_y)$-plane time-averaged 1D spectral
energies and stresses as a function of
$k_z$.}\label{fig:integrated_spectra}
\end{figure}

Figure \ref{fig:optimalgrowth} presents the growth factor $G$ at
different vertical wavenumbers $k_z=0,1,2,3$. It is plotted as a
function of the wavenumber $k_x(t_d)=k_x(0)+qk_yt_d$, which an
optimal mode with initial radial $k_x(0)$ and azimuthal $k_y$
wavenumbers has at time $t_d$ as a result of drift due to the shear.
Although the maximum of the nonmodal amplification in
$(k_x,k_y)$-plane, $G_m$, at fixed $k_z$, comes at axisymmetric
modes with $k_x=k_y=0$, there is a broad range of non-axisymmetric
modes achieving comparable growth (red and yellow areas). At
$k_z=0$, the linear magnetic and velocity perturbations are
decoupled: there are two separate amplification regions, the right
corresponding to magnetic perturbations and the left to kinetic
ones, both with growth factors smaller than that at $k_z=1$. The
maximum nonmodal growth $G_m$ vs $k_z$ is shown in Figure
\ref{fig:nonmodal_modal_vskz}. For the sake of comparison, this
figure also shows the modal (exponential) growth rate, $\gamma$, of
these horizontally uniform $k_z=k_y=0$ modes, which are known to
exhibit the fastest growth also in the modal theory of MRI
\citep[e.g.,][]{Pessah_Chan08}\footnote{Here, this growth rate has
been found by solving a standard dispersion relation of non-ideal
MRI with viscosity and Ohmic resistivity \citep[see
e.g.,][]{Pessah_Chan08,Longaretti_Lesur10}.}, and the corresponding
modal growth factor at $t_d$, $\exp(2\gamma t_d)$, as a function of
$k_z$ for the same values of $\beta, {\rm Re}, {\rm Rm}$. It is
evident that the nonmodal and modal growths exhibit different
dependencies on $k_z$: $G_m$ achieves a maximum at around $k_z=1$
and decreases with increasing $k_z$, whereas $\gamma$ and the
associated modal growth factor reaches a maximum near $k_z=3$ and
decrease at small and large $k_z$. Note also in the upper panel of
Figure \ref{fig:nonmodal_modal_vskz} that the nonmodal growth is
larger than or equal to the modal one at all $k_z$, especially at
the dynamically active small $k_z=0,1$. We will see below that the
axisymmetric mode $k_x=k_y=0, k_z=1$, which undergoes higher
nonmodal amplification and is referred to as \textit{the channel
mode}, is in fact the most active participant in the nonlinear
(turbulence) dynamics as well. Of course, in the turbulent state,
the nonlinear terms qualitatively modify the dynamical picture,
resulting in somewhat different behavior with $k_z$ (see e.g.,
Figure \ref{fig:integrated_spectra}), still the linear analysis
presented here gives a preliminary feeling/insight into energy
exchange process with the basic flow and, most importantly, vividly
shows substantial influence of the disk flow nonnormality on the
growth factors of spectrally/modally unstable modes.

We stress that the nonmodal growth phenomenon -- mathematically the
nonnormality (non-self-adjointness) of linear evolution operators --
is inherently caused by shear and, according to the above
calculations, is significant for both non-axisymmetric and
axisymmetric (channel) modes in MRI-active disks with a net vertical
field. Specifically, the channel modes, which are stronger in this
case, grow purely exponentially only at large times, whereas their
growth during finite/dynamical time is governed by nonmodal physics
and can have growth factors larger than the modal one during these
times (Figure \ref{fig:nonmodal_modal_vskz}). In the case of (net
vertical field) MRI-turbulence, the characteristic timescales of
dynamical processes are evidently of the order of orbital/shear time
\citep{Walker_etal16}. Therefore, it is necessary to take into
account nonmodal effects on the channel mode dynamics and not rely
solely on its modal growth in order to properly understand
saturation and energy balance processes of the turbulence. (This
point regarding the importance of nonmodal physics in the dynamics
of (axisymmetric) modes in MRI-turbulence was also emphasized by
\citet{Squire_Bhattacharjee14}). Thus, for a full representation of
the dynamical picture, it is important that a simulation box
sufficiently well encompasses these nonmodally most amplified modes.
The discrete modes contained in the adopted box $(4,4,1)$ are shown
as black dots on the map of $G$ in Figure \ref{fig:optimalgrowth}.
It is seen that they indeed sufficiently densely cover the yellow
and red regions of the largest nonmodal amplification, which, in
turn, justifies its choice in the present study.

\begin{table}[t]
\caption{Simulation parameters: box size, number of grid points,
stratification and the Reynolds number. In all the runs, ${\rm
Rm}={\rm Pe}={\rm Re}$ and $N^2=0.25$, except for the unstratified
run, where $N^2=0$ and accordingly ${\rm Pe}$ is absent.}
\begin{ruledtabular}
\begin{tabular}{ccccccccccccc}
$(L_x,L_y,L_z)$ & $(N_x, N_y, N_z)$ & stratification & ${\rm Re}$ \\
\hline
$(4,4,1)$ & $(256,256,128)$ & yes &  3000 \\
$(4,4,1)$ & $(256,256,128)$ & no  &  3000 \\
$(4,4,1)$ & $(256,256,128)$ & yes &  6000 \\
\end{tabular}
\end{ruledtabular}
\end{table}

Thus, the above-presented linear nonmodal stability analysis allows
us to optimize the simulation parameters for the nonlinear
(turbulent) regime, especially, the domain size such as to include
as fully as possible the dynamics of essential active modes defining
MRI-turbulence dynamics. In addition to the fiducial run described
above, we performed two other simulations with the same box aspect
ratio (Table I), the first of which is carried out in the
unstratified case in order to compare with the fiducial stratified
case and the second one is to confirm that increasing Reynolds
number does not change a qualitative picture of the dynamical
processes. The analysis presented in the following sections is based
on the fiducial model and it is stated explicitly wherever the
results from other runs appear.

\begin{figure}
\includegraphics[width=\columnwidth]{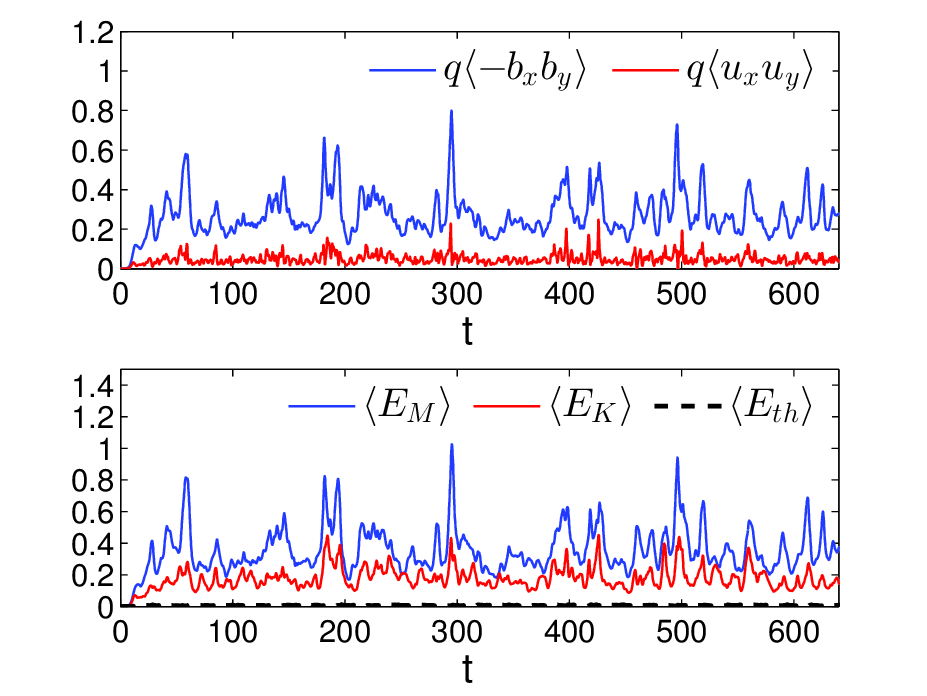}
\includegraphics[width=\columnwidth]{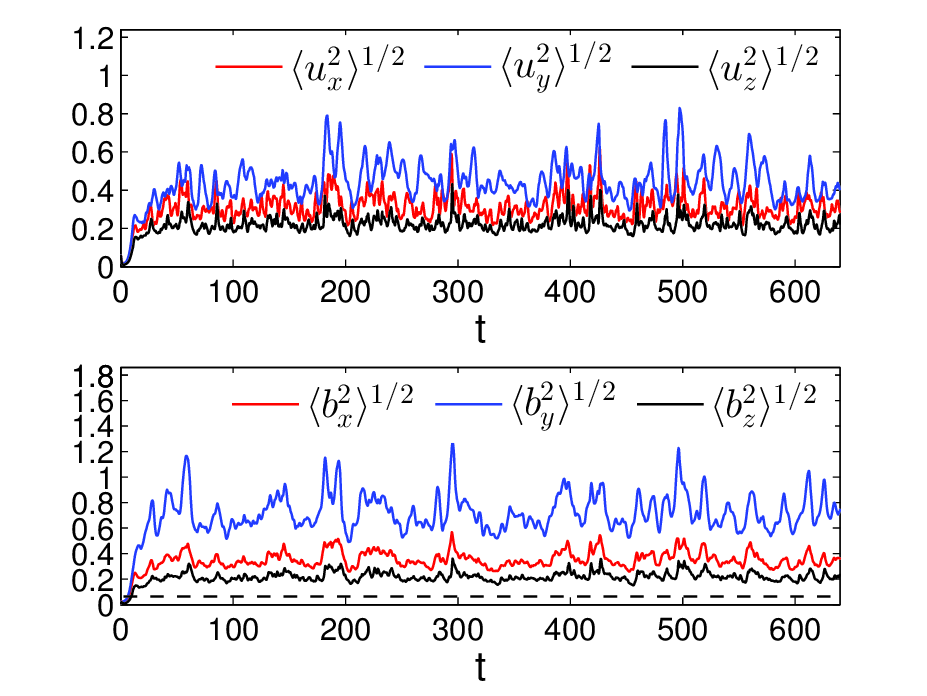}
\caption{Volume-averaged stresses (top row), energy densities
(second row) and the rms of velocity (third row) and magnetic field
(bottom row) components as a function of time. These quantities are
characterized by recurrent bursts all along the evolution, as is
characteristic for vertical field MRI-turbulence. Magnetic energy is
on average about twice larger than the kinetic one, while the
thermal energy is relatively small. The Maxwell stress is larger
than the Reynolds one on average by a factor of 5.5. The azimuthal
components of the turbulent velocity and magnetic field exhibit more
pronounced bursts with higher peaks than those for the respective
other two components. For reference, dashed line in the bottom panel
corresponds to the background vertical field
$B_{0z}=0.0447$.}\label{fig:timeevolution}
\end{figure}

\section{Numerical simulations and characteristic features of the turbulence}
\label{sec:DNS}

Initial perturbations grow mainly due to the nonmodal mechanism,
which later continues into exponential growth for axisymmetric
modes. After a few orbits, the flow breaks down into a fully
developed MHD turbulence. Figure \ref{fig:timeevolution} shows the
evolution of the volume-averaged perturbed kinetic and magnetic
quantities in the turbulent state. The magnetic energy is on average
larger than the kinetic one by a factor of about 2 and both dominate
the thermal energy. The Maxwell stress exceeds the Reynolds one on
average by a factor of about 5.5. The azimuthal components of the
velocity and magnetic field are always larger than the other two
respective ones due to the shear; the imposed vertical field is
dominated by the turbulent magnetic field with the time-averaged rms
values $\langle b_x^2\rangle^{1/2}=8B_{0z}$, $\langle
b_y^2\rangle^{1/2}=16B_{0z}$, $\langle
b_z^2\rangle^{1/2}=4.9B_{0z}$. As seen in Figure
\ref{fig:timeevolution}, the main characteristic feature of the
temporal evolution of all these quantities, which distinguishes the
vertical field case from the azimuthal one analyzed in Paper I, is a
burst-like behavior with intermittent peaks and quiescent intervals,
as also observed in other related studies on net vertical field
MRI-turbulence \citep[e.g.,][]{Sano_Inutsuka01,Lesur_Longaretti07,
Bodo_etal08, Longaretti_Lesur10,Murphy_Pessah15}. The peaks are most
pronounced/intensive in the azimuthal velocity and magnetic field,
inducing corresponding peaks in the magnetic energy and Maxwell
stress. As a result, these bursts yield enhanced rates of angular
momentum transport. On a closer examination, one can notice that the
peaks in the rms of $u_z$ and $b_z$ follow just after the respective
stronger peaks in the rms of $u_y$ and $b_y$ \citep[see
also][]{Murphy_Pessah15}. We show below that these bursts are in
fact closely related to the manifestations of the dynamics of
two modes:\\
-- the channel mode with wavenumber ${\bf k}_c=(0,0,\pm 1)$, which
is horizontally uniform varying only on the largest $z$-vertical
scale in the box, and\\
-- the zonal flow mode with wavenumber ${\bf k}_{zf}=(\pm 1,0,0)$
varying only on the largest $x$-radial scale in the box.\\
Namely, the burst-like growth of the channel mode due to MRI
amplifies $x$- and, especially, $y$-components of velocity and
magnetic field (the channel mode itself does not possess the
$z$-components of these quantities). In turn, its nonlinear
interaction with other dynamically active non-axisymmetric modes
(referred to as parasitic modes in the literature) shortly
afterwards gives rise to the peaks of the $z$-components of these
quantities. Below we show that this nonlinear interaction is mainly
manifested as the transverse cascade in ${\bf k}$-space.

The role of the channel mode and hence the intensity of bursts
depend on the box aspect ratio $L_x/L_z$, being more pronounced when
this aspect ratio is around unity, but becoming weaker as it
increases \citep{Bodo_etal08}. However, as we demonstrate in this
paper, the channel mode is a key participant in the dynamics of a
net vertical field MRI-turbulence. In particular, the adopted here
box $(4,4,1)$ would exhibit only relatively weak bursts according to
\cite{Bodo_etal08}, while as we found here, the amplification of the
channel mode can nevertheless influence the behavior of other active
modes and ultimately the total stress and energy.

\begin{figure}
\includegraphics[width=\columnwidth]{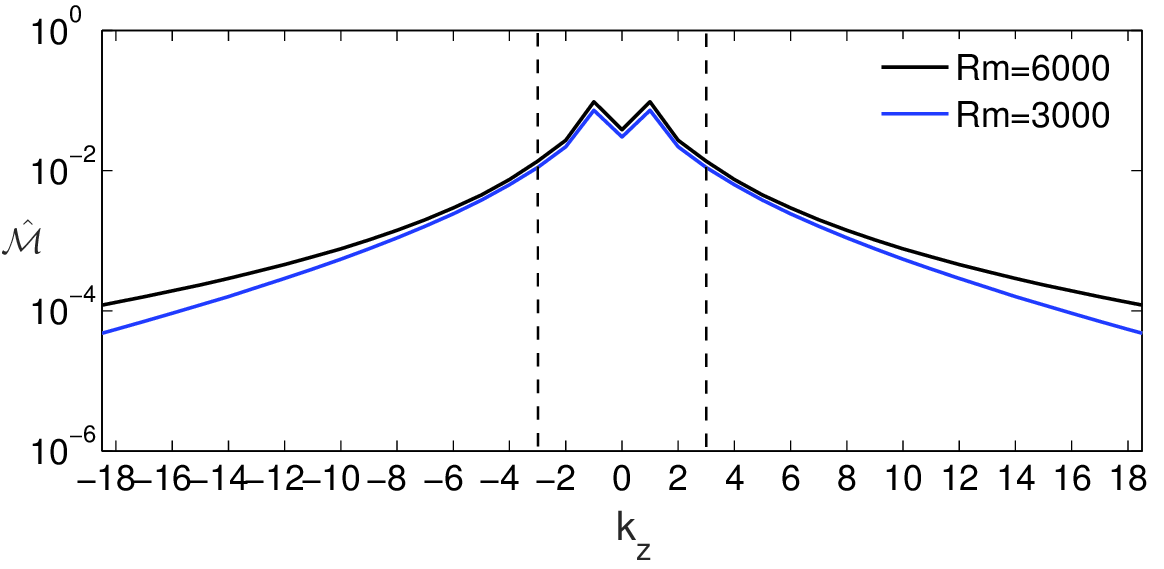}
\caption{$\hat{\cal M}$ vs. $k_z$ at ${\rm Re}={\rm Rm}=3000$ (blue)
for the fiducial model (from Figure \ref{fig:integrated_spectra})
and at ${\rm Re}={\rm Rm}=6000$ (black). The vital area along
$k_z$-axis lies within the vertical dashed lines. It is seen that
the difference between the lower and higher ${\rm Rm}$ cases are
noticeable only at larger $|k_z|$ outside the vital area, leaving
the main dynamical processes within the vital area at small $|k_z|$
nearly unaffected.}\label{fig:maxwell_stress_diffRm}
\end{figure}

\begin{figure*}
\includegraphics[width=0.34\textwidth, height=0.26\textwidth]{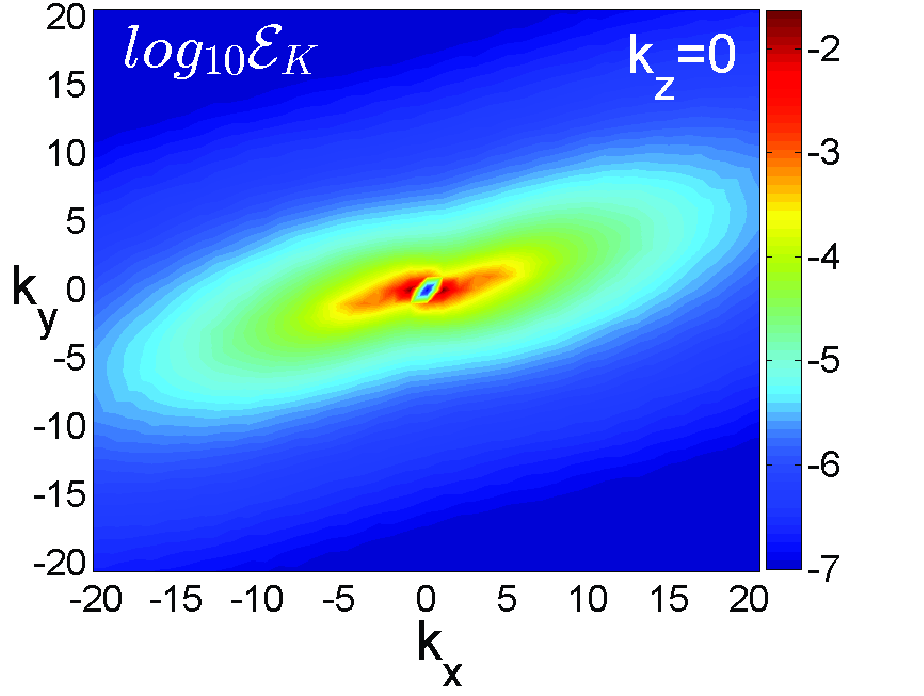}
\includegraphics[width=0.34\textwidth, height=0.26\textwidth]{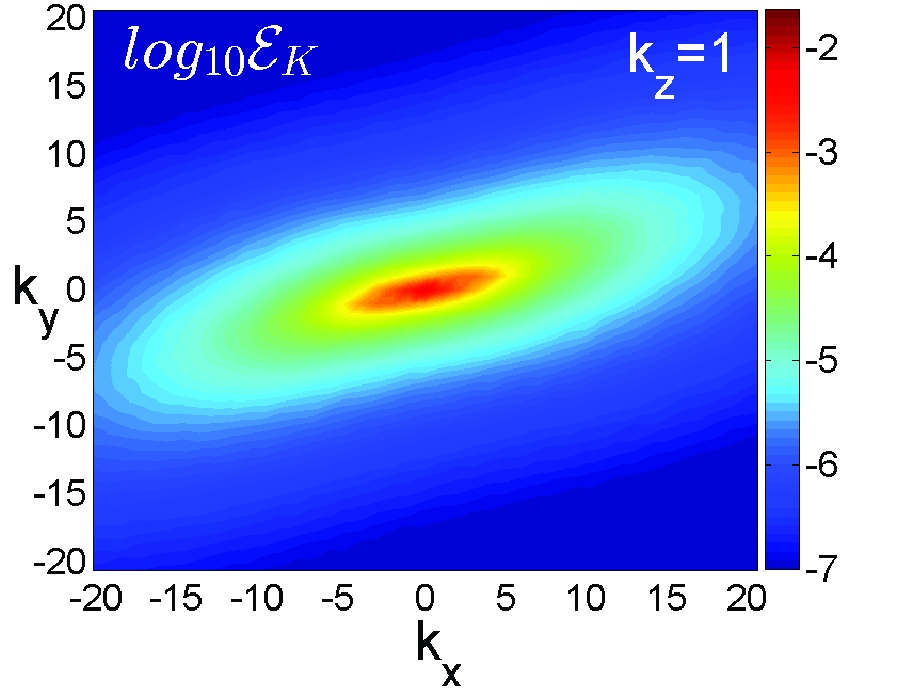}
\includegraphics[width=0.34\textwidth, height=0.26\textwidth]{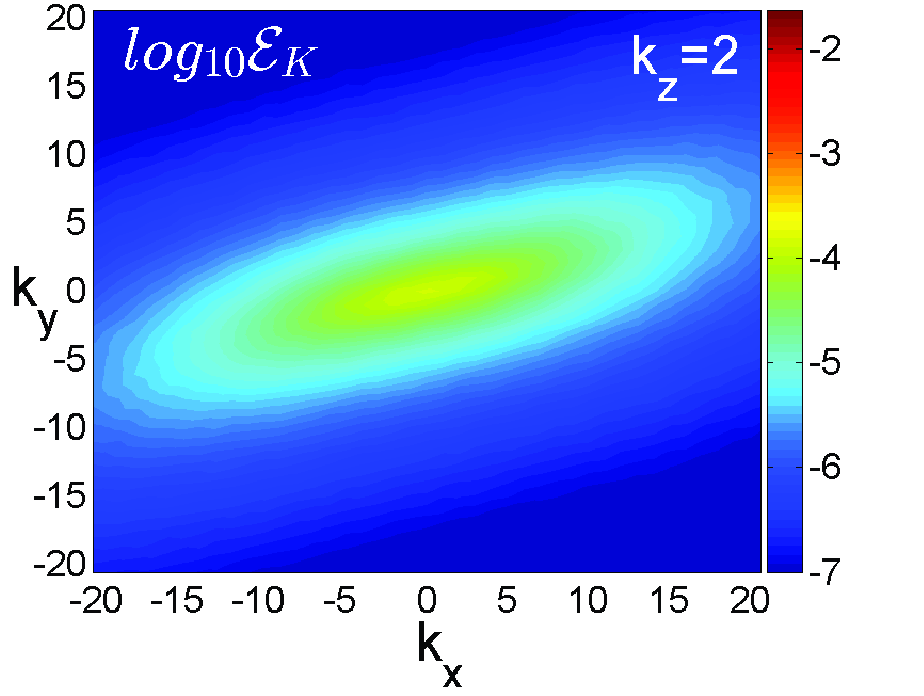}
\includegraphics[width=0.34\textwidth, height=0.26\textwidth]{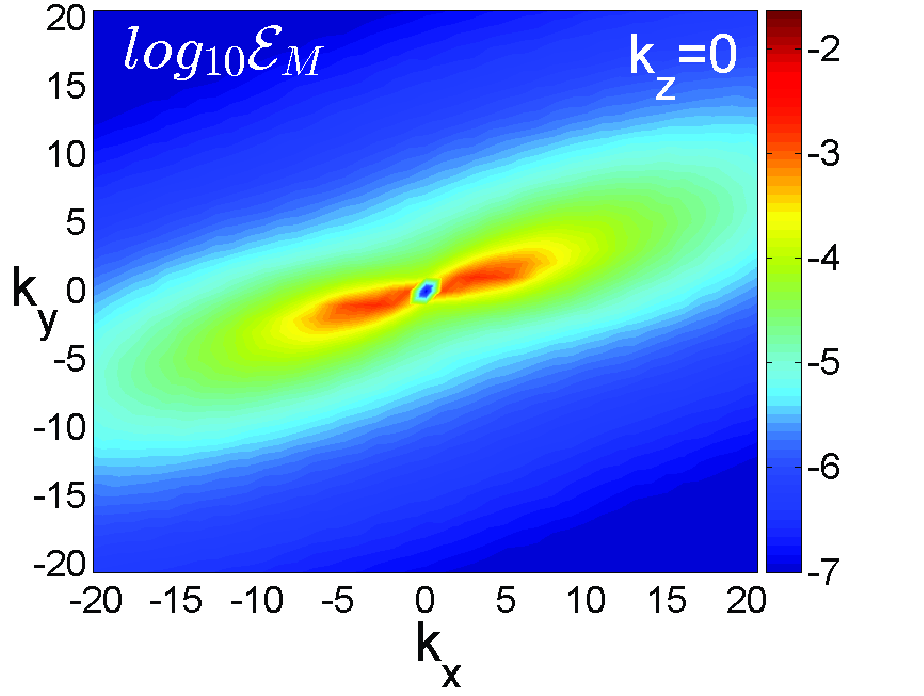}
\includegraphics[width=0.34\textwidth, height=0.26\textwidth]{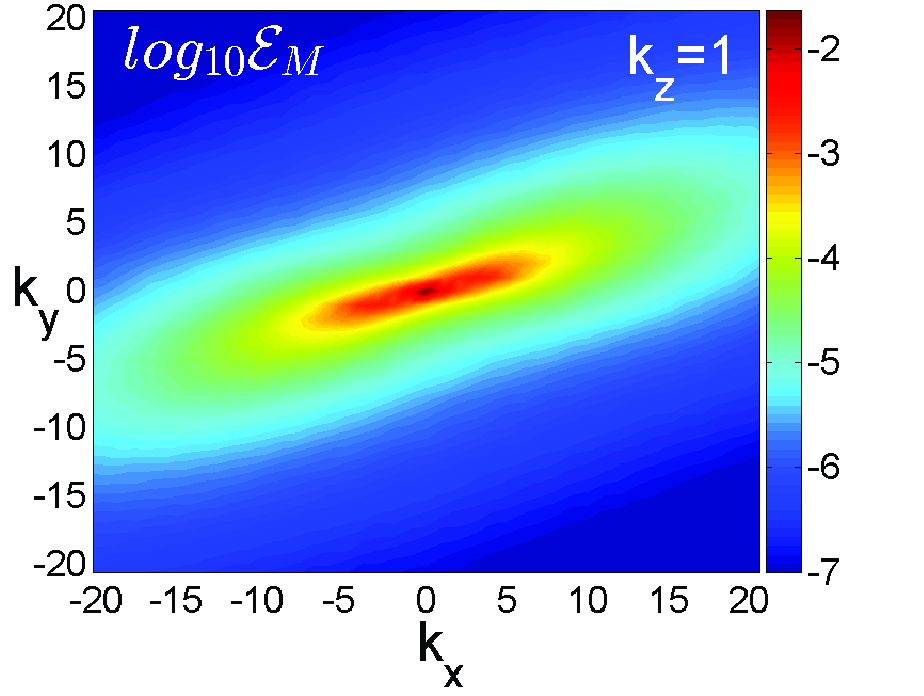}
\includegraphics[width=0.34\textwidth, height=0.26\textwidth]{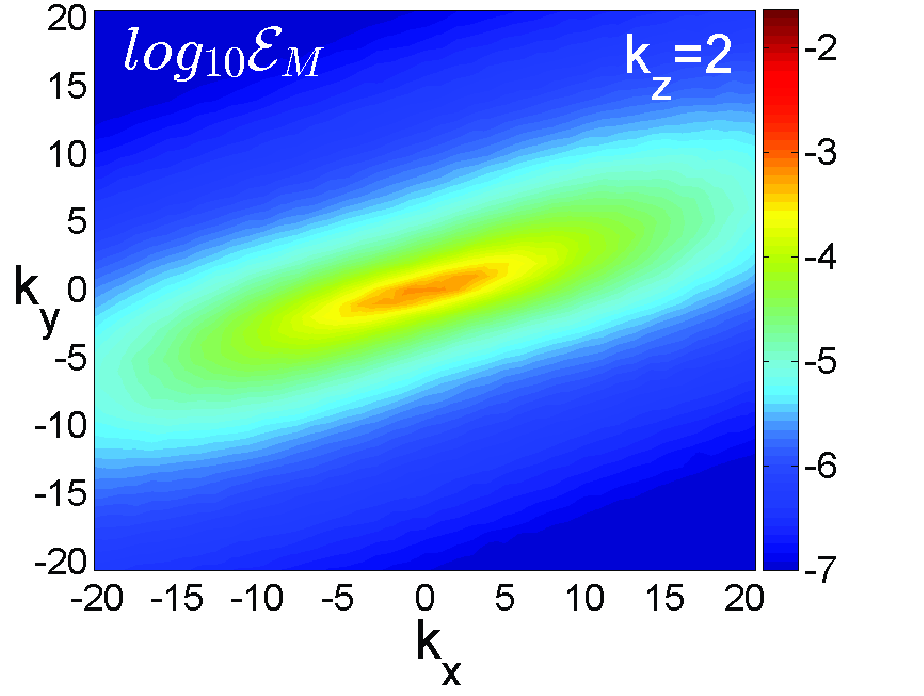}
\caption{Slices of the time-averaged logarithmic 3D spectra of the
kinetic, $log_{10}{\cal E}_K$ (upper row), and magnetic,
$log_{10}{\cal E}_M$ (lower row), energies in $(k_x,k_y)$-plane at
$k_z=0,1,2$. Both spectra are strongly anisotropic with similar
overall shape and inclination towards $k_x$-axis. The maximum of the
kinetic energy spectrum comes at the zonal flow mode ${\bf
k}_{zf}=(\pm 1,0,0)$, while that of the magnetic energy spectrum at
the channel mode ${\bf k}_c=(0,0,\pm 1)$.}
\label{fig:energy_spectra}
\end{figure*}

\subsection{Energy spectra}
\label{sec:Energyspectra}

We start the analysis of the 3D spectra of the energies by examining
first their dependence on the vertical wavenumber. For this purpose,
we integrate the spectra of the energies and the stresses in
$(k_x,k_y)$-plane, $\hat{\cal E}_{K,M,th}(k_z)=\int {\cal
E}_{K,M,th}dk_xdk_y$, $(\hat{\cal H}(k_z),\hat{\cal M}(k_z))=\int
({\cal H},{\cal M})dk_xdk_y$, then average over time from $t=100$
till $t=650$ and represent as a function of $k_z$ in Figure
\ref{fig:integrated_spectra}. The spectral magnetic energy is larger
than the kinetic one except at $k_z=0$, where they are comparable,
and both dominate the spectral thermal energy; the spectral Maxwell
stress, in turn, is higher than the Reynolds one. The maximum of the
magnetic energy as well as both stresses comes at $|k_z|=1$,
corresponding to the channel mode ${\bf k}_c=(0,0,\rm 1)$. Thus,
most of energy supply by the stresses occurs in fact at
$|k_z|=0,1,2$ rather than at $|k_z|=3$ that corresponds to the most
unstable axisymmetric mode of MRI in the modal approach for our
parameters (lower panel of Figure \ref{fig:nonmodal_modal_vskz}).
This is in agreement with the above linear nonmodal analysis, which
have demonstrated how the nonnormality eliminates the dominance of
the $|k_z|=3$ mode in preference to lower $k_z$ modes during finite
times (Figures \ref{fig:optimalgrowth} and
\ref{fig:nonmodal_modal_vskz}). This result supports again the
statement made above that the nonmodal growth over short,
dynamical/orbital timescale is more relevant in the turbulent state.
A maximum of the kinetic energy spectrum is instead at $k_z=0$ and
is a bit higher than the magnetic energy at this $k_z$. It
corresponds to the axisymmetric zonal flow mode with ${\bf
k}_{zf}=(\pm 1,0,0)$.

The main energy-carrying dynamical processes concentrated mostly at
small $|k_z|=0,1,2$ in the vital area are nearly unaffected by
dissipation, as demonstrated in Figure
\ref{fig:maxwell_stress_diffRm} showing the 1D spectra of the
Maxwell stress, $\hat{\cal M}$, for the fiducial run and at twice
higher Reynolds numbers. The difference between these two spectra is
only appreciable outside the vital area, where the stress as well as
the energies fall off by about three orders of magnitude and
therefore these wavenumbers are not anyway dynamically important.

Having examined the dependence of the spectra on the vertical
wavenumber, we now present in Figure \ref{fig:energy_spectra} slices
of the 3D spectra of the kinetic, ${\cal E}_K$, and magnetic, ${\cal
E}_M$, energies in $(k_x,k_y)$-plane. Both spectra are fairly
anisotropic due to the shear with a similar overall elliptical form.
We checked that this spectral anisotropy in fact extends to larger
wavenumbers, up to the dissipation scale. The maximum of the kinetic
energy spectrum comes at the wavenumber ${\bf k}_{zf}=(\pm 1,0,0)$
of the zonal flow mode, while the maximum of the magnetic energy
spectrum comes at the channel mode wavenumber ${\bf k}_c=(0,0,\pm
1)$. These spectra also indicate that there is a broad range of
trailing non-axisymmetric modes (red areas) with energies, on
average, comparable to that of the channel mode and hence taking an
active part in the turbulence dynamics \cite[see
also][]{Longaretti_Lesur10}. By contrast, in the case with azimuthal
field in Paper I, although the maximum of the magnetic energy
spectrum comes again at $k_z=1$, in the $(k_x,k_y)$-plane, this
maximum comes instead at nearby non-axisymmetric modes with
$|k_y|=1$, while the channel mode is not dynamically as significant.
A more detailed analysis of the relative roles and dynamics of the
channel, zonal flow and other (non-axisymmetric) modes is presented
below. Although not shown here, the slices of the 3D spectrum of the
Maxwell stress in $(k_x,k_y)$-plane share a similar shape and type
of anisotropy as the magnetic energy spectrum in Figure
\ref{fig:energy_spectra}. Finally, we note that analogous
anisotropic energy spectra were also observed in other local studies
of MRI-turbulence with a nonzero net vertical field
\citep{Hawley_etal95,Lesur_Longaretti11,Murphy_Pessah15}.

\begin{figure}
\includegraphics[width=\columnwidth]{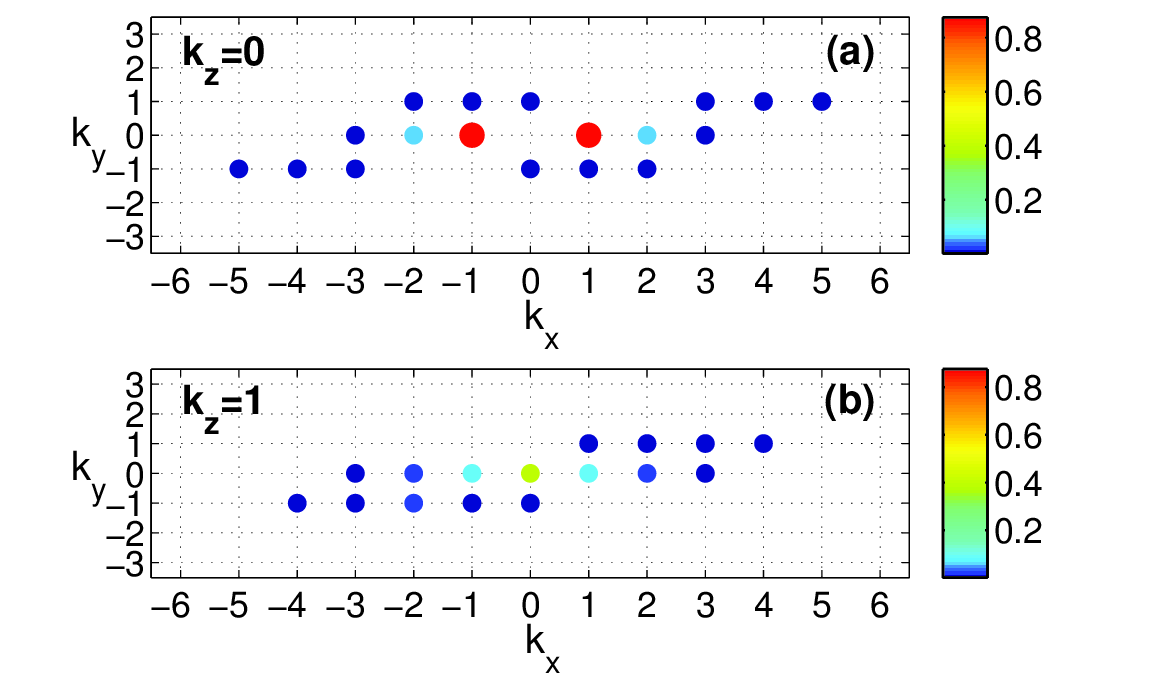}
\includegraphics[width=\columnwidth]{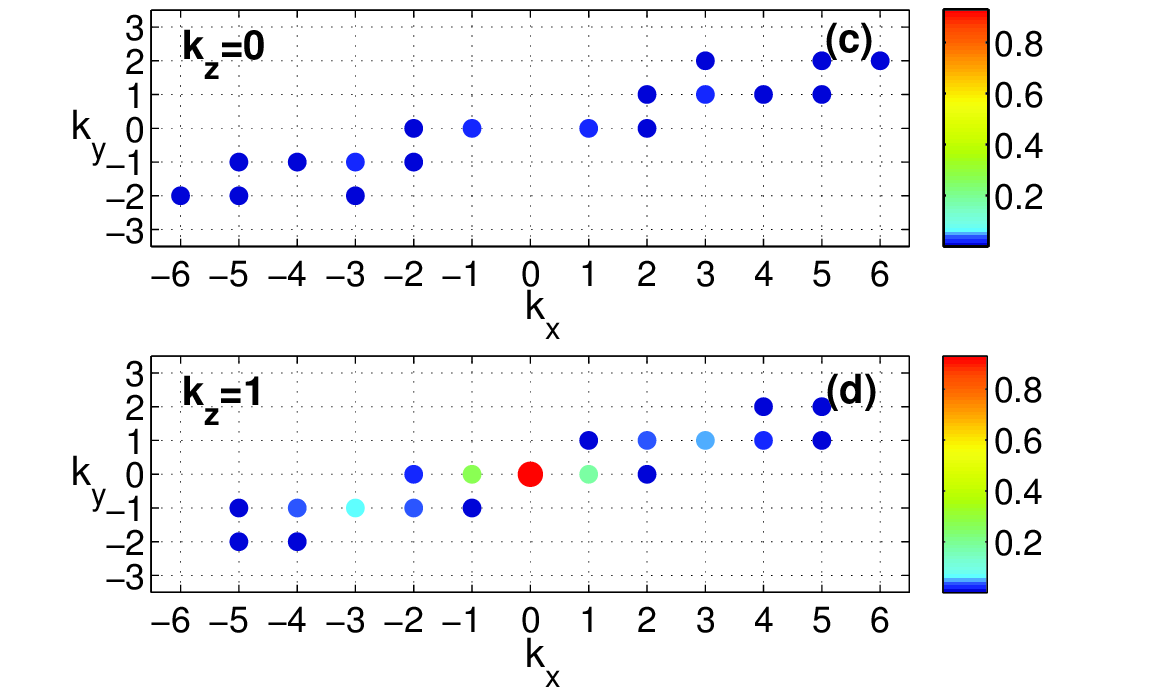}
\caption{Active modes (color dots) in {\textbf{k}}-space at
$k_z=0,1$, as defined in the text, computed separately for the
kinetic [panels (a) and (b)] and magnetic [panels (c) and (d)]
components. The mode colors denote the time, relative to the entire
duration of the run, during which each of these modes retains energy
higher than 50 \% of the maximum spectral energy. The zonal flow
mode ${\bf k}_{zf}=(\pm 1,0,0)$ and the channel mode ${\bf
k}_c=(0,0,\pm 1)$ [denoted by bigger red dots, respectively, in
panels (a) and (d)] clearly stand out against other modes, as they
retain, respectively, higher kinetic and magnetic energies most of
the time.}\label{fig:modes}
\end{figure}

\subsection{Active modes in ${\bf k}$-space}
\label{sec:Activemodes}

The dynamically important modes with most effective amplification
have been identified above based on the optimal nonmodal growth
calculations (Figure \ref{fig:optimalgrowth}), i.e., on the analysis
of the linear dynamics. However, the overall dynamical picture of
the turbulence is formed as a result of the interplay of linear and
nonlinear processes. So, it is reasonable to introduce the notion of
\textit{active modes} in {\textbf{k}}-space -- the energy-carrying
modes playing a major role in the dynamics -- separately for the
kinetic and magnetic components. The active kinetic (magnetic) modes
are labeled those modes whose spectral kinetic (magnetic) energy is
higher than $50 \%$ of the maximum spectral kinetic energy ${\cal
E}_{K,max}$ (magnetic energy ${\cal E}_{M,max}$) at the same time.
Both types of active modes in Fourier space at $k_z=0,1$ are shown
in Figure \ref{fig:modes} with color dots. The color of each mode
indicates the fraction of the total evolution time during which this
mode retains the higher energy. These dynamically active modes are
located anisotropically in $(k_x,k_y)$-plane in the region of
wavenumbers, $|k_x|\leq 6, |k_y|\leq 2$, forming the so-called
\emph{vital area} of the turbulence in {\textbf{k}}-space. In this
area, the most energetic dynamical processes are concentrated that
set the strength and statistical properties of the turbulence. There
are two -- the channel ${\bf k}_c$ and the zonal flow ${\bf k}_{zf}$
-- modes (bigger red dots, respectively, in panels (a) and (d) of
Figure \ref{fig:modes}) that clearly stand out against other modes,
as they retain, respectively, higher magnetic (the channel mode) and
kinetic (the zonal flow mode) energies most of the time. As a
result, these two modes play a key role in shaping the turbulence
dynamics and deserve a more detailed analysis, which is given in the
following subsections.

Apart from the channel and zonal flow modes, there are a sufficient
number of other active modes (blue dots in Figure \ref{fig:modes})
that individually are not so significant compared to these two
modes, since they retain the higher kinetic or magnetic energies for
much less a time. However, the combined action of the multitude of
these modes is competitive to the channel and zonal flow modes. We
label them as \textit{the rest modes} and describe in a devoted
subsection \ref{sec:Therestmodes} below. By contrast, in the case of 
a purely azimuthal field (Paper I), these rest non-axisymmetric
modes (mainly those with $|k_y|=1$ and $k_z=0,1$) prevail not only
collectively but also individually -- they carry energies higher or
comparable to that of the channel mode for longer times than those
in the present case with net vertical field. This results in
different properties of the turbulence defined instead by dominant
individual non-axisymmetric rest modes. The other modes (including
those with $|k_z|\geq 2$ not shown in Figure \ref{fig:modes}) lie
outside the vital area, playing only a minor role in the dynamics

\begin{figure*}
\centering
\includegraphics[width=0.85\textwidth]{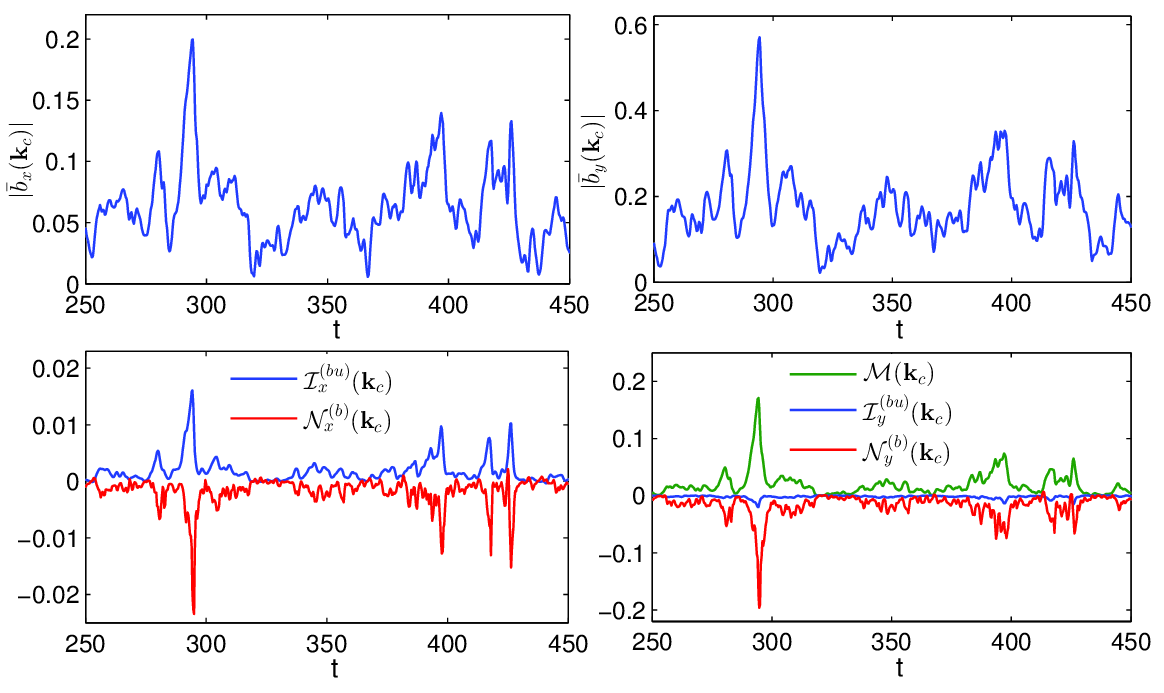}
\includegraphics[width=0.85\textwidth]{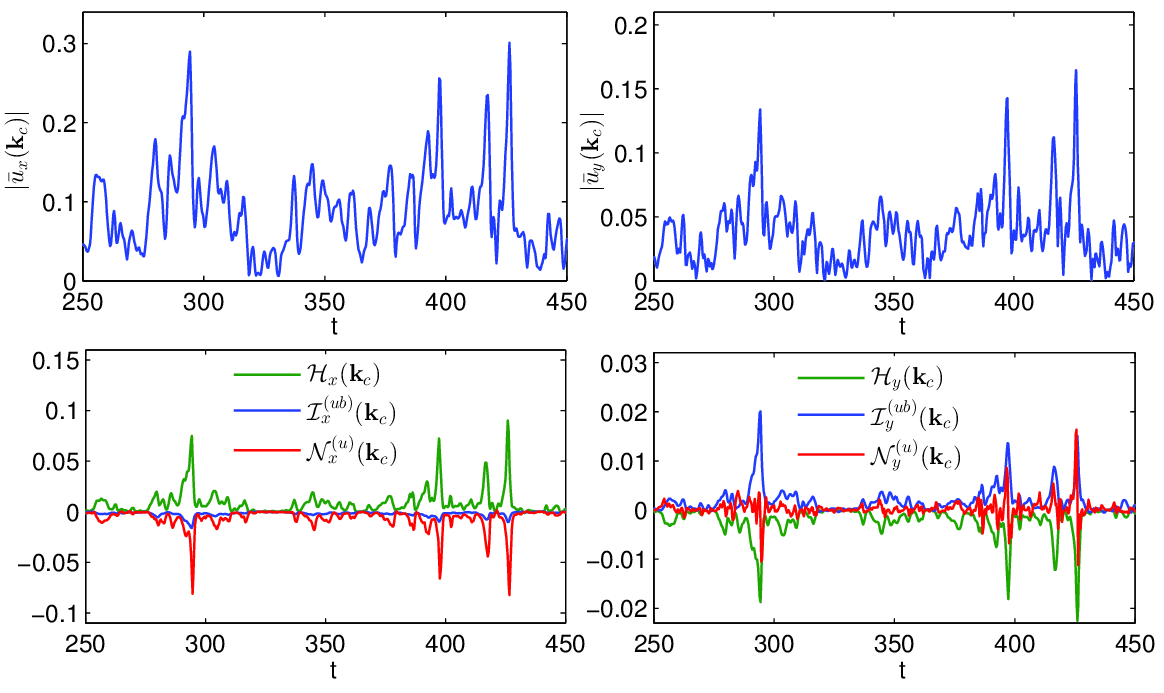}
\caption{Evolution of the main spectral amplitudes of the magnetic
field, $|\bar{b}_x({\bf k}_c)|, |\bar{b}_y({\bf k}_c)|$, and
velocity, $|\bar{u}_x({\bf k}_c)|,~|\bar{u}_y({\bf k}_c)|$, of the
channel mode with ${\bf k}_c=(0,0,\pm 1)$ as well as the
corresponding time-histories of the linear -- stresses, ${\cal
M}({\bf k}_c)$, ${\cal H}({\bf k}_c)$, and exchange, ${\cal
I}^{(ub)}({\bf k}_c)$, ${\cal I}^{(bu)}({\bf k}_c)$ -- and
nonlinear, ${\cal N}^{(u)}({\bf k}_c)$, ${\cal N}^{(b)}({\bf k}_c)$,
terms, which govern the dynamics of this mode. The channel mode is
supported by the linear processes -- the action of the stresses and
the linear exchange terms describe its amplification due to MRI,
while the nonlinear terms mostly oppose this growth, resulting in
the recurrent bursts in the evolution of the magnetic field and
velocity. $|\bar{b}_x|$ and $|\bar{b}_y|$ are amplified,
respectively, by ${\cal I}_x^{(bu)}$ and ${\cal M}$, as they are
always positive, and then drained, respectively, by the nonlinear
terms ${\cal N}_x^{(b)}$ and ${\cal N}_y^{(b)}$, which are always
negative, transferring energy to other wavenumbers and components
(see Figures \ref{fig:nonlinearterms_peak} and
\ref{fig:nonlinearterms_average}). Consequently, the negative peaks
of these terms a bit lag the corresponding peaks of the linear
terms. This azimuthal field component is also drained to a lesser
degree by ${\cal I}_y^{(bu)}$, giving its energy to $|\bar{u}_y|$.
$|\bar{u}_x|$ is amplified by positive ${\cal H}_x$ and drained by
negative ${\cal N}_x^{(u)}$ and, to a lesser degree, by ${\cal
I}_x^{(ub)}$ -- the peaks of the latter two sink terms lag those of
the linear source term ${\cal H}_x$. $|\bar{u}_y|$ is amplified by
${\cal I}_y^{(ub)}$ and drained by always negative ${\cal H}_y$.
${\cal N}_y^{(u)}$ alternates sign, as distinct from the other
nonlinear terms, providing for $|\bar{u}_y|$ either source, when
positive, or sink, when negative. The influence of the nonlinear
terms is maximal at the peaks of $|\bar{b}_x({\bf k}_c)|$ and
$|\bar{b}_y({\bf k}_c)|$ -- they halt the MRI-growth and lead to a
fast drain of the channel mode over a few orbital periods.}
\label{fig:channel_dynamics}
\end{figure*}

\subsection{The channel mode}
\label{sec:Channelmode}

We have seen above that the channel mode -- the harmonic with
wavenumber ${\bf k}_c=(0,0,\pm 1)$ that is uniform in the horizontal
$(x,y)$-plane and has the largest vertical wavelength
(correspondingly, the smallest wavenumber, $k_z=\pm 1$) in the
domain -- is a key participant in the turbulence dynamics. It
carries higher energy most of the time in the turbulent state among
other active modes in the vital area and corresponds to the maximum
of the spectral magnetic energy and stresses in Fourier space at any
moment (see Figure \ref{fig:timevars_channel_zonal} below). So,
first we analyze the dynamics of this mode and then move to the
zonal flow and the rest modes.

Figure \ref{fig:channel_dynamics} shows the evolution of the main
spectral amplitudes of the magnetic field, $|\bar{b}_x({\bf k}_c)|,
|\bar{b}_y({\bf k}_c)|$, and velocity, $|\bar{u}_x({\bf
k}_c)|,~|\bar{u}_y({\bf k}_c)|$ of the channel mode. Also plotted
are the time-histories of the corresponding linear -- stresses,
${\cal M}({\bf k}_c)$, ${\cal H}({\bf k}_c)$ and exchange ${\cal
I}^{(ub)}({\bf k}_c)$, ${\cal I}^{(bu)}({\bf k}_c)$ -- and
nonlinear, ${\cal N}^{(u)}({\bf k}_c)$, ${\cal N}^{(b)}({\bf k}_c)$,
terms in the above spectral Equations
(\ref{eq:uxk2})-(\ref{eq:uyk2}) and (\ref{eq:bxk2})-(\ref{eq:byk2}),
which govern the dynamics of this mode. The action of the stresses
and exchange terms describes together the (nonmodal) effect of MRI
on the channel mode and are mainly responsible for its
amplification. The action of the nonlinear terms describes, in turn,
interaction of the channel mode with other active modes with
different wavenumbers and components (this process is covered in
detail in subsection \ref{sec:Interdependence}, see also Figures
\ref{fig:nonlinearterms_peak} and \ref{fig:nonlinearterms_average}
below). These linear and nonlinear terms jointly determine the
temporal evolution of the channel mode's velocity and magnetic field
with characteristic recurrent bursts seen in Figure
\ref{fig:channel_dynamics}. It is clear from this figure that the
typical variation, or dynamical time of the nonlinear terms is
indeed of the order of the orbital time, indicating again that the
nonmodal physics of MRI is more relevant in the state of developed
turbulence. Analysis of these terms allows us to understand nuances
of the channel mode dynamics and ultimately the behavior of the
total stress and the energy (Figure \ref{fig:timeevolution}),
because, as we will see below (subsection \ref{sec:Therestmodes}),
the channel mode bursts manifest themselves in the time-development
of the volume-averaged magnetic energy and the Maxwell stress. These
terms operate differently for velocity and magnetic fields.

Let us first consider the magnetic field components in Figure
\ref{fig:channel_dynamics}. $|\bar{b}_x|$ and $|\bar{b}_y|$ are
amplified, respectively, by ${\cal I}_x^{(bu)}$ and ${\cal M}$, as
they are always positive, and then drained, respectively, by the
nonlinear terms ${\cal N}_x^{(b)}$ and ${\cal N}_y^{(b)}$, which are
always negative, transferring energy to other wavenumbers and
components. Consequently, the negative peaks of these nonlinear
terms a bit lag the corresponding peaks of the linear terms. The
azimuthal field, which is a dominant field component in the channel
mode, is also drained to a lesser degree by the negative exchange
term ${\cal I}_y^{(bu)}$, transferring its energy to $|\bar{u}_y|$.
As for the velocity components, $|\bar{u}_x|$ is amplified by
positive ${\cal H}_x$, then drained mostly by ${\cal N}_x^{(u)}$,
which is always negative and, to a lesser degree, by also negative
${\cal I}_x^{(ub)}$ -- the peaks of the latter two sink terms lag
corresponding peaks of the linear source term ${\cal H}_x$. Finally,
$|\bar{u}_y|$ is mostly amplified by ${\cal I}_y^{(ub)}$, taking
energy from $|\bar{b}_y|$, and is drained by always negative ${\cal
H}_y$. ${\cal N}_y^{(u)}$ acts differently from the above nonlinear
terms -- it alternates sign, providing for $|\bar{u}_y|$ either
source, when positive, or sink, when negative. So, other modes via
nonlinear interaction occasionally reinforce the growth of the
channel mode's azimuthal velocity, which is primarily due to MRI and
is represented by the exchange term ${\cal I}_y^{(ub)}$ which is
always positive.

\begin{figure*}
\centering
\includegraphics[width=0.85\textwidth]{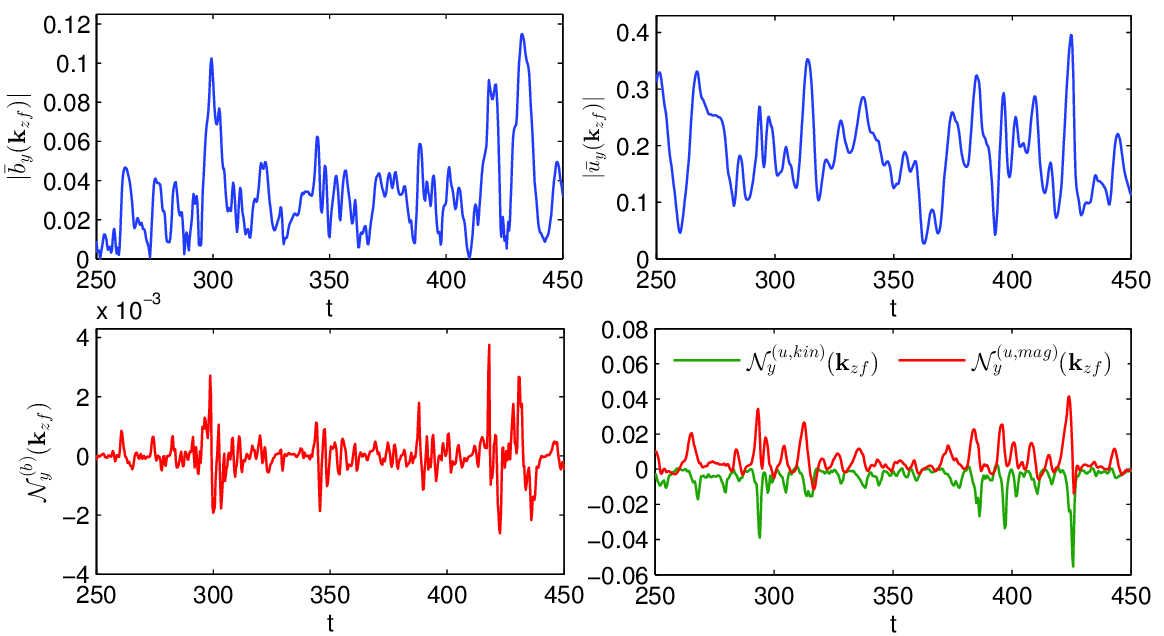}
\caption{Evolution of the azimuthal magnetic field, $\bar{b}_y({\bf
k}_{zf})$, and velocity, $\bar{u}_y({\bf k}_{zf})$, components of
the zonal flow mode with ${\bf k}_{zf}=(\pm 1,0,0)$ as well as the
corresponding nonlinear terms ${\cal N}_y^{(b)}({\bf k}_{zf})$,
${\cal N}_y^{(u,kin)}({\bf k}_{zf})$, ${\cal N}_y^{(u,mag)}({\bf
k}_{zf})$, which drive this mode. The azimuthal velocity $\bar{u}_y$
is dominant in this mode. It is produced by the magnetic part ${\cal
N}_y^{(u,mag)}$, as it is mostly positive, and drained by the
hydrodynamic part ${\cal N}_y^{(u,kin)}$, which is always negative.}
\label{fig:zonalflow_dynamics}
\end{figure*}

In conclusion, the channel mode is supported by the linear --
nonmodal MRI growth -- processes, while nonlinear processes mostly
oppose this growth. The effect of the nonlinear terms is maximal
during the peaks of $|\bar{b}_x({\bf k}_c)|$ and $|\bar{b}_y({\bf
k}_c)|$ -- they halt the growth and lead to a fast drain of the
channel mode energy, redistributing this energy to other modes.
Thus, in a strict self-consistent approach of the net vertical field
MRI-turbulence, this redistribution of the channel mode energy to
other modes is actually due to the nonlinear processes (in the form
of the transverse cascade, Figures \ref{fig:nonlinearterms_peak} and
\ref{fig:nonlinearterms_average}). However, in previous studies, the
channel mode has been considered, for simplicity, not as a
variable/perturbation mode itself, but as a part of the
basic/stationary dynamics \citep[see
e.g.,][]{Goodman_Xu94,Pessah_Goodman09,Latter_etal09,Pessah10}. In
those analysis, the redistribution of the channel mode energy to
other modes is classified as a linear process -- called
\textit{parasitic instability} -- and these modes (parasites, in our
terms ``the rest modes''), which feed on the channel mode, are
assumed to have smaller amplitudes than that of the latter. This
simplified approach definitely gives a good feeling of the
broadening of perturbation spectrum at the expense of the channel
mode. At the same time, this simplification might somewhat suffer
from inaccuracy when used for the turbulent case, since the channel
mode in fact is not stationary and consists of a chaotically
repeated processes of bursts and subsequent drains -- chaotic are
both the values of the peaks as well as time intervals between them.

\subsection{The zonal flow mode}
\label{sec:Zonalflowmode}

The second mode that also plays an important role in the turbulence
dynamics is the zonal flow mode -- the harmonic with wavenumber
${\bf k}_{zf}=(\pm 1,0,0)$ that is independent of the azimuthal and
vertical coordinates (i.e., is axisymmetric and vertically uniform)
and has the largest radial wavelength (correspondingly, the smallest
wavenumber, $k_x=\pm 1$) in the domain. Zonal flow in MRI-turbulence
with zero and nonzero net vertical field was found in several
studies \citet{Johansen_etal09, Bai_Stone14,Simon_Armitage14},
however, its dynamics was usually analyzed based on the simplified
models. Here we trace the zonal flow evolution in Fourier space
self-consistently (without invoking the simplified models) by
examining its governing dynamical terms in the main spectral
equations directly from the simulation data.

We have seen in Figure \ref{fig:modes} that in the turbulent state,
the zonal flow mode carries higher kinetic energy among other active
modes and corresponds to the maximum of the spectral kinetic energy
in Fourier space most of the evolution time (see Figure
\ref{fig:timevars_channel_zonal} below). However, this mode does not
contribute to the spectral Reynolds and Maxwell stresses, because
its radial velocity and magnetic field components are identically
zero, $\bar{u}_x({\bf k}_{zf})=\bar{b}_x({\bf k}_{zf})=0$, due to
incompressibility (\ref{eq:App-divvk}) and divergence-free
(\ref{eq:App-divbk}) conditions. Besides, we checked that the
vertical velocity, $\bar{u}_z({\bf k}_{zf})$, and magnetic field,
$\bar{b}_z({\bf k}_{zf})$, components are also much smaller than the
respective azimuthal ones. So, in Figure
\ref{fig:zonalflow_dynamics} we present only the evolution of the
dominant spectral amplitudes of the azimuthal magnetic field,
$|\bar{b}_y({\bf k}_{zf})|$, and velocity, $|\bar{u}_y({\bf
k}_{zf})|$, in this mode. $|u_y({\bf k}_{zf})|$ changes on longer
timescale corresponding to axisymmetric zonal flow in physical
space, while $|b_y({\bf k}_{zf})|$ together with longer also
exhibits shorter timescale variations. It is readily seen that the
linear -- stresses and exchange -- terms are identically zero for
the zonal flow mode in Equations (\ref{eq:uyk2}) and
(\ref{eq:byk2}), ${\cal H}_y({\bf k}_{zf})={\cal M}({\bf
k}_{zf})=0$, ${\cal I}_y^{(ub)}({\bf k}_{zf})={\cal
I}_y^{(u\theta)}({\bf k}_{zf})=0$. Thus, the linear processes (i.e.,
MRI) do not affect this mode. Therefore, it can be supported only by
the nonlinear terms ${\cal N}_y^{(u)}, {\cal N}_y^{(b)}$. Since
$\bar{u}_y$ is the dominant component in this mode, we look more
into its nonlinear term in order to pin down a mechanism of zonal
flow generation. This term consists of the magnetic, ${\cal
N}^{(u,mag)}_y$, and hydrodynamic, ${\cal N}^{(u,kin)}_y$,
contributions,
\begin{equation}
{\cal N}^{(u)}_y={\cal N}^{(u,mag)}_y + {\cal N}^{(u,kin)}_y,
\end{equation}
which for ${\bf k}_{zf}=(\pm 1,0,0)$, have the following forms:
\begin{multline}\nonumber
{\cal N}^{(u,mag)}_y({\bf k}_{zf})\\=\frac{\rm
i}{2}\bar{u}^{\ast}_y({\bf k}_{zf})\int d^3{\bf k'}\bar{b}_y({\bf
k'})\bar{b}_x({\bf k}_{zf}-{\bf k'}) + c.c.,
\end{multline}
with the integrand composed of the turbulent magnetic stresses and
\begin{multline}\nonumber
{\cal N}^{(u,kin)}_y({\bf k}_{zf})\\= - \frac{\rm
i}{2}\bar{u}^{\ast}_y({\bf k}_{zf})\int d^3{\bf k'}\bar{u}_y({\bf
k'})\bar{u}_x({\bf k}_{zf}-{\bf k'}) + c.c.,
\end{multline}
with the integrand composed of the turbulent hydrodynamic stresses.
The time-histories of these hydrodynamic and magnetic nonlinear
parts as well as ${\cal N}_y^{(b)}$ for the zonal flow mode are also
shown in Figure \ref{fig:zonalflow_dynamics}. It is seen that ${\cal
N}_y^{(b)}({\bf k}_{zf})$ changes sign, acting as a source for
$|\bar{b}_y({\bf k}_{zf})|$, when positive, and as a sink, when
negative. On the other hand, the magnetic part is positive most of
the time, producing and amplifying $|\bar{u}_y({\bf k}_{zf})|$.
Thus, the zonal flow mode is supported by ${\cal N}^{(u,mag)}_y$,
which describes the action of the total azimuthal magnetic tension
exerted by all the rest modes on this mode. By contrast, the
hydrodynamic part ${\cal N}^{(u,kin)}_y({\bf k}_{zf})$, which
describes the total azimuthal hydrodynamic tension exerted by all
other modes, is always negative and drains $|\bar{u}_y({\bf
k}_{zf})|$. As a result, the peaks of the hydrodynamic part ${\cal
N}^{(u,kin)}_y$ a bit lag the peaks of the magnetic part ${\cal
N}^{(u,mag)}_y$, which initiates the growth of the azimuthal
velocity.

\begin{figure*}
\includegraphics[width=\columnwidth]{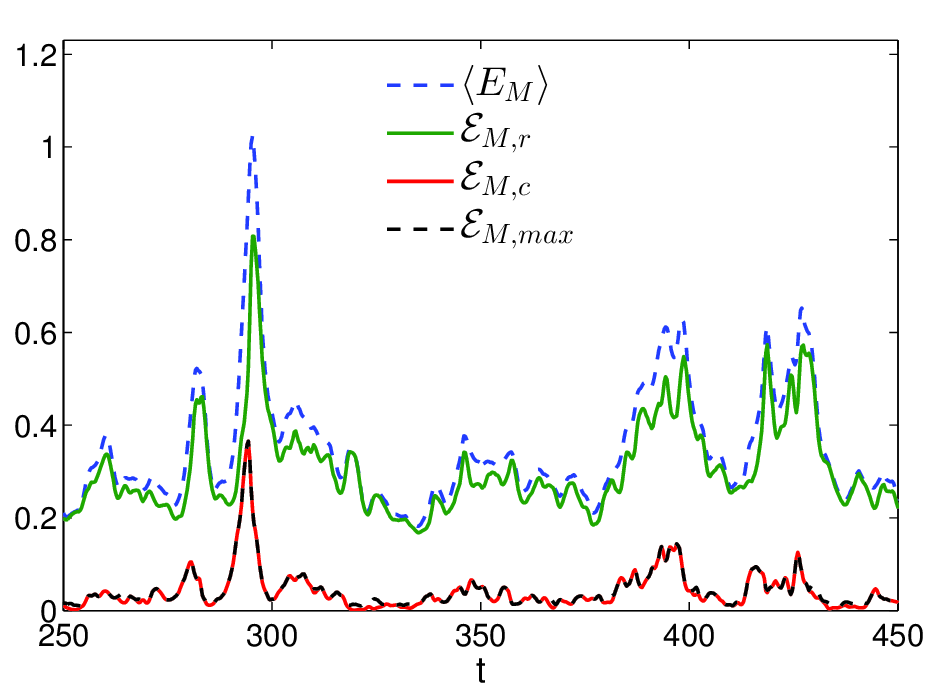}
\includegraphics[width=\columnwidth]{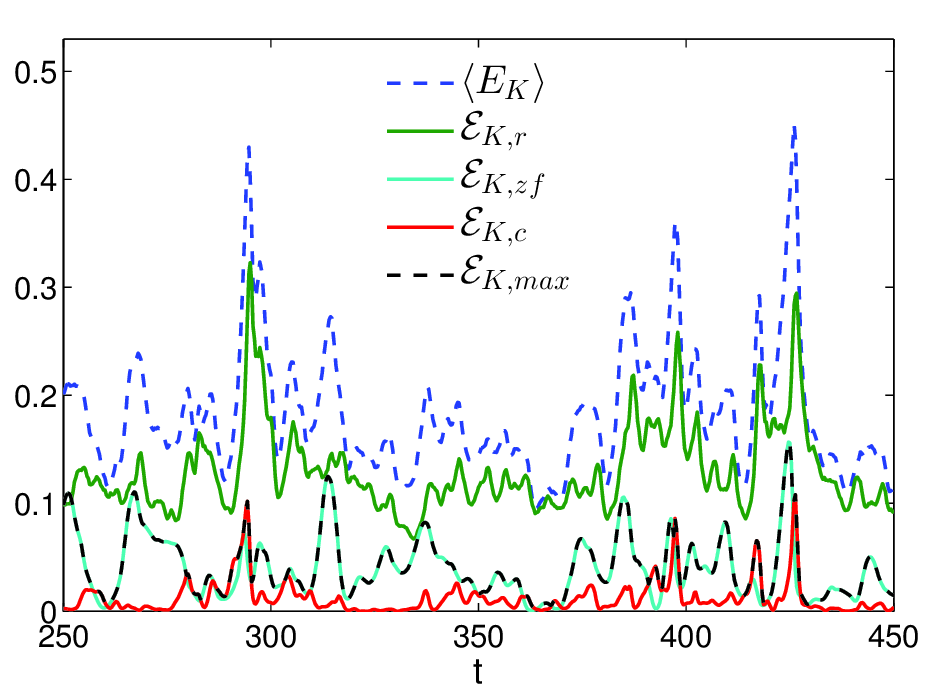}
\caption{The left panel shows the evolution of the total magnetic
energy, $\langle E_M\rangle$ (dashed blue), the magnetic energy of
the channel mode, ${\cal E}_{M,c}$ (red), the total magnetic energy
of the rest modes, ${\cal E}_{M,r}\approx \langle E_M\rangle-{\cal
E}_{M,c}$ (green) and the maximum value of the spectral magnetic
energy, ${\cal E}_{M,max}$ (dashed black), which always coincides
with the energy of the channel mode. The right panel shows the
evolution of the total kinetic energy, $\langle E_K\rangle$ (dashed
blue), the kinetic energy of the channel mode, ${\cal E}_{K,c}$
(red), and of the zonal flow mode, ${\cal E}_{K,zf}$ (cyan) the
total kinetic energy of the rest modes, ${\cal
E}_{K,r}\approx\langle E_K\rangle-{\cal E}_{K,c}-{\cal E}_{K,zf}$
(green) and the maximum value of the spectral kinetic energy, ${\cal
E}_{K,max}$ (dashed black), which always coincides with either the
kinetic energy of the channel mode or the zonal flow mode, whichever
is larger at a given time. It is also seen that the peaks of the
magnetic (kinetic) energy of the rest modes tend to mainly follow
the respective peaks of the magnetic (kinetic) energy of the channel
mode, because, this mode nonlinearly transfers energy to the rest
modes (Figure \ref{fig:nonlinearterms_peak}), causing their
subsequent amplification (peak). Note also that often the zonal flow
energy increases when the channel mode energy decreases and vice
versa.}\label{fig:timevars_channel_zonal}
\end{figure*}

\subsection{The rest modes}
\label{sec:Therestmodes}

So far we have described two -- the channel and zonal flow -- modes
that are the main participants of the dynamical processes. In
addition to these modes, as classified in subsection
\ref{sec:Activemodes}, there are a sufficient number of active modes
in the system (blue dots in Figure \ref{fig:modes}), which are not
individually significant, but collectively have sufficiently large
kinetic and magnetic energies - more than channel or/and zonal flow
modes. Consequently, the combined influence of these modes on the
overall dynamics of the turbulence becomes important. This set of
active modes has been labeled as \textit{the rest modes}, which, as
mentioned above, are also called parasitic modes in the studies of
net vertical field MRI-turbulence. As we checked, the contribution
of other larger wavenumber modes, whose kinetic (magnetic) energy
always remains less than 50\% of the maximum spectral kinetic
(magnetic) energy, in the energy balances is small compared to that
of these three main types of modes and therefore is neglected
below.\footnote{For this reason, in this subsection, these energy
balances are written with approximate equality sign.}

Figure \ref{fig:timevars_channel_zonal} allows us to compare the
total magnetic, ${\cal E}_{M,r}$ (left panel), and kinetic, ${\cal
E}_{K,r}$ (right panel), energies of the rest modes with the
corresponding energies of the channel and zonal flow modes at
different stages of the evolution. The magnetic energy of the zonal
flow mode, ${\cal E}_{M,zf}={\cal E}_M(-{\bf k}_{zf})+{\cal
E}_M({\bf k}_{zf})\approx 0$, is small compared to that of the
channel and the rest modes and is not shown in the left plot. In
this case, the total volume-averaged magnetic energy is
approximately the sum of the magnetic energies of the channel mode,
${\cal E}_{M,c}={\cal E}_M(-{\bf k}_c)+{\cal E}_M({\bf k}_c)$, and
the rest modes, $\langle E_M\rangle\approx {\cal E}_{M,c}+{\cal
E}_{M,r}$ (neglecting the very small contribution from the zonal
flow mode), whereas in the total kinetic energy, all the three types
of modes contribute: $\langle E_K\rangle \approx {\cal E}_{K,c} +
{\cal E}_{K,zf} + {\cal E}_{K,r}$ with ${\cal E}_{K,c}={\cal
E}_K(-{\bf k}_c)+{\cal E}_K({\bf k}_c)$ and ${\cal E}_{K,zf}={\cal
E}_K(-{\bf k}_{zf})+{\cal E}_K({\bf k}_{zf})$ being the kinetic
energies of the channel and zonal flow modes, respectively. In
Figure \ref{fig:timevars_channel_zonal}, ${\cal E}_{M,max}$ and
${\cal E}_{K,max}$ are, respectively, the maximum values of the
spectral magnetic and kinetic energies\footnote{Actually, Figure
\ref{fig:timevars_channel_zonal} shows these maximum energies with
factor 2 to match the definition of the channel and zonal flow mode
energies.}. The left panel shows that this maximum value of the
magnetic energy falls on the channel mode, i.e., ${\cal
E}_{M,max}={\cal E}_{M,c}$, during an entire course of the
evolution. In other words, the channel mode is always magnetically
the strongest among all the individual active rest modes \citep[see
also][]{Longaretti_Lesur10}, although the total magnetic energy of
the rest modes dominates the magnetic energy of the channel mode,
${\cal E}_{M,r} > {\cal E}_{M,c}$, especially in the quiescent
intervals between the bursts, when the magnetic energy of the
channel mode is relatively low. This trend is also seen in the
transport due to these modes (see Figure
\ref{fig:stresses_axi_nonaxi} below) and points to the importance of
the rest modes in the turbulence dynamics. Despite this dominance,
however, it is seen in the left panel of Figure
\ref{fig:timevars_channel_zonal} that the peaks of the rest modes'
magnetic energy always occur shortly after the corresponding peaks
of the channel mode's magnetic energy, indicating that the former
increases, or is driven by the latter (see subsection
\ref{sec:Interdependence}). Thus, the typical burst-like behavior of
the nonzero the net vertical field MRI-turbulence is closely related
to the manifestation of the channel mode dynamics.

The role of the rest modes is also seen in the evolution of the
kinetic energies in the right panel of Figure
\ref{fig:timevars_channel_zonal}. As is in the case of the magnetic
energy, the contribution of the total kinetic energy of the rest
modes, ${\cal E}_{K,r}$, in the volume-averaged total kinetic energy
is always dominant. The kinetic energy of the zonal flow mode,
${\cal E}_{K,zf}$, is comparable to the kinetic energy of the rest
modes only occasionally, while the contribution of the kinetic
energy of the channel mode, ${\cal E}_{K,c}$, in the overall balance
of the total kinetic energy is even less. However, also in this
case, the maximum value of the spectral kinetic energy, ${\cal
E}_{K,max}$, always coincides with either the kinetic energy of the
channel mode or the zonal flow mode, whichever is larger at a given
moment. Moreover, on a closer inspection of the kinetic energy
curves, it appears that the peaks of the total kinetic energy of the
rest modes in fact tends to mainly follow the corresponding peaks of
the channel mode. Note also that often the zonal flow energy
increases and has a peak when the channel mode energy decreases and
has a minimum, that is, the zonal flow tends to grow when the
channel mode is at its minimum and vice versa. This anti-correlation
between the zonal flow and magnetic activity, which was also
reported in \citet{Johansen_etal09}, is further explored in Fourier
space in the next subsection.

\begin{figure}
\includegraphics[width=\columnwidth]{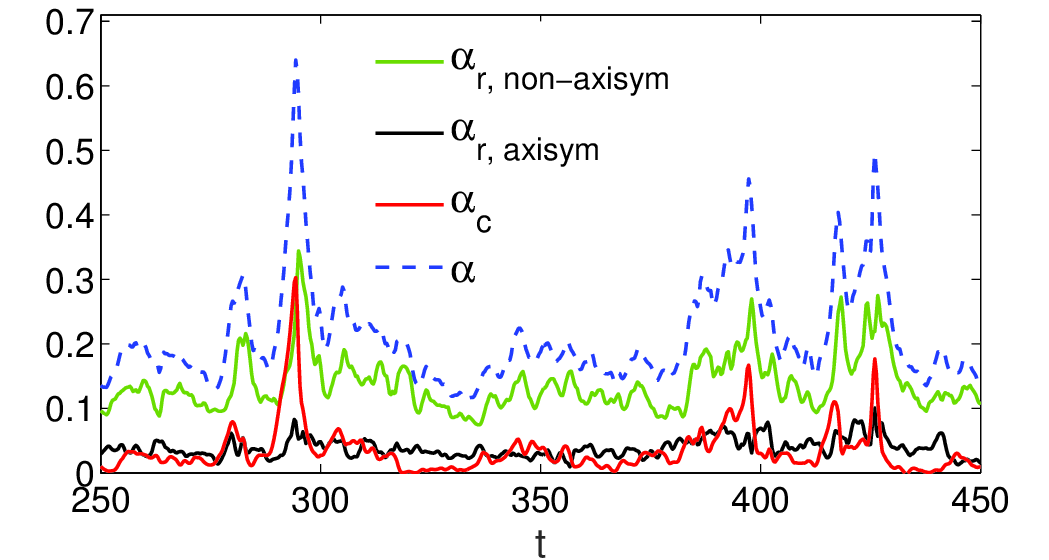}
\caption{The transport parameter for the non-axisymmetric rest
modes, $\alpha_{r, non-axisym}$ (green), for the axisymmetric rest
modes, $\alpha_{r, axisym}$ (black), for the channel mode,
$\alpha_{c}$ (red) and the total value $\alpha\approx \alpha_{r,
non-axisym}+\alpha_{r, axisym}+\alpha_c$ (dashed blue). Note that
the peaks in $\alpha_{r, non-axisym}$ occur shortly after the peaks
of the channel mode $\alpha_c$, although the former dominates the
latter.}\label{fig:stresses_axi_nonaxi}
\end{figure}

Finally, we characterize the angular momentum transport due to the
rest modes and compare it to that of the channel mode. The transport
is measured by the normalized sum of the Maxwell and Reynolds
stresses, i.e., by the usual $\alpha$ parameter which in our
non-dimensional units is defined for each harmonic as,
\[
\bar{\alpha}=\frac{1}{2}(\bar{u}_x\bar{u}_y^{\ast}-\bar{b}_x\bar{b}_y^{\ast})+c.c.
\]
Figure \ref{fig:stresses_axi_nonaxi} shows the time-histories of the
transport parameter for the rest modes, separately for
non-axisymmetric ones, $\alpha_{r, non-axisym}=\sum_{k_y\neq
0}\bar{\alpha}$ and axisymmetric ones, $\alpha_{r,
axisym}=\sum_{k_y=0}\bar{\alpha}$, as well as for the channel mode
$\alpha_c=\bar{\alpha}({\bf k}_c)$ and the total transport due to
all the active modes in the box, $\alpha=\sum_{\bf
k}\bar{\alpha}\approx \alpha_{r, non-axisym}+\alpha_{r,
axisym}+\alpha_c$. (The zonal flow mode does not contribute to the
transport, since $\bar{u}_x({\bf k}_{zf})=\bar{b}_x({\bf
k}_{zf})=0$.) Like that for the energies above, the transport due to
the non-axisymmetric rest modes always dominates the transport due
to the channel and axisymmetric rest modes, $\alpha_{r,
non-axisym}>\alpha_c, \alpha_{r, axisym}$. This is consistent with
the results of \citet{Longaretti_Lesur10}, who also found the
prevalence of non-axisiymmetric modes over the channel mode in the
transport, although they arrived on the same conclusion by
calculating 1D Fourier transform of both the Maxwell and Reynolds
stress as a function of $k_y$ at the minimum and maximum of the
transport, whereas we calculate the cumulative transport of the
non-axisymmetric rest modes. Note, however, that the peaks in
$\alpha_{r, non-axisym}$ follow shortly after the peaks of the
channel mode $\alpha_c$, despite the dominance of the former over
the latter. At these moments, the channel mode typically exhibits
more dramatic changes in the transport than the non-axiymmetric
modes. This again indicates that the channel mode, nonlinearly
interacting with the rest modes, acts as a trigger for the bursts in
the total transport which, in turn, are primarily due to the
collective activity of non-axisymmetric modes.

\begin{figure*}[t!]
\centering
\includegraphics[width=0.32\textwidth, height=0.2\textwidth]{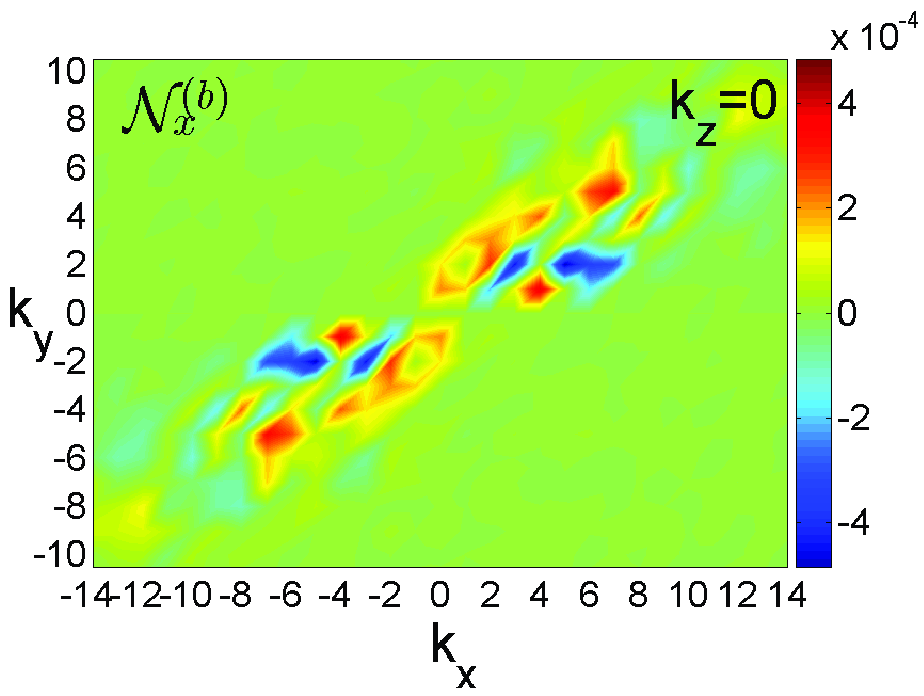}
\includegraphics[width=0.32\textwidth, height=0.2\textwidth]{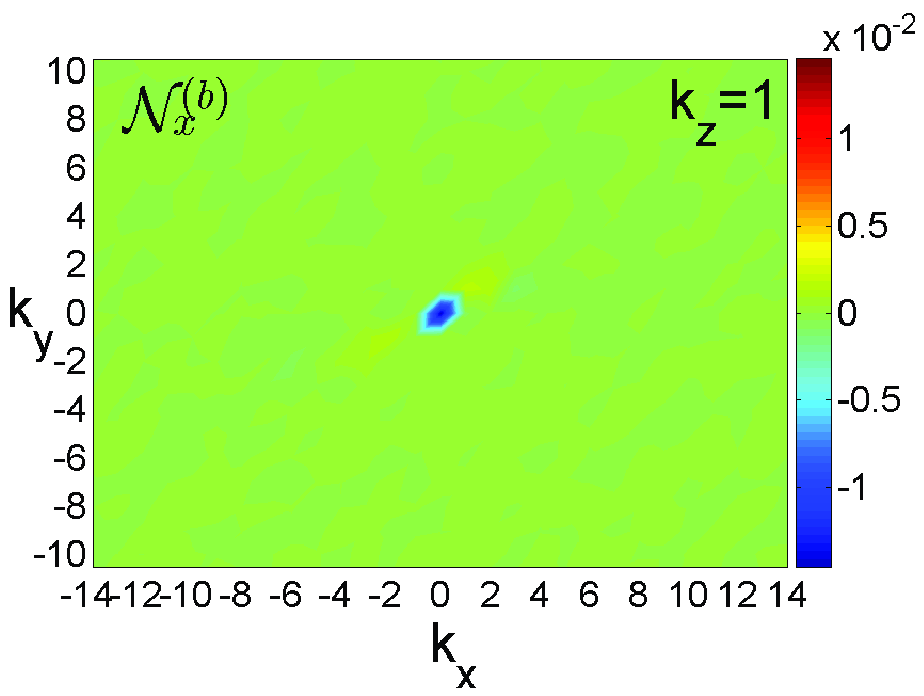}
\includegraphics[width=0.32\textwidth, height=0.2\textwidth]{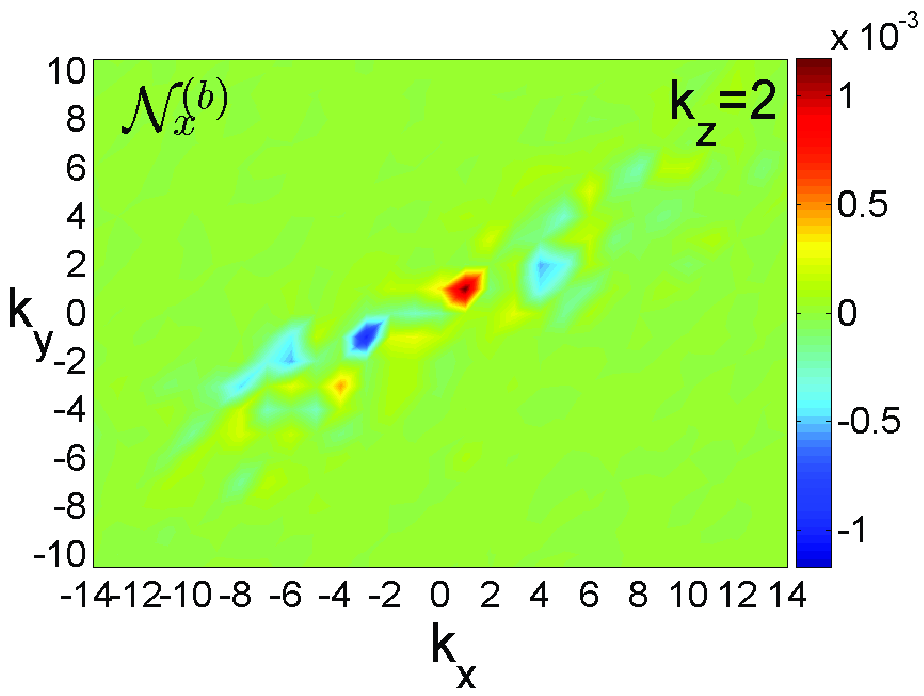}
\includegraphics[width=0.32\textwidth, height=0.2\textwidth]{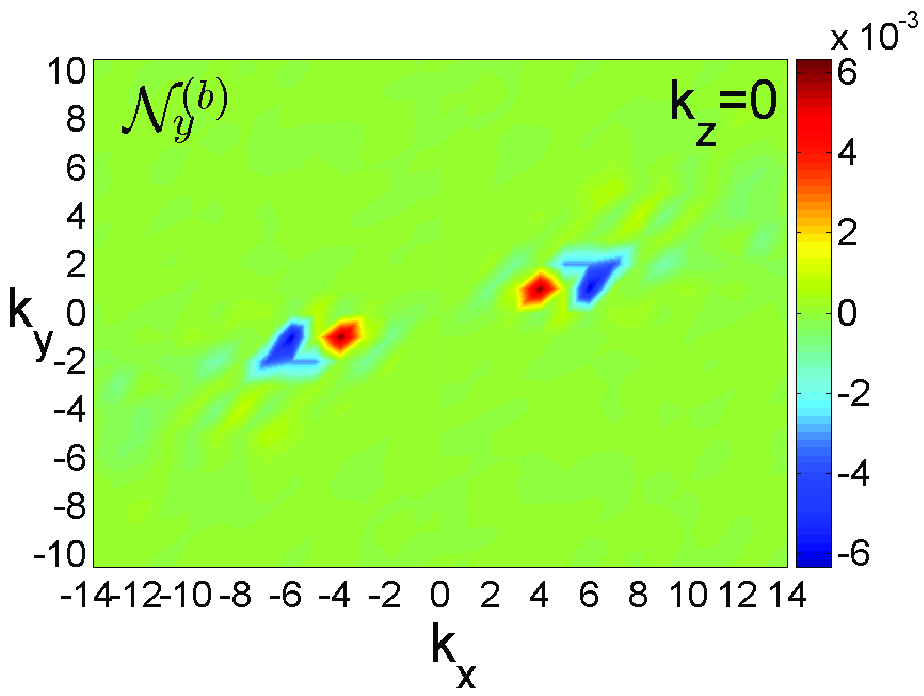}
\includegraphics[width=0.32\textwidth, height=0.2\textwidth]{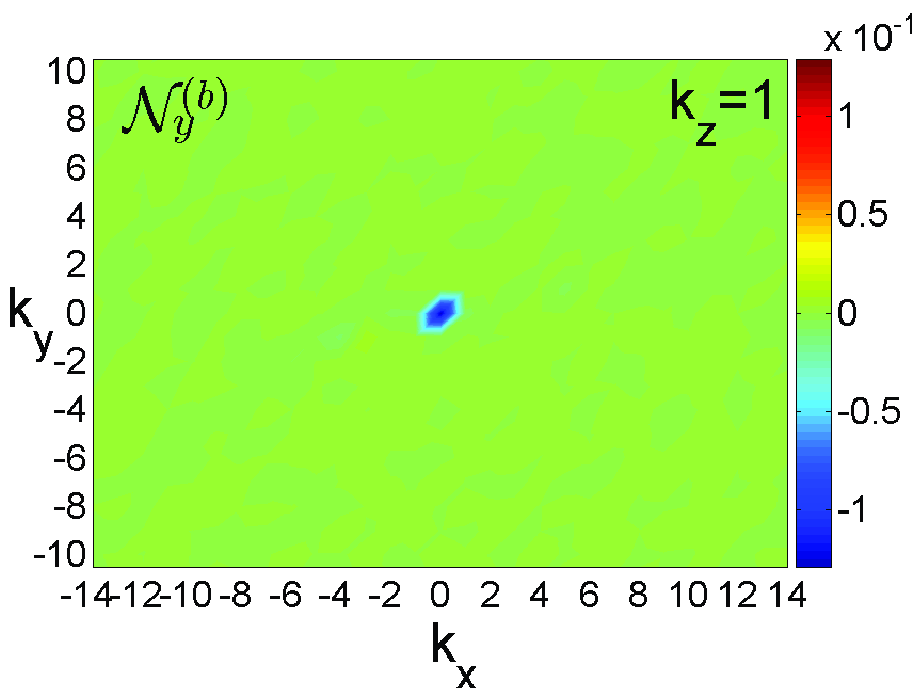}
\includegraphics[width=0.32\textwidth, height=0.2\textwidth]{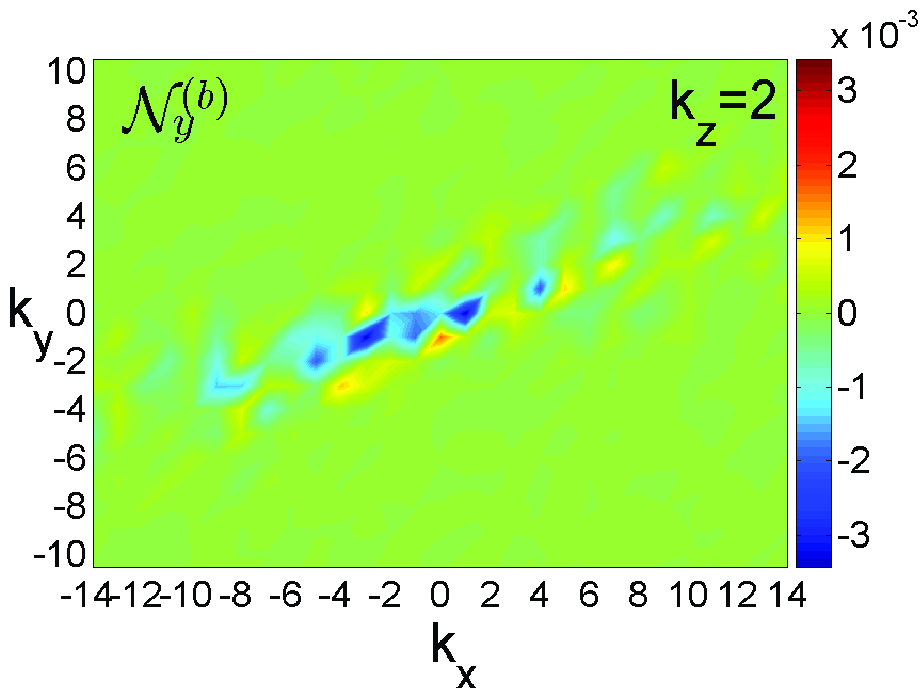}
\includegraphics[width=0.32\textwidth, height=0.2\textwidth]{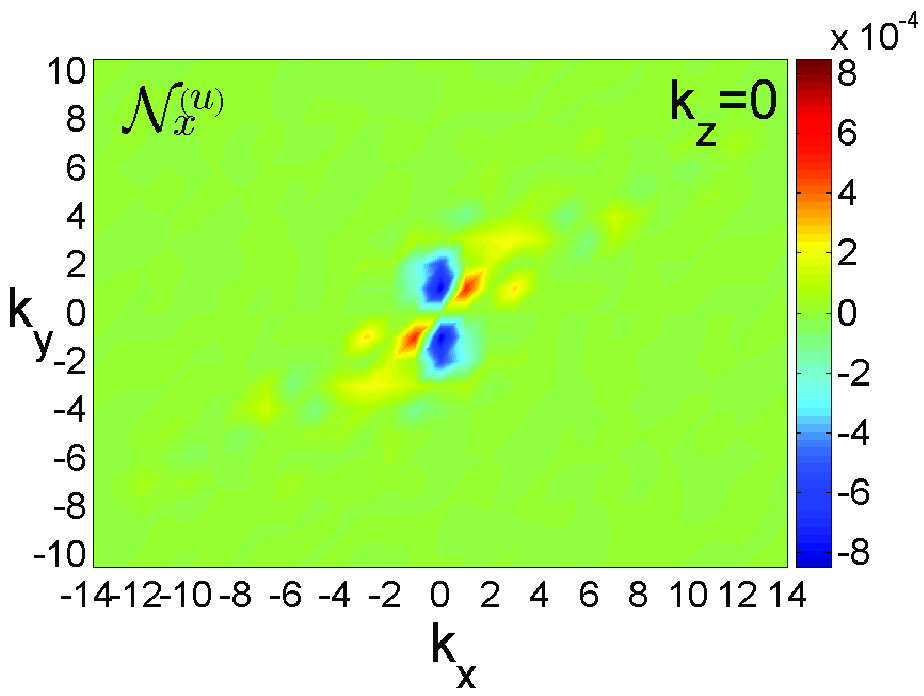}
\includegraphics[width=0.32\textwidth, height=0.2\textwidth]{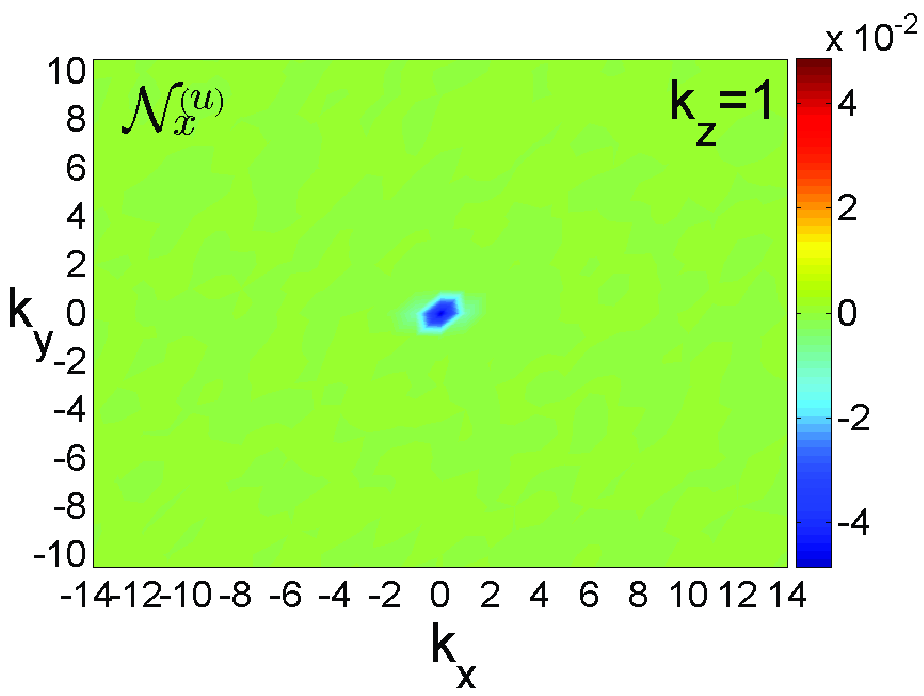}
\includegraphics[width=0.32\textwidth, height=0.2\textwidth]{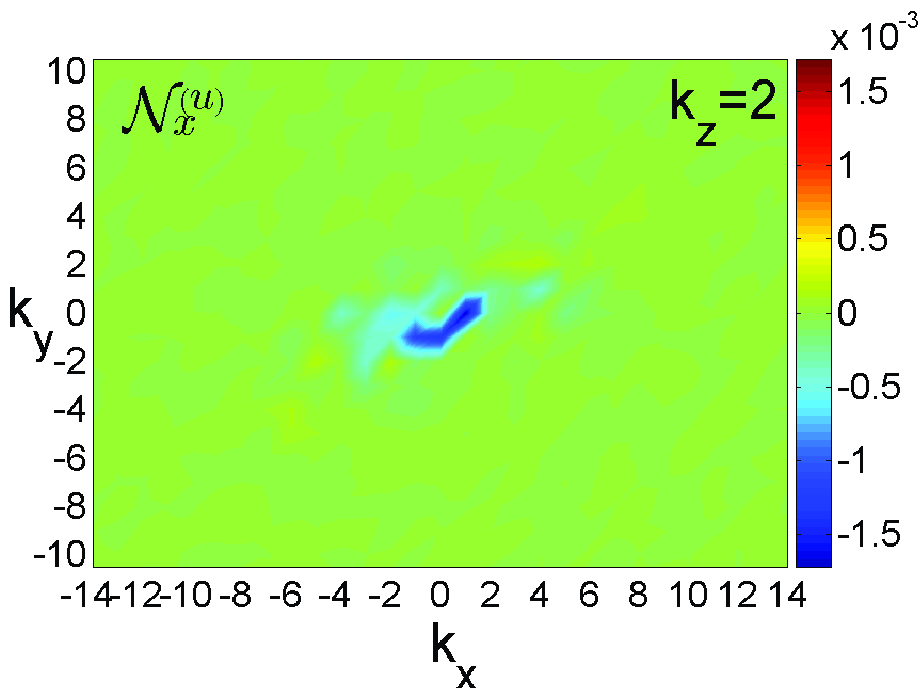}
\includegraphics[width=0.32\textwidth, height=0.2\textwidth]{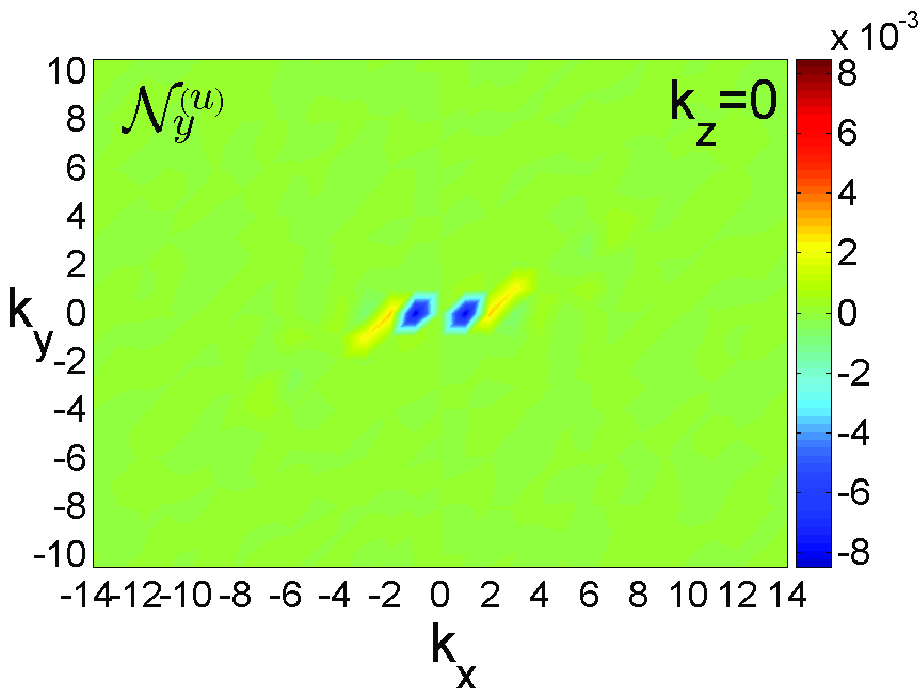}
\includegraphics[width=0.32\textwidth, height=0.2\textwidth]{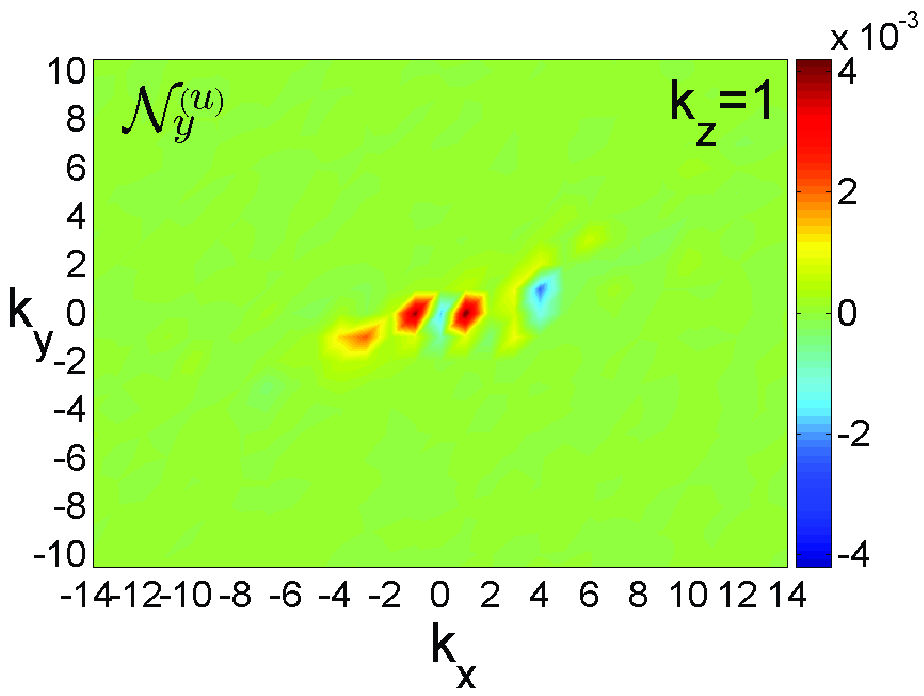}
\includegraphics[width=0.32\textwidth, height=0.2\textwidth]{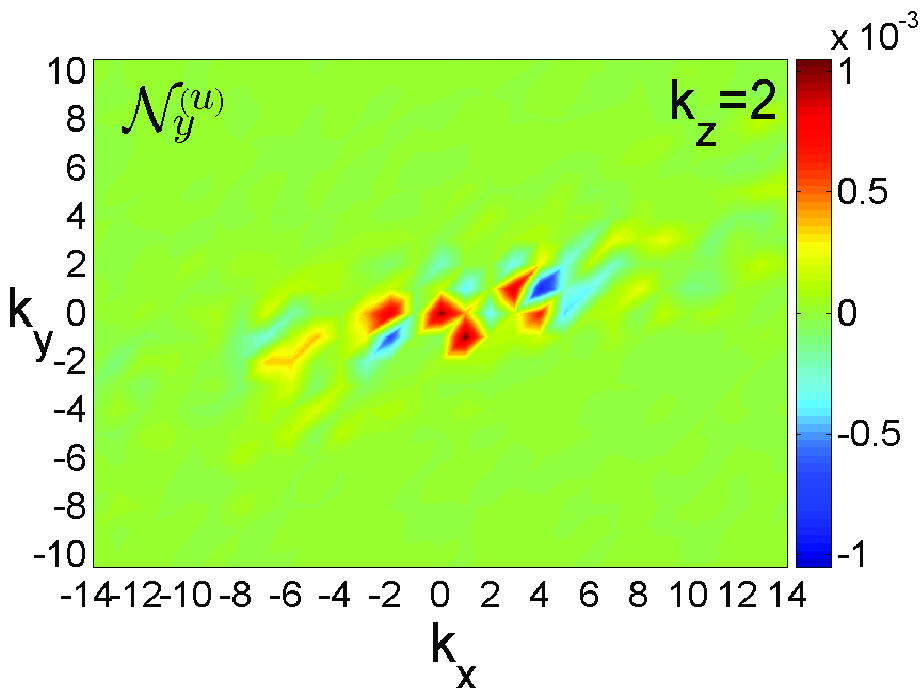}
\caption{Spectra of the nonlinear transfer terms in
$(k_x,k_y)$-plane at $k_z=0 (left), 1 (middle), 2 (right)$ at around
$t=294$, when the channel mode energy has a peak. At $k_z=1$ the
nonlinear transfer terms ${\cal N}_x^{(b)}$, ${\cal N}_y^{(b)}$,
${\cal N}_x^{(u)}$ strongly peak near $k_x=k_y=0$ and are negative,
while ${\cal N}_y^{(u)}$ has a positive peak at $k_x=\pm 1, k_y=0$,
but is negative for the channel mode. So, all these nonlinear terms
act as a sink for the channel mode at its highest point,
transferring its energy to a wide spectrum of the rest modes at
$k_z=0$ and higher $|k_z|>1$; however, the zonal flow does not
receive energy in this process. The yellow and red areas in each
plot show exactly which modes receive energy and hence are being
excited at this time due to the nonlinear transfers as a result of
the decline of the channel mode. This set of modes belong to the
class of \textit{the rest modes}, as discussed in the text. For
$k_z=0,2$ shown here, these transfers are notably anisotropic due to
shear, i.e., depend on the polar angle of wavevector, being mainly
concentrated in the first and third quadrants of $(k_x,k_y)$-plane
and are a consequence of the transverse cascade
process.}\label{fig:nonlinearterms_peak}
\end{figure*}

\begin{figure*}[t!]
\centering
\includegraphics[width=0.32\textwidth, height=0.2\textwidth]{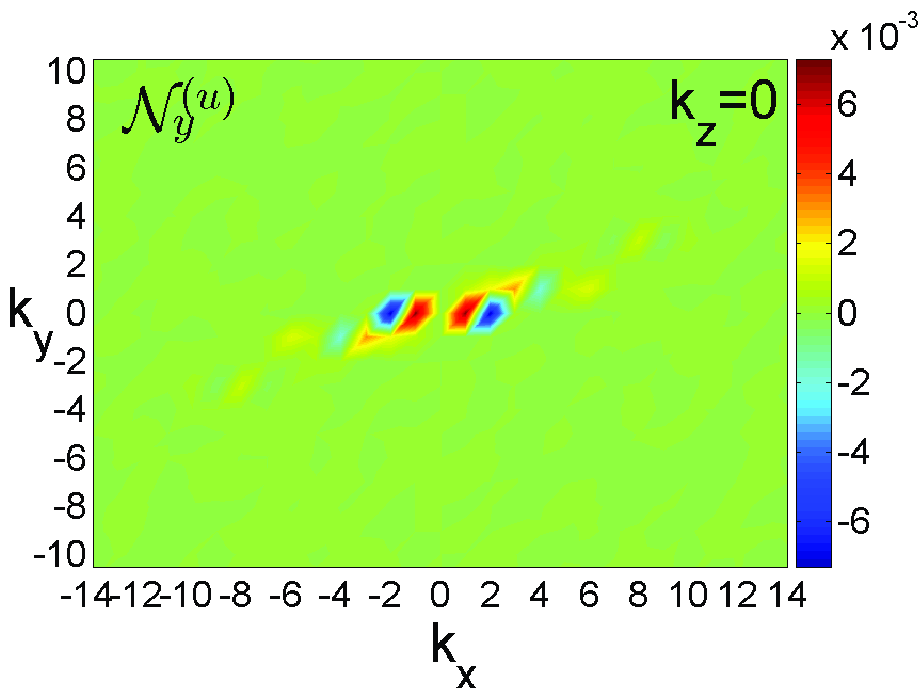}
\includegraphics[width=0.32\textwidth, height=0.2\textwidth]{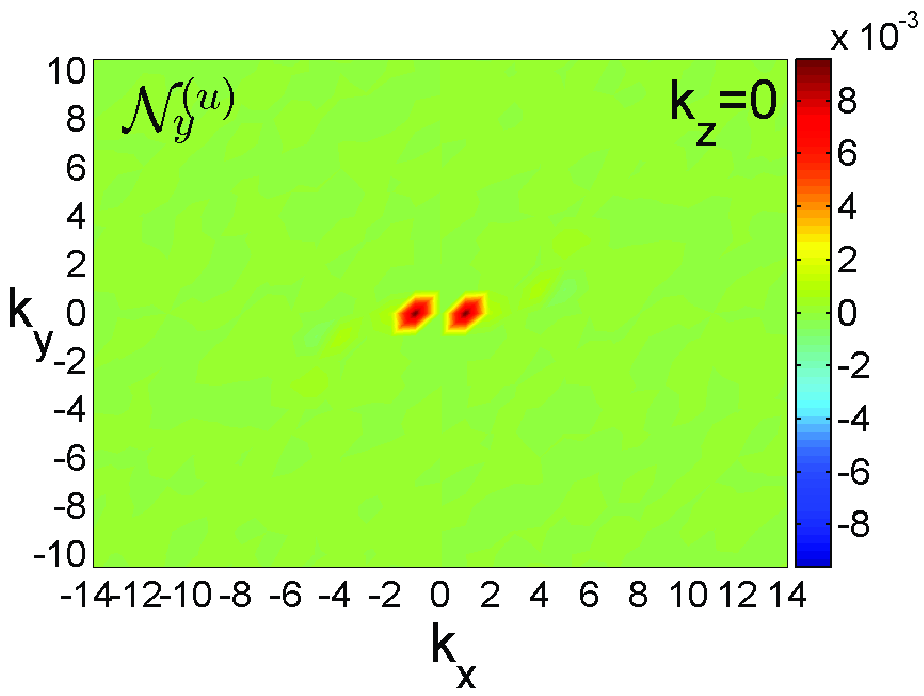}
\includegraphics[width=0.32\textwidth, height=0.2\textwidth]{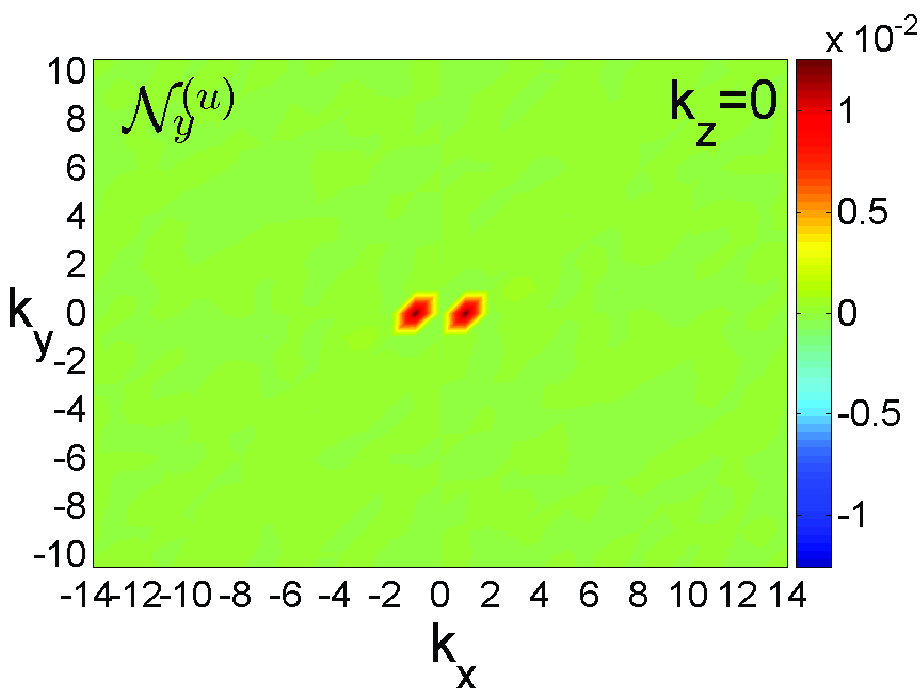}
\includegraphics[width=0.32\textwidth, height=0.2\textwidth]{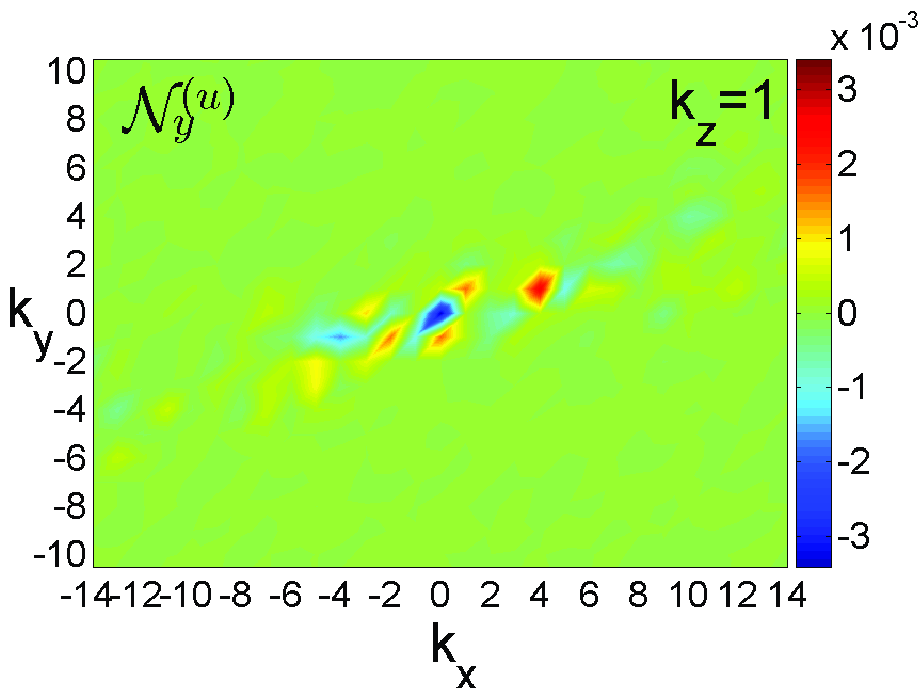}
\includegraphics[width=0.32\textwidth, height=0.2\textwidth]{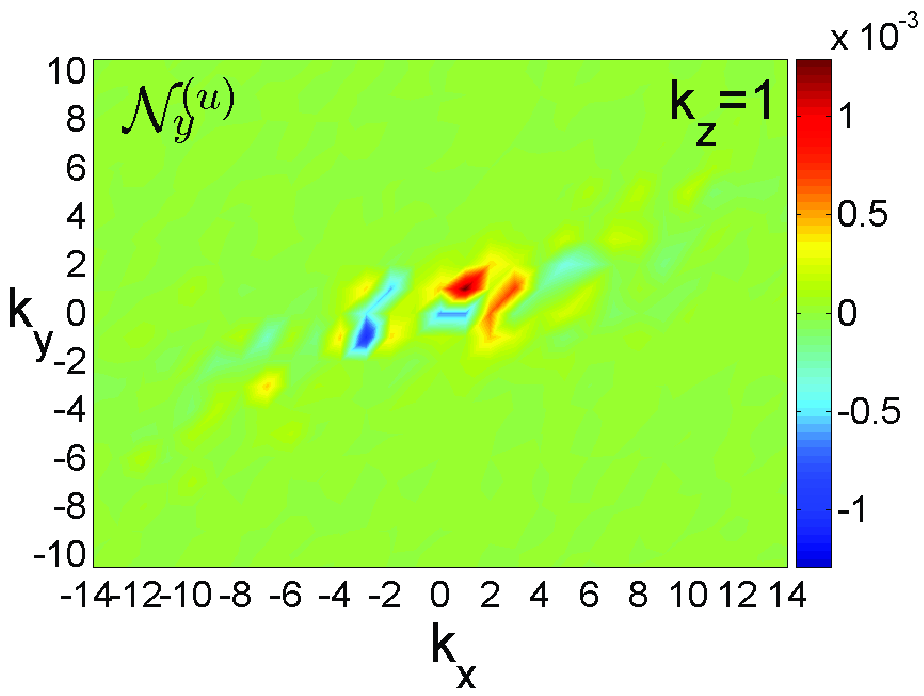}
\includegraphics[width=0.32\textwidth, height=0.2\textwidth]{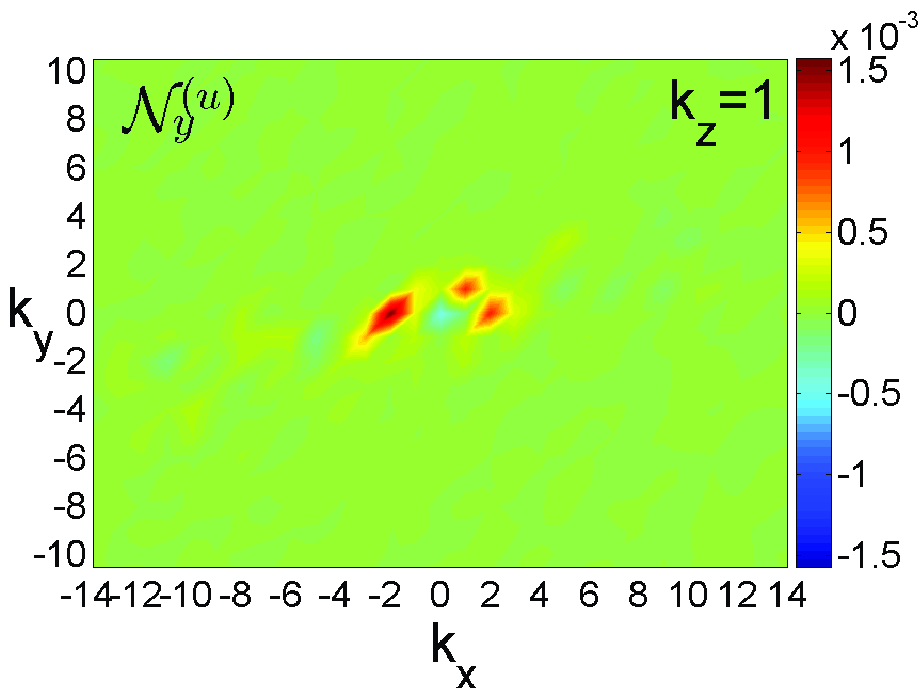}
\caption{The nonlinear transfer term ${\cal N}_y^{(u)}$ in
$(k_x,k_y)$-plane at $k_z=0,1$, when the channel mode energy is at
the minimum, i.e., at the end (``bottom'') of the burst at there
different moments, $t=296, 311, 383$. This term is positive at the
wavenumber of the zonal flow mode, ${\bf k}_{zf}$, but negative at
the wavenumber of the channel mode, ${\bf k}_c$, at all these
moments. Hence, ${\cal N}_y^{(u)}$ at the end of the channel mode
burst acts as the source for the zonal flow mode. In other words,
the formation of the zonal flow mode starts after the end of the
channel mode burst.}\label{fig:nonlinearterms_dip}
\end{figure*}
\begin{figure*}[t!]
\centering
\includegraphics[width=0.32\textwidth, height=0.2\textwidth]{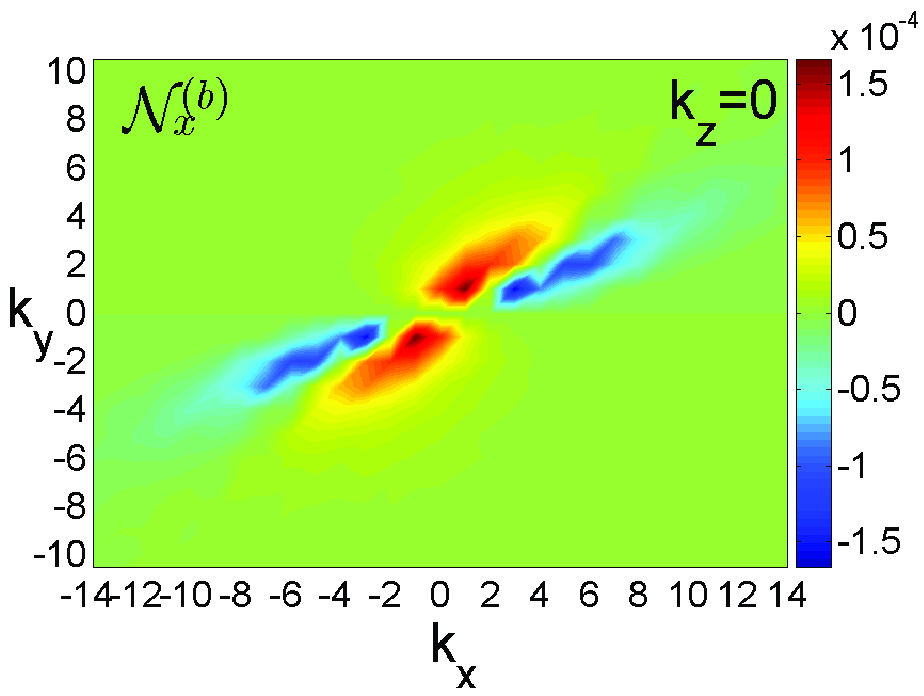}
\includegraphics[width=0.32\textwidth, height=0.2\textwidth]{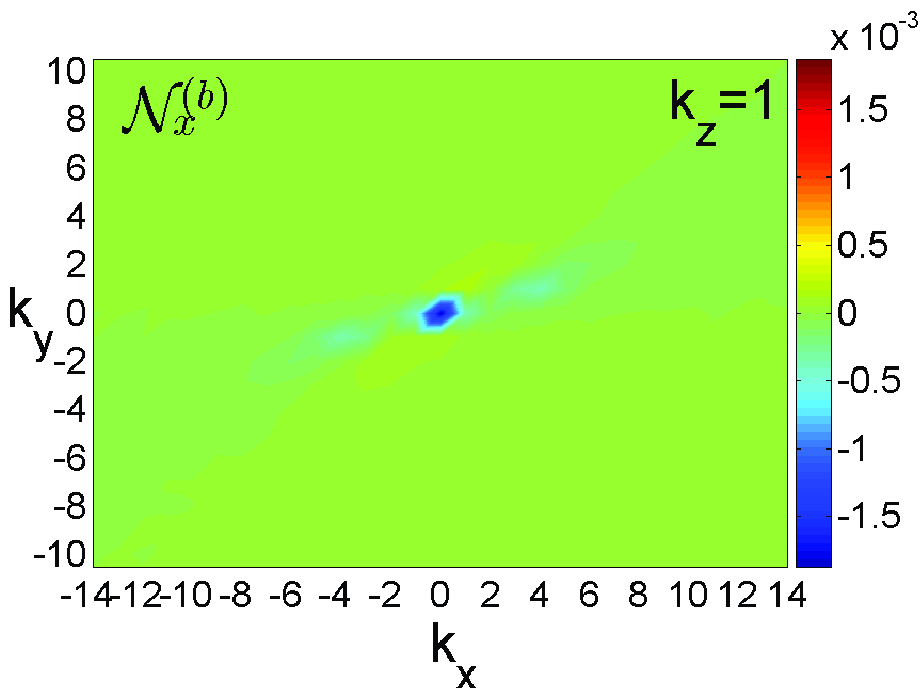}
\includegraphics[width=0.32\textwidth, height=0.2\textwidth]{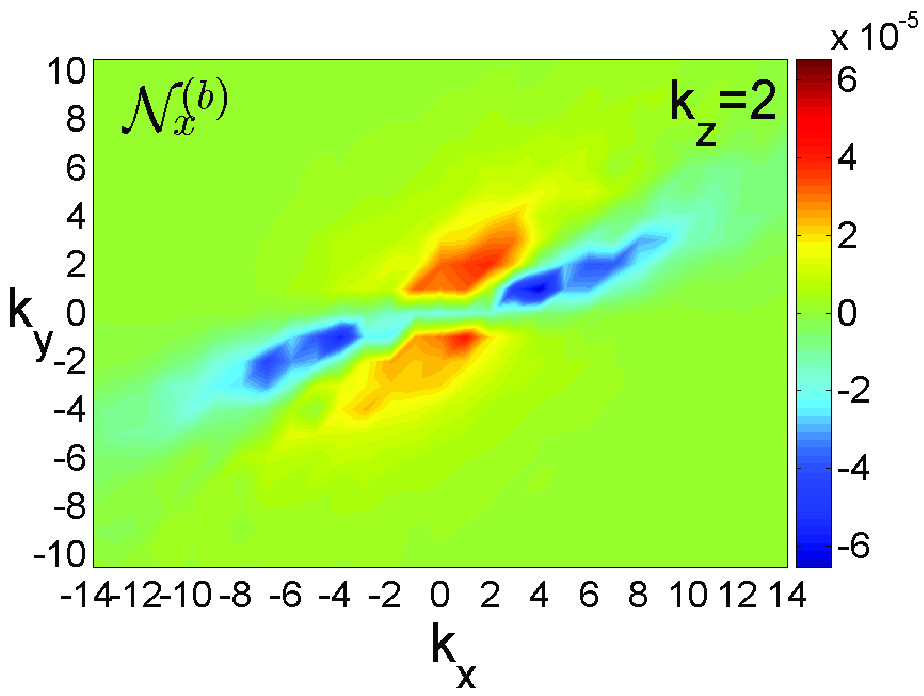}
\caption{Time-averaged nonlinear term ${\cal N}_x^{(b)}$ in
$(k_x,k_y)$-plane at $k_z=0 (left), 1 (middle), 2 (right)$. As in a
single instant shown in Figure \ref{fig:nonlinearterms_peak}, the
action of this term is highly anisotropic due to the shear, i.e.,
strongly depend on the wavevector polar angle. However, this
time-averaged distribution is much smoother than its instantaneous
counterpart and clearly shows transfer of modes over wavevector
angles, that is, \emph{the transverse cascade} of $|\bar{b}_x|^2$
from the blue regions, where ${\cal N}_x^{(b)}<0$ and acts as a sink
for it, to the red and yellow regions, where ${\cal N}_x^{(b)}>0$
and acts as a source/production. In light green regions, these terms
are small, although, as we checked, preserve the same anisotropic
shape. As in Figure \ref{fig:nonlinearterms_peak}, at $k_z=1$,
${\cal N}_x^{(b)}$ peaks again at $k_x=k_y=0$, implying that also in
the time-averaged sense, the dominant nonlinear process at $k_z=1$
is draining of the channel mode energy and transferring it to the
rest modes with other $k_z$ (mainly $k_z=0,1$).}
\label{fig:nonlinearterms_average}
\end{figure*}

\subsection{Interdependence of the channel, zonal flow and rest modes}
\label{sec:Interdependence}

So far we have separately described the dynamics of the main -- the
channel, the zonal flow and the rest modes. Now we move to the
analysis of their interlaced dynamics in Fourier space, which
determines the properties and balances of MRI-turbulence with net
vertical magnetic field. First of all it is a characteristic
burst-like behavior of the volume-averaged energies and stresses
(Figures \ref{fig:timeevolution}), which is related to the
manifestations of these modes' activity and nonlinear interaction.
As was discussed in subsection \ref{sec:Channelmode}, the short
bursts of the channel mode arise as a result of the competition
between the linear nonmodal MRI-amplification and nonlinear
redistribution of the channel mode's energy in ${\bf k}$-space. Each
such burst event starts with the linear amplification, which
initially overwhelms the nonlinear redistribution to other modes and
therefore the energy of the channel mode increases. At the same
time, the effect of the nonlinear terms on the channel mode,
described by ${\cal N}^{(b)}({\bf k}_c), {\cal N}^{(u)}({\bf k}_c)$,
increases as well, as it gradually loses its energy to the rest
modes due to nonlinear transfers. At a certain amplitude of the
channel mode, these nonlinear terms become comparable to the linear
ones that are responsible for MRI, halt the further growth of the
channel mode and start its fast drain (Figure
\ref{fig:channel_dynamics}). This moment, corresponding to the peak
of the channel mode energy and stress, is a critical stage in its
evolution and therefore it is important to examine in more detail
how nonlinear processes redistribute/transfer the energy of the
channel mode in Fourier space at this time. For this purpose, we
present the map of the nonlinear transfer terms in $(k_x,k_y)$-plane
at $k_z=0,1,2$ at two key stages of the burst dynamics -- first when
the channel mode energy has a peak (Figure
\ref{fig:nonlinearterms_peak}) and when its energy is at a minimum
(Figure \ref{fig:nonlinearterms_dip}). The nonlinear redistribution
quite differ from each other at these ``top'' and ``bottom'' points
of the channel mode bursts. In both moments they are notably
anisotropic due to the shear, i.e., depend on a wavenumber polar
angle, being preferably concentrated in the first and third
quadrants, and thus realize the transverse cascade of the modes in
Fourier space. Figure \ref{fig:nonlinearterms_peak}, corresponding
to the ``top'' in the vicinity of the moment $t=294$ of the highest
magnetic energy peak of the channel mode, shows strong negative
peaks of the nonlinear transfer terms ${\cal N}_x^{(b)}$, ${\cal
N}_y^{(b)}$, ${\cal N}_x^{(u)}$ for the channel mode (i.e., at
$k_x=k_y=0, k_z=1$), as the nonlinearity leads its fast drain at
this moment. The red and yellow areas in this figure represent the
subset of the rest modes, mostly with $k_z=0,2$, which receive
energy via nonlinear transfers at this moment of the channel mode's
peak activity. Note that at this time, the nonlinearity transfers
the channel mode energy not to the zonal flow mode (since ${\cal
N}_y^{(u)}({\bf k}_{zf})<0$ at this moment), but to the rest modes.
In other words, the rest modes can be considered as the main
``culprits/parasites'' in the decline of the channel mode in the
developed turbulent state. The action of the above nonlinear terms
in Fourier space during other peaks of the channel mode are
qualitatively similar to that given in Figure
\ref{fig:nonlinearterms_peak} at around $t=294$, however, concretely
which modes, from the total set of the rest modes, receive energy
is, of course, different for each burst.

Thus, this unifying approach -- the analysis of the dynamics and the
nonlinear transfer functions in Fourier space -- enables us to
self-consistently describe the interaction between the channel and
the rest (parasitic) modes when the standard assumptions in the
usual treatment of these modes \citep{Goodman_Xu94,Pessah_Goodman09}
break down: the amplitudes of these two mode types become
comparable, so that one can no longer separate the channel as a
primary background and parasites as small perturbations on top of
that. \emph{Overall, the rest (parasitic) modes play a major role in
the channel mode decline. After each its burst, the transverse
cascade determines a specific spectrum of the parasitic modes in the
turbulent state, which mostly gain energy as a result of draining
the channel mode}.

Quite differently acts the nonlinearity at the ``bottom'' point when
the channel mode is at its minimum. In this case, the main effect is
the production of the zonal flow mode, so that in Figure
\ref{fig:nonlinearterms_dip} we show only the nonlinear transfer
term for the azimuthal velocity, ${\cal N}_y^{(u)}$, at the end
(``bottom'') of the channel burst at there different moments. It is
seen that this term is positive at the wavenumber of the zonal flow
mode, ${\bf k}_{zf}$, at all there moments (red dots at $k_x=\pm 1,
k_y=0$ on the plots with $k_z=0$), acting as the source for the
zonal flow mode. As a result, the formation of the zonal flow mode
starts after the end of the channel mode burst. For the channel mode
wavenumber ${\bf k}_c$, ${\cal N}_y^{(u)}$ is negative at these
moments, acting as a sink. As for the other nonlinear terms at the
``bottom'' points (not shown here), they look qualitatively similar
to those in Figure \ref{fig:nonlinearterms_peak}, except that they
no longer have a prominent negative peak at ${\bf k}_c$, since the
rest and the zonal flow modes dominate instead in the dynamics at
these times.

We have seen from Figures \ref{fig:nonlinearterms_peak} and
\ref{fig:nonlinearterms_dip} that the shear-induced anisotropy of
nonlinear transfers in Fourier space lie at the basis of the
balances of dynamical processes in vertical field MRI-turbulence. In
particular, they determine the drain of the channel mode and the
resulting anisotropic spectrum of the rest modes which gain energy
after each peak of the channel mode. However, the snapshots of the
spectral dynamics of the modes shown in these figures, being at
separate moments, are still irregular, somewhat obscuring this
generic spectral anisotropy of the MRI-turbulence due to the shear.
They convey only a short-time behavior of the turbulence. The
spectral anisotropy of MRI-turbulence with a net vertical field was
demonstrated by \citet{Hawley_etal95,Lesur_Longaretti11} and
recently in more detail by \citet{Murphy_Pessah15}, who, however,
characterized only the anisotropic spectra of the magnetic energy
and Maxwell stress in ${\bf k}$-space, but not the nonlinear
processes. This inherent anisotropic character of the dynamical
process of shear flows in Fourier space is best identified and
described by considering a long-term evolution of the turbulence,
i.e., by averaging the spectral quantities over time, as was done in
the above studies. Such an averaging in the present case with a net
vertical field indeed shows the dominance of the nonlinear
transverse cascade, i.e., the angular (i.e., over wavevector angles)
redistribution of the power in Fourier space in the overall
dynamical balances (nonlinear interactions) between the channel and
the rest modes, which is a main process underlying the turbulence.
Previously, \citet{Lesur_Longaretti11} also investigated the
nonlinear transfers in a net vertical field MRI-turbulence, however,
they applied, together with time-averaging, also shell-averaging of
the linear and nonlinear dynamical terms in ${\bf k}$-space that
naturally smears out the spectral anisotropy and hence the
transverse cascade.

We demonstrate the anisotropic nature of the nonlinear transfers on
longer time-scale on the example of the radial component of the
magnetic field, for which it is most apparent. Figure
\ref{fig:nonlinearterms_average} shows the corresponding nonlinear
term ${\cal N}_x^{(b)}$ in $(k_x,k_y)$-plane at different $k_z=0, 1,
2$ averaged in time from $t=100$ till $t=650$. Like in a single
instant shown in Figure \ref{fig:nonlinearterms_peak}, the action of
this term is highly anisotropic, that is, depends strongly on the
azimuthal angle in $(k_x,k_y)$-plane due to the shear. However, this
time-averaged distribution is much smoother than its instantaneous
counterpart and clearly shows transfer of modes over wavevector
angles, i.e., the transverse cascade of $|\bar{b}_x|^2$ from the
blue regions, where ${\cal N}_x^{(b)}<0$ and acts as a sink for it,
to the red and yellow regions, where ${\cal N}_x^{(b)}>0$ and acts
as a source/production. In light green regions (outside the vital
area) these terms are small, although preserve the same anisotropic
shape. As in Figure \ref{fig:nonlinearterms_peak}, at $k_z=1$,
${\cal N}_x^{(b)}$ has a pronounced negative peak again at
$k_x=k_y=0$ in Figure \ref{fig:nonlinearterms_average}, implying
that in this time-averaged sense too the dominant nonlinear process
at $k_z=1$ is draining of the channel mode energy. This energy is
transferred transversely to the rest modes with other $k_z$ (mainly
$k_z=0,1$) and a range of $k_x,k_y$ indicated by the red and yellow
areas in Figure \ref{fig:nonlinearterms_average} where ${\cal
N}_x^{(b)}>0$. Since time-averaging procedure spans the whole
duration of the turbulence in the simulations, these smooth regions
in fact enclose all the rest modes ever excited during the entire
evolution.

\begin{figure}
\includegraphics[width=\columnwidth, height=0.6\columnwidth]{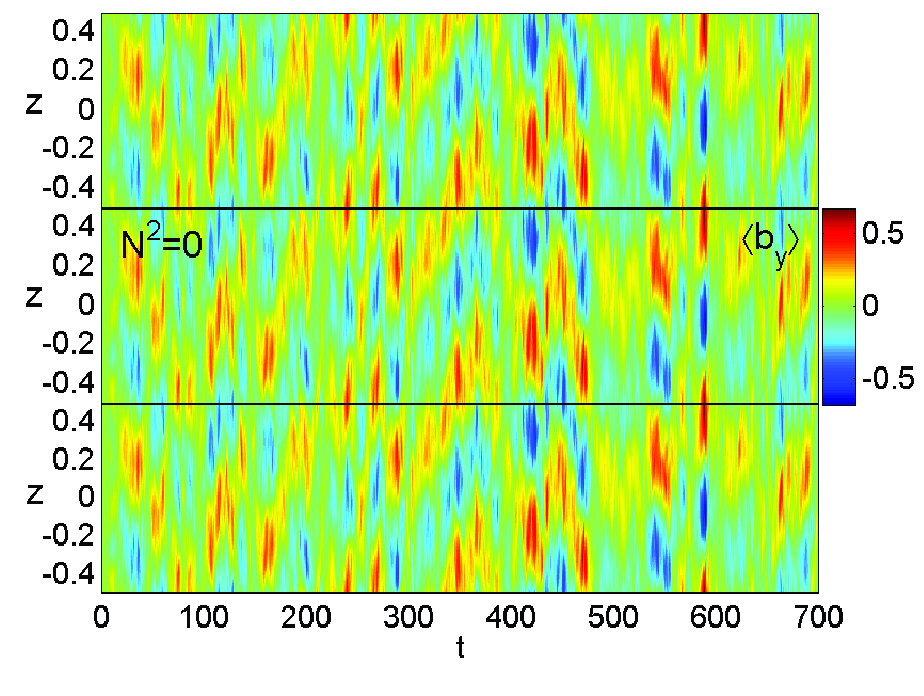}
\includegraphics[width=\columnwidth, height=0.6\columnwidth]{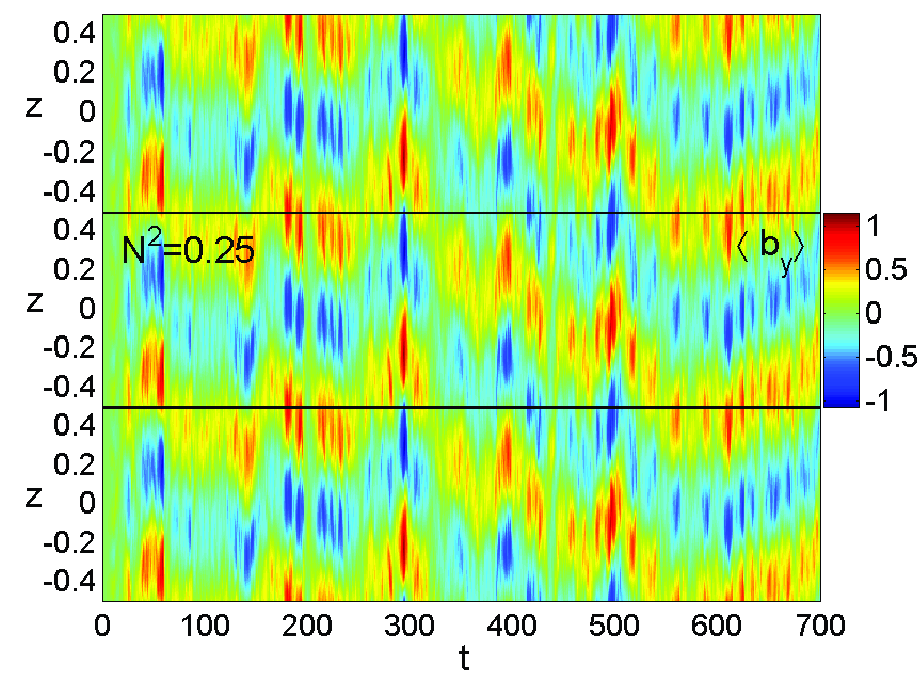}
\caption{Space-time diagram of the azimuthal field averaged in the
horizontal $(x,y)$-plane, $\langle b_y\rangle$, in the unstratified
$N^2=0$ (top) and stratified $N^2=0.25$ (bottom) boxes. In contrast
to the unstratified case, the stratification results in a remarkable
organization of the average azimuthal field, which is dominated by
the $k_z=\pm 1$ modes, into a relatively coherent pattern in the
$(t,z)$-plane with a regular phase variation. Each box has been
stacked with identical boxes from top and bottom to ease perception
of this pattern.} \label{fig:Bya_zt_strat_nostrat}
\end{figure}
\begin{figure}
\includegraphics[width=\columnwidth, height=0.6\columnwidth]{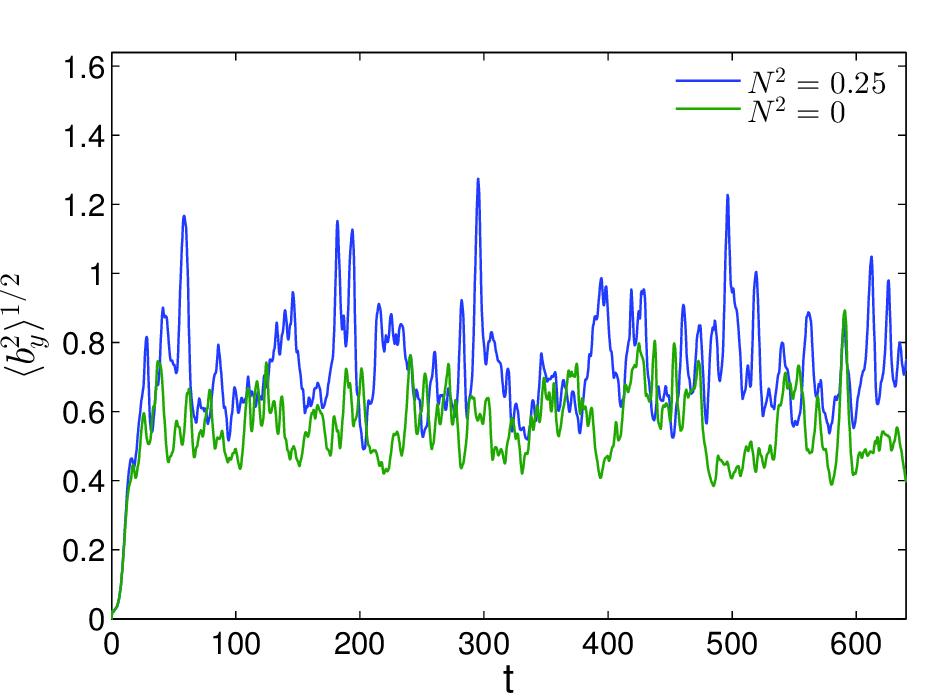}
\includegraphics[width=\columnwidth, height=0.6\columnwidth]{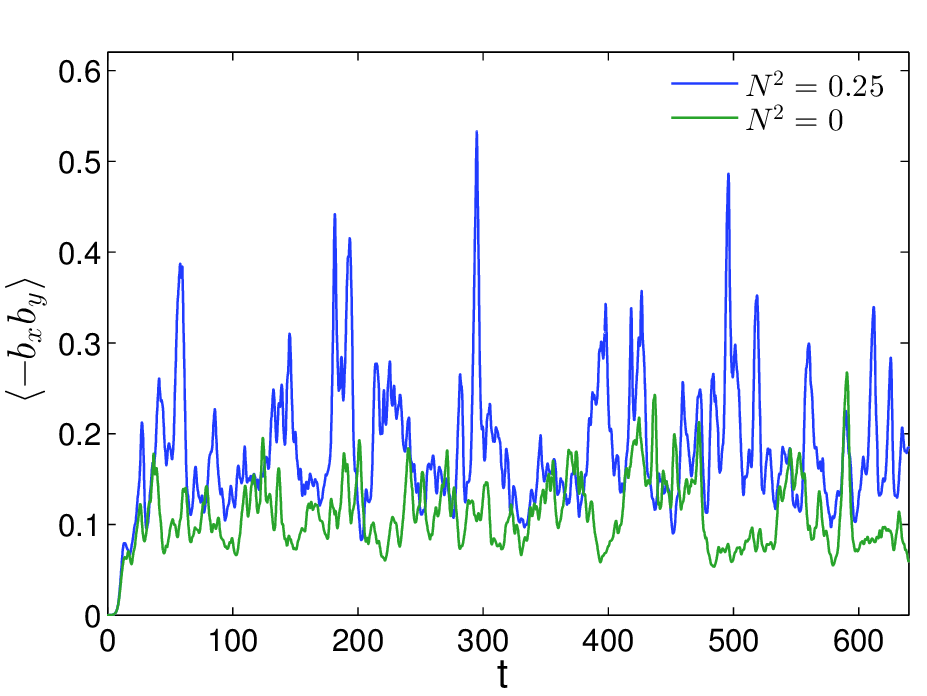}
\caption{Evolution of the volume-averaged rms of $b_y$ (upper panel)
and Maxwell stress $\langle -b_xb_y\rangle$ (lower panel) in the
stratified fiducial run at $N^2=0.25$ (blue) and unstratified
$N^2=0$ run (green). In the presence of stratification, both these
quantities are larger and exhibit more pronounced burst-like
behavior compared to those in the unstratified
case.}\label{fig:by2_strat_nostrat}
\end{figure}

\subsection{On the effect of stratification}
\label{sec:Stratification}

It is well-known from simulations of zero net flux MRI-turbulence
that vertical stratification favors notable dynamo action in
magnetized disks, whereby a large-scale azimuthal magnetic field is
generated and exhibits a ``butterfly'' diagram -- regular
quasi-periodic spatio-temporal oscillations (reversals)
\citep[e.g.,][]{Brandenburg_etal95,Johansen_etal09,Davis_etal10,Gressel10,
Shi_etal10,Simon_etal11,Bodo_etal12,Bodo_etal14,Gressel_Pessah15}.
By contrast, although large-scale azimuthal field generation can
still take place in unstratified zero net flux MRI-turbulence in
vertically sufficiently extended boxes, the spatio-temporal
variation of this field is still more incoherent/irregular in the
form of wandering patches \citep{Lesur_Ogilvie08,Shi_etal16}. A
detailed physical mechanism of how stratification brings about such
an organization of the magnetic field is not fully understood and
requires further studies. This is an important topical question, as
it closely bears on the convergence problem in zero net flux
MRI-turbulence simulations in stratified and unstratified shearing
boxes, which is currently under debate \citep[see e.g.,][and
references therein]{Bodo_etal14,Shi_etal16,Ryan_etal17}. Despite the
fact that the present problem, containing a nonzero net vertical
flux, differs from the case of zero net flux usually considered in
the context of MRI-dynamo, here we also observe a notable
spatio-temporal organization of the large-scale azimuthal field in
the turbulent state (see below).

An ordered cyclic behavior of characteristic quantities in
MRI-turbulence with a net vertical magnetic flux in the shearing box
model, but with more complex isothermal vertical stratification and
outflow boundary conditions in the vertical direction, was also
found. In particular, \citet{Suzuki_Inutsuka09} analyzed
quasi-periodic variations of the horizontally averaged vertical mass
flux (wind) in MRI-turbulent disks and
\citet{Bai_Stone13,Fromang_etal13,Simon_etal13,Salvesen_etal16}
showed a similar oscillatory behavior for mean azimuthal magnetic
field, which is somewhat analogous to that found in our setup.

The considered here disk flow configuration is based on the
Boussinesq approximation, where vertical stratification enters the
governing perturbation Equations
(\ref{eq:App-uxk1})-(\ref{eq:App-uzk1}) in a simpler manner than in
the above studies -- through buoyancy-induced terms, proportional to
the vertically uniform Brunt-V$\ddot{\rm a}$is$\ddot{\rm
a}$l$\ddot{\rm a}$ frequency squared, $N^2$. Still, this
simplification has an advantage in that one could simply put only
$N^2$ to zero in these dynamical equations without altering other
parameters and equilibrium configuration of the system. This allows
us to directly compare to the unstratified case and more
conveniently isolate the basic role of stratification in the dynamo
action. In the linear regime, stratification has only a little
influence on the dynamics of vertical field MRI
\citep{Mamatsashvili_etal13}. Moreover, it is easily seen that the
effect of stratification identically vanishes for the most
effectively amplified channel mode with $k_x=k_y=0$, whose azimuthal
field, equal to the averaged in the horizontal $(x,y)$-plane total
azimuthal field, in fact constitutes the large-scale dynamo field.
So, the effect of stratification on the dynamo action is solely
attributable to nonlinearity, which modifies the electromotive
force.

Figure \ref{fig:Bya_zt_strat_nostrat} shows the space-time diagrams
of the horizontally averaged azimuthal magnetic field, $\langle
b_y\rangle$, in the stratified fiducial case with $N^2=0.25$ as used
in the paper and unstratified, $N^2=0$, case for the same other
parameters. It is clearly seen that stratification causes a
remarkable organization of the azimuthal field into a coherent
wave-like pattern in contrast to the unstratified case. In both
cases, however, this horizontally averaged field is dominated by the
$k_z=\pm 1$ harmonics,\footnote{The magnetic field of the
large-scale modes $k_x=k_y=0, k_z=\pm 1$ gives the dominant
contribution to the horizontally averaged dynamo field and therefore
is a central focus of study in the zero net flux MRI-dynamo problem
\citep[see
e.g.,][]{Lesur_Ogilvie08,Davis_etal10,Herault_etal11,Shi_etal16,Riols_etal17}.}
i.e., it is associated with the dominant channel mode analyzed
above, but its phase variation with $t$ and $z$ in the stratified
case is much more regular than that in the unstratified one. The
amplitude of this averaged azimuthal field is about an order of
magnitude larger than that of the background vertical field
$B_{0z}$. Figure \ref{fig:by2_strat_nostrat} shows that the rms of
the azimuthal field and the volume-averaged Maxwell stress in the
stratified case are larger, with stronger/higher and more frequent
bursts, than those in the unstratified case. Clearly, the space-time
diagram shown in Figure \ref{fig:Bya_zt_strat_nostrat} in the
stratified case with periodic vertical boundary conditions differs
from a typical ``butterfly'' observed in the case of isothermal
stratification in the above mentioned studies. This difference can
be attributable to the different nature of the buoyancy term and
vertical boundary conditions used, which jointly cause elevation of
the azimuthal flux from the midplane. Nevertheless, this comparison
clearly demonstrates the capability of stratification to
order/regularize the spatio-temporal variation of the mean azimuthal
field -- the main field component in accretion disk dynamo. Here, we
have presented only a primary manifestation of the effect of
stratification in the net vertical field MRI-turbulence. In its own
right, it is a subject of a detailed and refined investigation.

\section{Summary and discussions}
\label{sec:Conclusion}

In this paper, we investigated the dynamical balances underlying
MRI-driven turbulence in Keplerian disks with a nonzero net vertical
magnetic field and vertically uniform thermal stratification using
shearing box simulations. Focusing on the analysis of the turbulence
dynamics in Fourier (${\bf k}$-) space, we identified three key
types of modes -- \emph{the channel, the zonal flow mode and the
rest modes} -- that are the main ``players'' in the turbulence
dynamics. We described the dynamics of these modes separately and
then their interdependence, which sets the properties of the nonzero
net vertical field MRI-turbulence. The processes of linear origin
are defined primarily by nonmodal, rather than modal, growth of MRI
due to disk flow nonnormality/shear. This is because the dynamical
time of the turbulence is of the order of the orbital/shear time
during which the nonmodal effects are important. In the turbulent
state, higher values of the stresses and magnetic energy fall just
on those active modes that exhibit high nonmodal MRI growth and
\emph{not} on the modally (exponentially) most unstable modes. In
other words, the properties of the turbulence are determined mostly
by nonmodal physics of MRI rather than by the modal one. From all
the active modes, the one that exhibits the maximum nonmodal growth
is the channel mode, which is horizontally uniform with the largest
vertical scale in the domain. As for the nonlinear processes, it can
be confidently stated that the decisive agent in forming and
maintaining the statistical characteristics of the net vertical
field MRI-turbulence is the transverse cascade -- nonlinear
redistribution of modes in Fourier space that changes the
orientation (angle) of their wavevectors -- arising from the
presence of the shear and hence being a generic phenomenon in shear
flows. Specifically, the nonlinear transverse cascade redistributes
the energy of the channel mode to the rest modes and, subsequently,
the energy of the rest modes to the zonal flow mode. (One has to
note that the rest modes receive energy not only due to the
nonlinear transfers, but they themselves also undergo nonmodal
transient growth, however, less than the channel mode does). The
combined action of these linear and nonlinear processes leads to the
channel mode exhibiting recurrent bursts of the energy and stresses.
The nonlinear transfer of its energy to the rest modes causes the
decline of the channel mode after each burst and subsequent abrupt
increase of the energy of the rest modes. This, in turn, induces
similar, burst-like evolution of the integral characteristics of the
turbulence -- the total volume-averaged energies, stresses and
transport.

The rest modes here, playing the main role in draining the channel
mode, were referred to as parasitic modes in previous studies of net
vertical field MRI. However, there is an important distinction.
These parasitic modes are often assumed to be small compared with
the channel mode and are treated as linear perturbations imposed on
the latter
\citep{Goodman_Xu94,Pessah_Goodman09,Latter_etal09,Latter_etal10,Pessah10}.
Besides, the effect of shear and hence the nonmodal (transient)
physics are neglected with respect to parasitic modes. In the
turbulent state, however, the rest modes can reach energies
comparable to the channel mode, as has been demonstrated in this
paper, so one can no longer separate the channel as a primary
background and parasites as small perturbations on top of that. As a
result, the complex interaction between these two mode types belongs
to the domain of nonlinearity. Our general/unifying approach -- the
analysis of the dynamics in 3D Fourier space -- allows us to
self-consistently characterize this interaction of the modes. In
this case, the transverse cascade defines the spectrum of the rest
modes, which contribute to the decline of the channel mode after
each burst and subsequently acquire energy in this process.

As we found in this study, the net vertical field MRI-turbulence is
robust and, in addition, multifarious -- determined by the
interdependent/interlaced dynamics of three qualitatively different
modes. Consequently, in order to properly quantify the relative
contribution of each of these modes in the turbulence
characteristics, one has to capture the main aspects of their
dynamics in numerical simulations. First of all, this concerns the
selection of relevant sizes (aspect ratio) of the simulation box
(which is actually arbitrary in the shearing box framework) and
resolutions, so that the discrete modes in the selected box densely
enough cover the vital area in $(k_x,k_y)$-plane and maximally
comprise effectively growing (optimal) modes (see Figure
\ref{fig:optimalgrowth}). Besides, one should also avoid
artificial/numerical anisotropyzation of nonlinear processes. As was
shown in Paper I, the anisotropy of the simulation box in
$(k_x,k_y)$-plane introduces artificial anisotropy of nonlinear
processes and somewhat ``deforms'' the overall dynamical picture of
MRI-turbulence in Fourier space. In the present case, this
artificial deformation could result in a change of the relative
importance of the above-classified modes in the overall dynamics,
for instance, could reduce the effectiveness of the channel mode
compared to other modes and hence weaken its manifestation in the
turbulence dynamics \citep[see e.g.,][]{Bodo_etal08}. In particular,
changing the role of the large-scale channel mode likely affects the
generation of the mean azimuthal magnetic field, or the dynamo
action, since this field is directly associated with this mode.
Susceptibility of the latter to specific factors of the dynamics is
clearly shown in Figure \ref{fig:Bya_zt_strat_nostrat}: although the
vertical stratification makes only negligible contribution to the
turbulence energy, it remarkably enhances the generation of the mean
azimuthal magnetic field and regularizes its spatio-temporal
variation. The latter process is significant and represents the
subject of a special detailed investigation.

\subsection{On types of anisotropy}

Apart from the analyzed in our paper shear-induced anisotropy of the
nonlinear processes, a net nonzero background magnetic field itself,
by definition, gives rise to anisotropy of the nonlinear dynamics to
its parallel and perpendicular directions. However, these two types
of anisotropy, having different origin, essentially differ from each
other.

Anisotropy of nonlinear cascade processes due to magnetic field
(directed along $z$-axis) in plasmas with static equilibrium is
analyzed in \cite{Goldreich_Sridhar95}. The source of energy is
assumed to be isotropic in both physical and Fourier spaces with
some outer scale of turbulence, $L$, corresponding to the smallest
wavenumber, $k_0\sim 1/L$. Nonlinear cascade processes take place in
the inertial range, at perpendicular and parallel to the field
wavenumbers $k_{\perp}, k_z\gtrsim k_0$, which is in fact the region
of activity of these processes only. Overall, the regions of the
energy supply and nonlinear cascade in Fourier space are separated
from each other. As a result, the nonlinear processes do not affect
the turbulence energy supply -- they only transfer energy mainly to
small perpendicular (to the magnetic field) wavelength, i.e., to
large $k_{\perp}$ rather than along $k_z$, thereby forming
anisotropic spectrum of the turbulence in $(k_{\perp}, k_z)$-plane.
This nonlinear cascade mostly to large $k_{\perp }$, besides
increasing the magnitude $k=\sqrt{k_{\perp}^2 + k_z^2}$ of the total
wavevector, ${\bf k}=(k_{\perp},k_z)$ (direct cascade), by
definition, implies also a change of its orientation that results in
some angular redistribution of harmonics (i.e., transverse cascade).
At $k_{\perp}/k_z \sim 1$, the direct and transverse cascades are
comparable. As the cascade proceeds, the ratio $k_{\perp }/ k_z$
increases, leading ${\bf k}$ to change primarily in magnitude rather
than in orientation and hence the direct cascade becomes dominant.
In any case, the angular redistribution (transverse cascade) in this
case is of secondary importance -- does not affect the turbulence
energy supply.

A completely different situation arises in the presence of the disk
flow shear. Linear (nonmodal) energy supply processes of
perturbations/turbulence in shear flows are strongly anisotropic in
Fourier space and generally occur over a broad range of wave numbers
without leaving a free room (i.e., inertial range) for the action of
nonlinear processes only. Due to the inherent anisotropy of the
linear dynamics due to the shear, the nonlinear processes in this
range of wavenumbers become strongly anisotropic, i.e., the
transverse cascade is the dominant nonlinear process. Overall,
dynamical picture of the turbulence is formed as a result of the
interplay of the linear and nonlinear (transverse cascade)
processes. The area of the most intensive interplay of the linear
and nonlinear processes we call the vital area of the turbulence. Of
course, the nonlinear processes leads to energy exchange between
different modes, redistributing perturbation energy in Fourier space
while leaving the total energy unchanged. However, the nonlinear
transverse cascade in shear flows repopulates the (transiently)
growing harmonics and, in this way, indirectly contributes to the
energy supply to turbulence. As we have shown above, in the present
problem of net vertical field MRI-turbulence in Keplerian shear
flow, it significantly affects the dynamical ``design'' of the
turbulence and determines the interplay of its ``building blocks''
-- channel, zonal flow and the rest modes. This is clearly
illustrated by Figures
\ref{fig:nonlinearterms_peak}-\ref{fig:nonlinearterms_average},
showing resulting anisotropic dynamics (transfers) in
$(k_x,k_y)$-plane perpendicular to the field. By contrast, in the
absence of basic flow velocity shear, nonlinear cascades in this
plane are isotropic in Goldreich \& Sridhar theory.

\acknowledgments

This project has received funding from the US Department of Energy
under grant No. DE-FG02-04ER54742, the European Union's Horizon 2020
research and innovation programme under the Marie
Sk{\l}odowska-Curie Grant Agreement No. 795158, the Shota Rustaveli
National Science Foundation of Georgia (SRNSFG; grant number
FR17-107), and the Space and Geophysics Laboratory at the University
of Texas at Austin. G.M. acknowledges support from the Alexander von
Humboldt Foundation (Germany). The simulations were performed at the
Texas Advanced Computing Center in Austin (TX) and on the
high-performance Linux cluster Hydra at the Helmholtz-Zentrum
Dresden-Rossendorf (Germany).

\appendix

\section{Perturbation equations in physical and Fourier space}
\label{sec:AppendixA}

From the main Equations (\ref{eq:mom})-(\ref{eq:divb}) one can
derive equations for the perturbations, ${\bf u}, p, {\bf b}$, of
arbitrary amplitude about the Keplerian shear flow ${\bf
U}_0=-q\Omega x{\bf e}_y$ with a net vertical field ${\bf
B}_0=B_{0z}{\bf e}_z$:
\begin{equation}\label{eq:App-ux}
\frac{Du_x}{Dt} = 2\Omega u_y- \frac{1}{\rho_0}\frac{\partial
p}{\partial x}+\frac{B_{0z}}{4\pi\rho_0}\frac{\partial b_x}{\partial
z}+\frac{\partial}{\partial
x}\left(\frac{b_x^2}{4\pi\rho_0}-u_x^2\right)+\frac{\partial}{\partial
y}\left(\frac{b_xb_y}{4\pi\rho_0}-u_xu_y\right)
+\frac{\partial}{\partial
z}\left(\frac{b_xb_z}{4\pi\rho_0}-u_xu_z\right)+\nu\nabla^2u_x,
\end{equation}
\begin{equation}\label{eq:App-uy}
\frac{Du_y}{Dt} = (q-2)\Omega u_x-\frac{1}{\rho_0}\frac{\partial
p}{\partial y}+\frac{B_{0z}}{4\pi\rho_0}\frac{\partial b_y}{\partial
z}+\frac{\partial}{\partial
x}\left(\frac{b_xb_y}{4\pi\rho_0}-u_xu_y\right)+\frac{\partial}{\partial
y}\left(\frac{b_y^2}{4\pi\rho_0}-u_y^2\right)+\frac{\partial}{\partial
z}\left(\frac{b_zb_y}{4\pi\rho_0}-u_zu_y\right)+\nu\nabla^2u_y
\end{equation}
\begin{equation}\label{eq:App-uz}
\frac{Du_z}{Dt} = -\frac{1}{\rho_0}\frac{\partial p}{\partial
z}-\theta+\frac{B_{0z}}{4\pi\rho_0}\frac{\partial b_z}{\partial
z}+\frac{\partial}{\partial
x}\left(\frac{b_xb_z}{4\pi\rho_0}-u_xu_z\right)+\frac{\partial}{\partial
y}\left(\frac{b_yb_z}{4\pi\rho_0}-u_yu_z\right)+\frac{\partial}{\partial
z}\left(\frac{b_z^2}{4\pi\rho_0}-u_z^2\right)+\nu\nabla^2u_z
\end{equation}
\begin{equation}\label{eq:App-theta}
\frac{D\theta}{Dt} = N^2u_z -\frac{\partial}{\partial
x}(u_x\theta)-\frac{\partial}{\partial
y}(u_y\theta)-\frac{\partial}{\partial
z}(u_z\theta)+\chi\nabla^2\theta
\end{equation}
\begin{equation}\label{eq:App-bx}
\frac{Db_x}{Dt}= B_{0z}\frac{\partial u_x}{\partial
z}+\frac{\partial}{\partial y} \left(u_xb_y-u_yb_x
\right)-\frac{\partial}{\partial z}(u_zb_x-u_xb_z)+\eta\nabla^2b_x,
\end{equation}
\begin{equation}\label{eq:App-by}
\frac{Db_y}{Dt}= -q\Omega b_x + B_{0z}\frac{\partial u_y}{\partial
z}-\frac{\partial}{\partial x}
\left(u_xb_y-u_yb_x\right)+\frac{\partial}{\partial
z}(u_yb_z-u_zb_y)+\eta\nabla^2b_y,
\end{equation}
\begin{equation}\label{eq:App-bz}
\frac{Db_z}{Dt}= B_{0z}\frac{\partial u_z}{\partial
z}+\frac{\partial}{\partial x}
\left(u_zb_x-u_xb_z\right)-\frac{\partial}{\partial
y}(u_yb_z-u_zb_y)+\eta\nabla^2b_z,
\end{equation}
\begin{equation}\label{eq:App-divperu}
\frac{\partial u_x}{\partial x}+\frac{\partial u_y}{\partial
y}+\frac{\partial u_z}{\partial z}=0,
\end{equation}
\begin{equation}\label{eq:App-divperb}
\frac{\partial b_x}{\partial x}+\frac{\partial b_y}{\partial
y}+\frac{\partial b_z}{\partial z}=0,
\end{equation}
where $D/Dt=\partial/\partial t-q\Omega x\partial/\partial y$ is the
total derivative along the background flow.

Substituting decomposition (\ref{eq:fourier}) into Equations
(\ref{eq:App-ux})-(\ref{eq:App-divperb}) and taking into account the
normalization, we obtain the evolution equations for the Fourier
modes:
\begin{equation}\label{eq:App-uxk}
\left(\frac{\partial}{\partial t}+qk_y\frac{\partial}{\partial
k_x}\right)\bar{u}_x=2\bar{u}_y-{\rm i}k_x\bar{p}+{\rm
i}k_zB_{0z}\bar{b}_x-\frac{k^2}{\rm Re}\bar{u}_x+{\rm i}k_x
N^{(u)}_{xx}+{\rm i}k_yN^{(u)}_{xy}+{\rm i}k_zN^{(u)}_{xz},
\end{equation}
\begin{equation}\label{eq:App-uyk}
\left(\frac{\partial}{\partial t}+qk_y\frac{\partial}{\partial
k_x}\right)\bar{u}_y=(q-2)\bar{u}_x-{\rm i}k_y\bar{p}+{\rm
i}k_zB_{0z}\bar{b}_y-\frac{k^2}{\rm Re}\bar{u}_y+{\rm
i}k_xN^{(u)}_{xy}+{\rm i}k_yN^{(u)}_{yy}+{\rm i}k_zN^{(u)}_{yz},
\end{equation}
\begin{equation}\label{eq:App-uzk}
\left(\frac{\partial}{\partial t}+qk_y\frac{\partial}{\partial
k_x}\right)\bar{u}_z=-{\rm i}k_z\bar{p}-\bar{\theta}+{\rm
i}k_zB_{0z}\bar{b}_z-\frac{k^2}{\rm Re}\bar{u}_z+{\rm
i}k_xN^{(u)}_{xz}+{\rm i}k_yN^{(u)}_{yz}+{\rm i}k_zN^{(u)}_{zz},
\end{equation}
\begin{equation}\label{eq:App-thetak}
\left(\frac{\partial}{\partial t}+qk_y\frac{\partial}{\partial
k_x}\right)\bar{\theta}=N^2\bar{u}_z-\frac{k^2}{\rm
Pe}\bar{\theta}+{\rm i}k_xN^{(\theta)}_{x}+{\rm
i}k_yN^{(\theta)}_{y}+{\rm i}k_zN^{(\theta)}_{z},
\end{equation}
\begin{equation}\label{eq:App-bxk}
\left(\frac{\partial}{\partial t}+qk_y\frac{\partial}{\partial
k_x}\right)\bar{b}_x={\rm i}k_zB_{0z}\bar{u}_x-\frac{k^2}{\rm
Rm}\bar{b}_x+{\rm i}k_y\bar{F}_z-{\rm i}k_z\bar{F}_y,
\end{equation}
\begin{equation}\label{eq:App-byk}
\left(\frac{\partial}{\partial t}+qk_y\frac{\partial}{\partial
k_x}\right)\bar{b}_y=-q\bar{b}_x+{\rm
i}k_zB_{0z}\bar{u}_y-\frac{k^2}{\rm Rm}\bar{b}_y+{\rm
i}k_z\bar{F}_x-{\rm i}k_x\bar{F}_z
\end{equation}
\begin{equation}\label{eq:App-bzk}
\left(\frac{\partial}{\partial t}+qk_y\frac{\partial}{\partial
k_x}\right)\bar{b}_z={\rm i}k_zB_{0z}\bar{u}_z-\frac{k^2}{\rm
Rm}\bar{b}_z+{\rm i}k_x\bar{F}_y-{\rm i}k_y\bar{F}_x
\end{equation}
\begin{equation}\label{eq:App-divvk}
k_x\bar{u}_x+k_y\bar{u}_y+k_z\bar{u}_z=0,
\end{equation}
\begin{equation}\label{eq:App-divbk}
k_x\bar{b}_x+k_y\bar{b}_y+k_z\bar{b}_z=0,
\end{equation}
where $k^2=k_x^2+k_y^2+k_z^2$ and $B_{0z}=\sqrt{2/\beta}$ is the
normalized constant background field. These spectral equations
involve the linear and nonlinear ($N^{(u)}_{ij}({\bf k},t),
N^{(\theta)}_i({\bf k},t), \bar{F}_i({\bf k},t)$, where $i,j=x,y,z$)
terms that are the Fourier transforms of the corresponding linear
and nonlinear terms of Equations
(\ref{eq:App-ux})-(\ref{eq:App-divperb}) in physical space. These
nonlinear terms are given by convolutions
\begin{equation}\label{eq:App-Nuij}
N^{(u)}_{ij}({\bf k},t)=\int d^3{\bf k'}\left[\bar{b}_i({\bf
k'},t)\bar{b}_j({\bf k}-{\bf k'},t)-\bar{u}_i({\bf
k'},t)\bar{u}_j({\bf k}-{\bf k'},t)\right],
\end{equation}
\begin{equation}\label{eq:App-Nth}
N^{(\theta)}_{i}({\bf k},t)=-\int d^3{\bf k'}\bar{u}_i({\bf
k'},t)\bar{\theta}({\bf k}-{\bf k'},t)
\end{equation}
and $\bar{F}_x, \bar{F}_y, \bar{F}_z$, which are the Fourier
transforms of the respective components of the perturbed
electromotive force ${\bf F}={\bf u}\times {\bf b}$,
\begin{equation}\label{eq:App-Fxk}
\bar{F}_x({\bf k},t)=\int d^3{\bf k'}\left[\bar{u}_y({\bf
k'},t)\bar{b}_z({\bf k}-{\bf k'},t)-\bar{u}_z({\bf
k'},t)\bar{b}_y({\bf k}-{\bf k'},t)\right]
\end{equation}
\begin{equation}\label{eq:App-Fyk}
\bar{F}_y({\bf k},t)=\int d^3{\bf k'}\left[\bar{u}_z({\bf
k'},t)\bar{b}_x({\bf k}-{\bf k'},t)-\bar{u}_x({\bf
k'},t)\bar{b}_z({\bf k}-{\bf k'},t)\right]
\end{equation}
\begin{equation}\label{eq:App-Fzk}
\bar{F}_z({\bf k},t)=\int d^3{\bf k'}\left[\bar{u}_x({\bf
k'},t)\bar{b}_y({\bf k}-{\bf k'},t)-\bar{u}_y({\bf
k'},t)\bar{b}_x({\bf k}-{\bf k'},t)\right]
\end{equation}
describe the effect of nonlinearity on the magnetic field
perturbations. From Equations (\ref{eq:App-uxk})-(\ref{eq:App-uzk})
and the divergence-free conditions (\ref{eq:App-divvk}) and
(\ref{eq:App-divbk}) we can express pressure
\begin{equation}\label{eq:App-pk}
\bar{p}=2{\rm i}(1-q)\frac{k_y}{k^2}\bar{u}_x-2{\rm
i}\frac{k_x}{k^2}\bar{u}_y+{\rm
i}\frac{k_z}{k^2}\bar{\theta}+\sum_{(i,j)=(x,y,z)}\frac{k_ik_j}{k^2}N^{(u)}_{ij}
\end{equation}
Inserting it back into Equations
(\ref{eq:App-uxk})-(\ref{eq:App-uzk}) we have
\begin{equation}\label{eq:App-uxk1}
\left(\frac{\partial}{\partial t}+qk_y\frac{\partial}{\partial
k_x}\right)\bar{u}_x=2\left(1-\frac{k_x^2}{k^2}\right)\bar{u}_y+
2(1-q)\frac{k_xk_y}{k^2}\bar{u}_x+\frac{k_xk_z}{k^2}\bar{\theta}+{\rm
i}k_zB_{0z}\bar{b}_x-\frac{k^2}{\rm Re}\bar{u}_x+Q_x,
\end{equation}
\begin{equation}\label{eq:App-uyk1}
\left(\frac{\partial}{\partial t}+qk_y\frac{\partial}{\partial
k_x}\right)\bar{u}_y=\left[q-2-2(q-1)\frac{k_y^2}{k^2}\right]\bar{u}_x-
2\frac{k_xk_y}{k^2}\bar{u}_y+\frac{k_yk_z}{k^2}\bar{\theta}+{\rm
i}k_zB_{0z}\bar{b}_y-\frac{k^2}{\rm Re}\bar{u}_y+Q_y,
\end{equation}
\begin{equation}\label{eq:App-uzk1}
\left(\frac{\partial}{\partial t}+qk_y\frac{\partial}{\partial
k_x}\right)\bar{u}_z=2(1-q)\frac{k_yk_z}{k^2}\bar{u}_x-2\frac{k_xk_z}{k^2}\bar{u}_y-
\left(1-\frac{k_z^2}{k^2}\right)\bar{\theta}+{\rm
i}k_zB_{0z}\bar{b}_z-\frac{k^2}{\rm Re}\bar{u}_z+Q_z,
\end{equation}
where
\begin{equation}\label{eq:App-Qi}
Q_i={\rm i}\sum_jk_jN^{(u)}_{ij}-{\rm
i}k_i\sum_{m,n}\frac{k_mk_n}{k^2}N^{(u)}_{mn}, ~~~~~ i,j,m,n=x,y,z.
\end{equation}

\bibliographystyle{aasjournal}
\bibliography{biblio}

\providecommand{\noopsort}[1]{}\providecommand{\singleletter}[1]{#1}%
\begin{thebibliography}{}
\expandafter\ifx\csname natexlab\endcsname\relax\def\natexlab#1{#1}\fi
\providecommand{\url}[1]{\href{#1}{#1}}

\bibitem[{{Afshordi} {et~al.}(2005){Afshordi}, {Mukhopadhyay}, \&
  {Narayan}}]{Afshordi_etal05}
{Afshordi}, N., {Mukhopadhyay}, B., \& {Narayan}, R. 2005, \apj, 629, 373

\bibitem[{{Alexakis} {et~al.}(2007){Alexakis}, {Mininni}, \&
  {Pouquet}}]{Alexakis_etal07}
{Alexakis}, A., {Mininni}, P.~D., \& {Pouquet}, A. 2007, New J. Phys., 9, 298

\bibitem[{{Bai} \& {Stone}(2013)}]{Bai_Stone13}
{Bai}, X.-N., \& {Stone}, J.~M. 2013, \apj, 767, 30

\bibitem[{{Bai} \& {Stone}(2014)}]{Bai_Stone14}
---. 2014, \apj, 796, 31

\bibitem[{{Balbus} \& {Hawley}(1991)}]{Balbus_Hawley91}
{Balbus}, S.~A., \& {Hawley}, J.~F. 1991, \apj, 376, 214

\bibitem[{{Balbus} \& {Hawley}(1992)}]{Balbus_Hawley92}
---. 1992, \apj, 400, 610

\bibitem[{{Balbus} \& {Hawley}(1998)}]{Balbus_Hawley98}
---. 1998, Rev. Mod. Phys., 70, 1

\bibitem[{{Bodo} {et~al.}(2012){Bodo}, {Cattaneo}, {Mignone}, \&
  {Rossi}}]{Bodo_etal12}
{Bodo}, G., {Cattaneo}, F., {Mignone}, A., \& {Rossi}, P. 2012, \apj, 761, 116

\bibitem[{{Bodo} {et~al.}(2014){Bodo}, {Cattaneo}, {Mignone}, \&
  {Rossi}}]{Bodo_etal14}
---. 2014, \apjl, 787, L13

\bibitem[{{Bodo} {et~al.}(2008){Bodo}, {Mignone}, {Cattaneo}, {Rossi}, \&
  {Ferrari}}]{Bodo_etal08}
{Bodo}, G., {Mignone}, A., {Cattaneo}, F., {Rossi}, P., \& {Ferrari}, A. 2008,
  \aap, 487, 1

\bibitem[{{Brandenburg} \& {Dintrans}(2006)}]{Brandenburg_Dintrans06}
{Brandenburg}, A., \& {Dintrans}, B. 2006, \aap, 450, 437

\bibitem[{{Brandenburg} {et~al.}(1995){Brandenburg}, {Nordlund}, {Stein}, \&
  {Torkelsson}}]{Brandenburg_etal95}
{Brandenburg}, A., {Nordlund}, A., {Stein}, R.~F., \& {Torkelsson}, U. 1995,
  \apj, 446, 741

\bibitem[{{Chagelishvili} {et~al.}(2003){Chagelishvili}, {Zahn}, {Tevzadze}, \&
  {Lominadze}}]{Chagelishvili_etal03}
{Chagelishvili}, G.~D., {Zahn}, J.-P., {Tevzadze}, A.~G., \& {Lominadze}, J.~G.
  2003, \aap, 402, 401

\bibitem[{{Chandrasekhar}(1960)}]{Chandrasekhar60}
{Chandrasekhar}, S. 1960, Proceedings of the National Academy of Science, 46,
  253

\bibitem[{{Davis} {et~al.}(2010){Davis}, {Stone}, \& {Pessah}}]{Davis_etal10}
{Davis}, S.~W., {Stone}, J.~M., \& {Pessah}, M.~E. 2010, \apj, 713, 52

\bibitem[{{Farrell} \& {Ioannou}(1996)}]{Farrell_Ioannou96}
{Farrell}, B.~F., \& {Ioannou}, P.~J. 1996, Journal of Atmospheric Sciences,
  53, 2025

\bibitem[{{Flock} {et~al.}(2012){Flock}, {Henning}, \& {Klahr}}]{Flock_etal12a}
{Flock}, M., {Henning}, T., \& {Klahr}, H. 2012, \apj, 761, 95

\bibitem[{{Fromang} {et~al.}(2013){Fromang}, {Latter}, {Lesur}, \&
  {Ogilvie}}]{Fromang_etal13}
{Fromang}, S., {Latter}, H., {Lesur}, G., \& {Ogilvie}, G.~I. 2013, \aap, 552,
  A71

\bibitem[{{Fromang} \& {Nelson}(2006)}]{Fromang_Nelson06}
{Fromang}, S., \& {Nelson}, R.~P. 2006, \aap, 457, 343

\bibitem[{{Fromang} \& {Papaloizou}(2007)}]{Fromang_Papaloizou07}
{Fromang}, S., \& {Papaloizou}, J. 2007, \aap, 476, 1113

\bibitem[{{Gogichaishvili} {et~al.}(2017){Gogichaishvili}, {Mamatsashvili},
  {Horton}, {Chagelishvili}, \& {Bodo}}]{Gogichaishvili_etal17}
{Gogichaishvili}, D., {Mamatsashvili}, G., {Horton}, W., {Chagelishvili}, G.,
  \& {Bodo}, G. 2017, \apj, 845, 70 (Paper I)

\bibitem[{{Goldreich} \& {Sridhar}(1995)}]{Goldreich_Sridhar95}
{Goldreich}, P., \& {Sridhar}, S. 1995, \apj, 438, 763

\bibitem[{{Goodman} \& {Xu}(1994)}]{Goodman_Xu94}
{Goodman}, J., \& {Xu}, G. 1994, \apj, 432, 213

\bibitem[{{Gressel}(2010)}]{Gressel10}
{Gressel}, O. 2010, \mnras, 405, 41

\bibitem[{{Gressel} \& {Pessah}(2015)}]{Gressel_Pessah15}
{Gressel}, O., \& {Pessah}, M.~E. 2015, \apj, 810, 59

\bibitem[{{Guan} \& {Gammie}(2011)}]{Guan_Gammie11}
{Guan}, X., \& {Gammie}, C.~F. 2011, \apj, 728, 130

\bibitem[{{Guan} {et~al.}(2009){Guan}, {Gammie}, {Simon}, \&
  {Johnson}}]{Guan_etal09}
{Guan}, X., {Gammie}, C.~F., {Simon}, J.~B., \& {Johnson}, B.~M. 2009, \apj,
  694, 1010

\bibitem[{Hawley {et~al.}(1995)Hawley, Gammie, \& Balbus}]{Hawley_etal95}
Hawley, J., Gammie, C., \& Balbus, S. 1995, \apj, 440, 742

\bibitem[{{Hawley} \& {Balbus}(1991)}]{Hawley_Balbus91}
{Hawley}, J.~F., \& {Balbus}, S.~A. 1991, \apj, 376, 223

\bibitem[{{Hawley} \& {Balbus}(1992)}]{Hawley_Balbus92}
---. 1992, \apj, 400, 595

\bibitem[{{Herault} {et~al.}(2011){Herault}, {Rincon}, {Cossu}, {Lesur},
  {Ogilvie}, \& {Longaretti}}]{Herault_etal11}
{Herault}, J., {Rincon}, F., {Cossu}, C., {et~al.} 2011, \pre, 84, 036321

\bibitem[{{Horton} {et~al.}(2010){Horton}, {Kim}, {Chagelishvili}, {Bowman}, \&
  {Lominadze}}]{Horton_etal10}
{Horton}, W., {Kim}, J.-H., {Chagelishvili}, G.~D., {Bowman}, J.~C., \&
  {Lominadze}, J.~G. 2010, \pre, 81, 066304

\bibitem[{{Johansen} {et~al.}(2009){Johansen}, {Youdin}, \&
  {Klahr}}]{Johansen_etal09}
{Johansen}, A., {Youdin}, A., \& {Klahr}, H. 2009, \apj, 697, 1269

\bibitem[{{Latter} {et~al.}(2015){Latter}, {Fromang}, \&
  {Faure}}]{Latter_etal15}
{Latter}, H.~N., {Fromang}, S., \& {Faure}, J. 2015, \mnras, 453, 3257

\bibitem[{{Latter} {et~al.}(2010){Latter}, {Fromang}, \&
  {Gressel}}]{Latter_etal10}
{Latter}, H.~N., {Fromang}, S., \& {Gressel}, O. 2010, \mnras, 406, 848

\bibitem[{{Latter} {et~al.}(2009){Latter}, {Lesaffre}, \&
  {Balbus}}]{Latter_etal09}
{Latter}, H.~N., {Lesaffre}, P., \& {Balbus}, S.~A. 2009, \mnras, 394, 715

\bibitem[{{Lesur} \& {Longaretti}(2007)}]{Lesur_Longaretti07}
{Lesur}, G., \& {Longaretti}, P.-Y. 2007, \mnras, 378, 1471

\bibitem[{{Lesur} \& {Longaretti}(2011)}]{Lesur_Longaretti11}
---. 2011, \aap, 528, A17

\bibitem[{{Lesur} \& {Ogilvie}(2008)}]{Lesur_Ogilvie08}
{Lesur}, G., \& {Ogilvie}, G.~I. 2008, \aap, 488, 451

\bibitem[{{Lesur} \& {Ogilvie}(2010)}]{Lesur_Ogilvie10}
---. 2010, \mnras, 404, L64

\bibitem[{{Lominadze} {et~al.}(1988){Lominadze}, {Chagelishvili}, \&
  {Chanishvili}}]{Lominadze_etal88}
{Lominadze}, D.~G., {Chagelishvili}, G.~D., \& {Chanishvili}, R.~G. 1988, Sov.
  Astron. Lett., 14, 364

\bibitem[{{Longaretti} \& {Lesur}(2010)}]{Longaretti_Lesur10}
{Longaretti}, P.-Y., \& {Lesur}, G. 2010, \aap, 516, A51

\bibitem[{{Lynden-Bell} \& {Pringle}(1974)}]{Lynden-Bell_Pringle74}
{Lynden-Bell}, D., \& {Pringle}, J.~E. 1974, \mnras, 168, 603

\bibitem[{{Mamatsashvili} {et~al.}(2016){Mamatsashvili}, {Khujadze},
  {Chagelishvili}, {Dong}, {Jim{\'e}nez}, \& {Foysi}}]{Mamatsashvili_etal16}
{Mamatsashvili}, G., {Khujadze}, G., {Chagelishvili}, G., {et~al.} 2016, \pre,
  94, 023111

\bibitem[{{Mamatsashvili} {et~al.}(2013){Mamatsashvili}, {Chagelishvili},
  {Bodo}, \& {Rossi}}]{Mamatsashvili_etal13}
{Mamatsashvili}, G.~R., {Chagelishvili}, G.~D., {Bodo}, G., \& {Rossi}, P.
  2013, \mnras, 435, 2552

\bibitem[{{Mamatsashvili} {et~al.}(2014){Mamatsashvili}, {Gogichaishvili},
  {Chagelishvili}, \& {Horton}}]{Mamatsashvili_etal14}
{Mamatsashvili}, G.~R., {Gogichaishvili}, D.~Z., {Chagelishvili}, G.~D., \&
  {Horton}, W. 2014, \pre, 89, 043101

\bibitem[{{Meheut} {et~al.}(2015){Meheut}, {Fromang}, {Lesur}, {Joos}, \&
  {Longaretti}}]{Meheut_etal15}
{Meheut}, H., {Fromang}, S., {Lesur}, G., {Joos}, M., \& {Longaretti}, P.-Y.
  2015, \aap, 579, A117

\bibitem[{{Murphy} \& {Pessah}(2015)}]{Murphy_Pessah15}
{Murphy}, G.~C., \& {Pessah}, M.~E. 2015, \apj, 802, 139

\bibitem[{{Nauman} \& {Blackman}(2014)}]{Nauman_Blackman14}
{Nauman}, F., \& {Blackman}, E.~G. 2014, \mnras, 441, 1855

\bibitem[{{Papaloizou} \& {Terquem}(1997)}]{Papaloizou_Terquem97}
{Papaloizou}, J.~C.~B., \& {Terquem}, C. 1997, \mnras, 287, 771

\bibitem[{{Pessah}(2010)}]{Pessah10}
{Pessah}, M.~E. 2010, \apj, 716, 1012

\bibitem[{{Pessah} \& {Chan}(2008)}]{Pessah_Chan08}
{Pessah}, M.~E., \& {Chan}, C.-K. 2008, \apj, 684, 498

\bibitem[{{Pessah} \& {Chan}(2012)}]{Pessah_Chan12}
---. 2012, \apj, 751, 48

\bibitem[{{Pessah} {et~al.}(2006){Pessah}, {Chan}, \&
  {Psaltis}}]{Pessah_etal06}
{Pessah}, M.~E., {Chan}, C.-K., \& {Psaltis}, D. 2006, \mnras, 372, 183

\bibitem[{{Pessah} \& {Goodman}(2009)}]{Pessah_Goodman09}
{Pessah}, M.~E., \& {Goodman}, J. 2009, \apjl, 698, L72

\bibitem[{{Razdoburdin} \& {Zhuravlev}(2017)}]{Razdoburdin_Zhuravlev17}
{Razdoburdin}, D.~N., \& {Zhuravlev}, V.~V. 2017, \mnras, 467, 849

\bibitem[{{Riols} {et~al.}(2017){Riols}, {Rincon}, {Cossu}, {Lesur}, {Ogilvie},
  \& {Longaretti}}]{Riols_etal17}
{Riols}, A., {Rincon}, F., {Cossu}, C., {et~al.} 2017, \aap, 598, A87

\bibitem[{{Ryan} {et~al.}(2017){Ryan}, {Gammie}, {Fromang}, \&
  {Kestener}}]{Ryan_etal17}
{Ryan}, B.~R., {Gammie}, C.~F., {Fromang}, S., \& {Kestener}, P. 2017, \apj,
  840, 6

\bibitem[{{Salhi} {et~al.}(2012){Salhi}, {Lehner}, {Godeferd}, \&
  {Cambon}}]{Salhi_etal12}
{Salhi}, A., {Lehner}, T., {Godeferd}, F., \& {Cambon}, C. 2012, \pre, 85,
  026301

\bibitem[{{Salvesen} {et~al.}(2016){Salvesen}, {Simon}, {Armitage}, \&
  {Begelman}}]{Salvesen_etal16}
{Salvesen}, G., {Simon}, J.~B., {Armitage}, P.~J., \& {Begelman}, M.~C. 2016,
  \mnras, 457, 857

\bibitem[{{Sano} \& {Inutsuka}(2001)}]{Sano_Inutsuka01}
{Sano}, T., \& {Inutsuka}, S.-I. 2001, \apjl, 561, L179

\bibitem[{Schmid \& Henningson(2001)}]{Schmid_Henningson01}
Schmid, P.~J., \& Henningson, D.~S. 2001, Stability and Transition in Shear
  Flows (Springer)

\bibitem[{{Shakura} \& {Postnov}(2015)}]{Shakura_Postnov15}
{Shakura}, N., \& {Postnov}, K. 2015, \mnras, 448, 3697

\bibitem[{{Shakura} \& {Sunyaev}(1973)}]{Shakura_Sunyaev73}
{Shakura}, N.~I., \& {Sunyaev}, R.~A. 1973, \aap, 24, 337

\bibitem[{{Shi} {et~al.}(2010){Shi}, {Krolik}, \& {Hirose}}]{Shi_etal10}
{Shi}, J.-M., {Krolik}, J.~H., \& {Hirose}, S. 2010, \apj, 708, 1716

\bibitem[{{Shi} {et~al.}(2016){Shi}, {Stone}, \& {Huang}}]{Shi_etal16}
{Shi}, J.-M., {Stone}, J.~M., \& {Huang}, C.~X. 2016, \mnras, 456, 2273

\bibitem[{{Shtemler} {et~al.}(2011){Shtemler}, {Mond}, \&
  {Liverts}}]{Shtemler_etal11}
{Shtemler}, Y.~M., {Mond}, M., \& {Liverts}, E. 2011, \mnras, 413, 2957

\bibitem[{{Simon} \& {Armitage}(2014)}]{Simon_Armitage14}
{Simon}, J.~B., \& {Armitage}, P.~J. 2014, \apj, 784, 15

\bibitem[{{Simon} {et~al.}(2013){Simon}, {Bai}, {Armitage}, {Stone}, \&
  {Beckwith}}]{Simon_etal13}
{Simon}, J.~B., {Bai}, X.-N., {Armitage}, P.~J., {Stone}, J.~M., \& {Beckwith},
  K. 2013, \apj, 775, 73

\bibitem[{{Simon} {et~al.}(2012){Simon}, {Beckwith}, \&
  {Armitage}}]{Simon_etal12}
{Simon}, J.~B., {Beckwith}, K., \& {Armitage}, P.~J. 2012, \mnras, 422, 2685

\bibitem[{{Simon} \& {Hawley}(2009)}]{Simon_Hawley09}
{Simon}, J.~B., \& {Hawley}, J.~F. 2009, \apj, 707, 833

\bibitem[{{Simon} {et~al.}(2009){Simon}, {Hawley}, \&
  {Beckwith}}]{Simon_etal09}
{Simon}, J.~B., {Hawley}, J.~F., \& {Beckwith}, K. 2009, \apj, 690, 974

\bibitem[{{Simon} {et~al.}(2011){Simon}, {Hawley}, \&
  {Beckwith}}]{Simon_etal11}
---. 2011, \apj, 730, 94

\bibitem[{{Squire} \& {Bhattacharjee}(2014)}]{Squire_Bhattacharjee14}
{Squire}, J., \& {Bhattacharjee}, A. 2014, \apj, 797, 67

\bibitem[{{Suzuki} \& {Inutsuka}(2009)}]{Suzuki_Inutsuka09}
{Suzuki}, T.~K., \& {Inutsuka}, S.-I. 2009, \apjl, 691, L49

\bibitem[{{Tevzadze} {et~al.}(2008){Tevzadze}, {Chagelishvili}, \&
  {Zahn}}]{Tevzadze_etal08}
{Tevzadze}, A.~G., {Chagelishvili}, G.~D., \& {Zahn}, J.-P. 2008, \aap, 478, 9

\bibitem[{{Velikhov}(1959)}]{Velikhov59}
{Velikhov}, E. 1959, Zh. Eksp. Teor. Fiz., 36, 1398

\bibitem[{{Verma}(2004)}]{Verma04}
{Verma}, M.~K. 2004, Phys. Rep., 401, 229

\bibitem[{{Walker} {et~al.}(2016){Walker}, {Lesur}, \&
  {Boldyrev}}]{Walker_etal16}
{Walker}, J., {Lesur}, G., \& {Boldyrev}, S. 2016, \mnras, 457, L39

\bibitem[{{Wardle}(1999)}]{Wardle99}
{Wardle}, M. 1999, \mnras, 307, 849

\bibitem[{{Yecko}(2004)}]{Yecko04}
{Yecko}, P.~A. 2004, \aap, 425, 385

\bibitem[{{Zhuravlev} \& {Razdoburdin}(2014)}]{Zhuravlev_Razdoburdin14}
{Zhuravlev}, V.~V., \& {Razdoburdin}, D.~N. 2014, \mnras, 442, 870

\end{thebibliography}
\end{document}